# An Exhaustive Survey on P4 Programmable Data Plane Switches: Taxonomy, Applications, Challenges, and Future Trends

**ELIE F. KFOURY[1], (Student Member, IEEE), JORGE CRICHIGNO[1], (Member, IEEE) AND ELIAS BOU-HARB[2], (Member, IEEE)**

[1]College of Engineering and Computing, University of South Carolina, Columbia, SC 29201 USA (e-mail: ekfoury@email.sc.edu, jcrichigno@cec.sc.edu)

[2]The Cyber Center For Security and Analytics, University of Texas at San Antonio, TX 78249 USA, CO 80523 USA (e-mail: elias.bouharb@utsa.edu)

Corresponding author: Elie F. Kfoury (e-mail: ekfoury@email.sc.edu).

This material is based upon work supported by the National Science Foundation under grant numbers 1925484 and 1829698, funded by the Office of Advanced Cyberinfrastructure (OAC).

**ABSTRACT** Traditionally, the data plane has been designed with fixed functions to forward packets using a small set of protocols. This closed-design paradigm has limited the capability of the switches to proprietary implementations which are hard-coded by vendors, inducing a lengthy, costly, and inflexible process. Recently, data plane programmability has attracted significant attention from both the research community and the industry, permitting operators and programmers in general to run customized packet processing functions. This open-design paradigm is paving the way for an unprecedented wave of innovation and experimentation by reducing the time of designing, testing, and adopting new protocols; enabling a customized, top-down approach to develop network applications; providing granular visibility of packet events defined by the programmer; reducing complexity and enhancing resource utilization of the programmable switches; and drastically improving the performance of applications that are offloaded to the data plane. Despite the impressive advantages of programmable data plane switches and their importance in modern networks, the literature has been missing a comprehensive survey. To this end, this paper provides a background encompassing an overview of the evolution of networks from legacy to programmable, describing the essentials of programmable switches, and summarizing their advantages over Software-defined Networking (SDN) and legacy devices. The paper then presents a unique, comprehensive taxonomy of applications developed with P4 language; surveying, classifying, and analyzing more than 200 articles; discussing challenges and considerations; and presenting future perspectives and open research issues.

## I. INTRODUCTION

SINCE the emergence of the world wide web and the explosive growth of the Internet in the 1990s, the networking industry has been dominated by closed and proprietary hardware and software. Consider the observations made by McKeown [1] and the illustration in Fig. 1, which shows the cumulative number of Request For Comments (RFCs) [2]. While at first an increase in RFCs may appear encouraging, it has actually represented an entry barrier to the network market. The progressive reduction in the flexibility of protocol design caused by standardized requirements, which cannot be easily removed to enable protocol changes, has perpetuated the status quo. This protocol ossification [3, 4] has been characterized by a slow innovation pace at the hand of few network vendors. As an example, after being initially conceived by Cisco and VMware [5], the Application Specific Integrated Circuit (ASIC) implementation of the Virtual Extensible LAN (VXLAN) [6], a simple frame encapsulation protocol, took several years, a process that could have been reduced to weeks by software implementations[1].

Protocol ossification has been challenged first by Software-defined Networking (SDN) [7, 8] and then by the recent advent of programmable switches. SDN fostered major advances by explicitly separating the control and data planes, and by implementing the control plane intelligence as a software outside of the switches. While SDN reduced network complexity and spurred control plane innovation at the speed of software development, it did not wrest control of the actual packet processing functions away from network vendors. Traditionally, the data plane has been designed with fixed functions to forward packets using a small set of protocols (e.g., IP, Ethernet). The design cycle of switch ASICs has been characterized by a lengthy, closed, and proprietary process that usually takes years. Such process contrasts with the agility of the software industry.

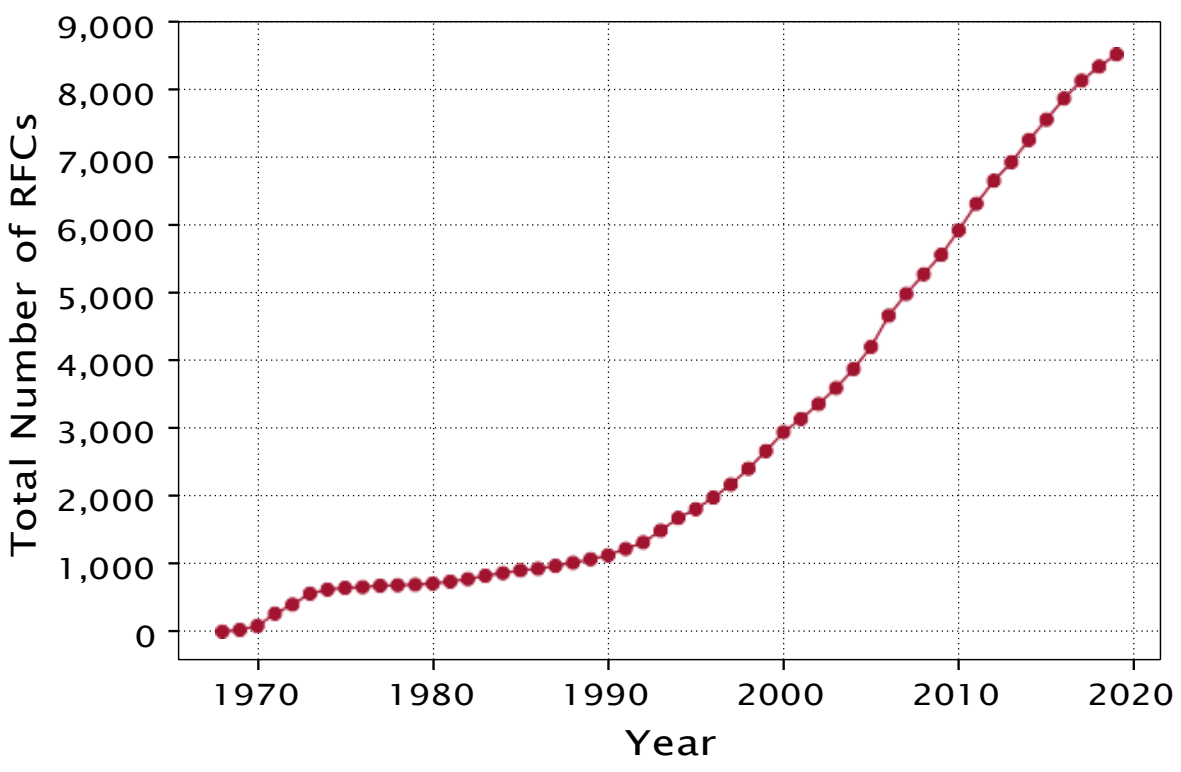

**FIGURE 1:** Cumulative number of RFCs.

The programmable forwarding can be viewed as a natural evolution of SDN, where the software that describes the behavior of how packets are processed can be conceived, tested, and deployed in a much shorter time span by operators, engineers, researchers, and practitioners in general. The de-facto standard for defining the forwarding behavior is the P4 language [9], which stands for Programming Protocol-independent Packet Processors. Essentially, P4 programmable switches have removed the entry barrier to network design, previously reserved to network vendors.

The momentum of programmable switches is reflected in the global ecosystem around P4. Operators such as ATT [10], Comcast [11], NTT [12], KPN [13], Turk Telekom [14], Deutsche Telekom [15], and China Unicom [14], are now using P4-based platforms and applications to optimize their networks. Companies with large data centers such as Facebook [16], Alibaba [17], and Google [18] operate on programmable platforms running customized software, a contrast from the fully proprietary implementations of just a few years ago [19]. Switch manufacturers such as Edgecore [20], Stordis [21], Cisco [22], Arista [23], Juniper [24], and Interface Masters [25] are now manufacturing P4 programmable switches with multiple deployment models, from fully programmable or white boxes to hybrid schemes. Chip manufactures such as Barefoot Networks (Intel) [26], Xilinx [27], Pensando [28], Mellanox [29], and Innovium [30] have embraced programmable data planes without compromising performance. The availability of tools and the agility of software development have opened an unprecedented possibility of experimentation and innovation by enabling network owners to build custom protocols and process them using protocol-independent primitives, reprogram the data plane in the field, and run P4 codes on diverse platforms. Main agencies supporting engineering research and education worldwide are investing in programmable networks as well. For example, the U.S. National Science Foundation (NSF) has funded FABRIC [31, 32], a national research backbone based on P4 programmable switches. Another project funded by the NSF operates an international Software Defined Exchange (SDX) which includes a P4 testbed that enables international research and education institutions to share P4 resources [33]. Similarly, an European consortium has recently built 2STiC [34], a P4 programmable network that interconnects universities and research centers.

[1] The RFC and VXLAN observations are extracted from Dr. McKeown's presentation in [1].

### A. CONTRIBUTION

Despite the increasing interest on P4 switches, previous work has only partially covered this technology. As shown in Table 1, currently, there is no updated and comprehensive material. Thus, this paper addresses this gap by providing an overview of the evolution of networks from legacy to programmable; describing the essentials of programmable switches and P4; and summarizing the advantages of programmable switches over SDN and legacy devices. The paper continues by presenting a taxonomy of applications developed with P4; surveying, classifying, and analyzing and comparing more than 200 articles; discussing challenges and considerations; and putting forward future perspectives and open research issues.

### B. PAPER ORGANIZATION

The road-map of this survey is illustrated in Fig. 2. Section II studies and compares existing surveys on various P4-related topics and demonstrates the added value of the offered work. Section III describes the traditional and SDN devices, and the evolution toward programmable data planes. Section IV introduces programmable switches and their features and explains the Protocol Independent Switch Architecture (PISA), a pipeline forwarding model. Section V describes the survey methodology and the proposed taxonomy. Subsequent sections (from Section VI to Section XII) explore the works pertaining to various categories proposed in the taxonomy, and compare the P4 approaches in each category, as well as with the legacy-enabled solutions. Section XIII outlines challenges and considerations extracted and induced from the literature, and pinpoints directions that can be explored in the future to ameliorate the state-of-the-art solutions. Finally, Section XIV concludes the survey. The abbreviations used in this article are summarized in Table 36, at the end of the article.

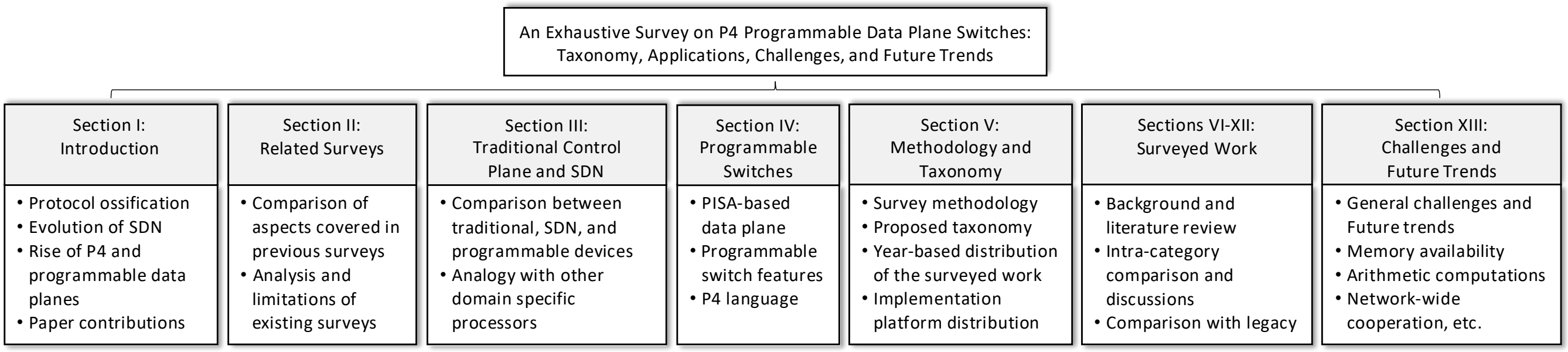

**FIGURE 2:** Paper roadmap.

## II. RELATED SURVEYS

The advantages of programmable switches attracted considerable attention from the research community. They were described in previous surveys.

Stubbe et al. [35] discussed various P4 compilers and interpreters in a short survey. This work provided a short background on the P4 language and demonstrated the main building blocks that describe packet processing in a programmable switch. It outlined reference hardware and software programmable switch implementations. The survey lacks critical discussions on the evolution of programmable switches, the features of P4 language, the existing applications, challenges, and the potential future work.

Dargahi et al. [36] focused on stateful data planes and their security implications. There are two main objectives of this survey. First, it introduces the reader to recent trends and technologies pertaining to stateful data planes. Second, it discusses relevant security issues by analyzing selected use cases. The scope of the survey is not limited to P4 for programming the data plane. Instead, it describes other schemes such as OpenState [44], Flow-level State Transitions (FAST) [45], etc. When reviewing the security properties of stateful data planes, the authors described a mapping between potential attacks and corresponding vulnerabilities. The survey lacks critical discussions on the P4 language and its features, the existing applications beyond security, the challenges, and the potential future work.

Cordeiro et al. [37] discussed the evolution of SDN from OpenFlow to data plane programmability. The survey briefly explained the layout of a P4 program and how it is mapped to the abstract forwarding model. It then listed various compilers, tools, simulators, and frameworks for P4 development. The authors categorized the literature into two categories: 1) programmable security and dependability management; 2) enhanced accounting and performance management. In the first category, the authors listed works pertaining to policy modeling, analysis, and verification, as well as intrusion detection and prevention, and network survivability. In the second category, the authors focused on network monitoring, traffic engineering, and load balancing. The survey only lists a limited set of papers without providing much details or how papers differ from each other. Moreover, the survey was published in 2017, and since then, a significant percentage of P4-related works are missing.

Satapathy et al. [38] presented a limited description about the pitfalls of traditional networks and the evolution of SDN. The report briefly described elements of the P4 language. The authors then discussed the control plane and P4Runtime [46], and enumerated three use cases of P4 applications. The report concludes with potential future work. This work lacks critical discussions on the P4 language and its features, the existing applications, and challenges.

The short survey presented by Bifulco et al. [39] reviews the trends and issues of abstractions and architectures that realize programmable networks. The authors discussed the motivation of packet processing devices in the networking

**TABLE 1:** Comparison with related surveys.

| Paper | Programmable switches and P4 language | | | Taxonomy | | | | Discussions | |
|---|---|---|---|---|---|---|---|---|---|
| | Evolution | Description | Features | Background | Literature | Intra-category comparison | Comparison with legacy | Challenges | Future directions |
| [35] | | | | | | | | | |
| [36] | ⓾ | | | | | | | | |
| [37] | ⓾ | | | | ⓾ | | | | |
| [38] | | | | | | | | | |
| [39] | ⓾ | | | | | | | | |
| [40] | ⓾ | | | | | | | | |
| [41] | ⓾ | | | ⓾ | | | | | |
| [42] | | | | | | | | | |
| [43] | ⓾ | | | | | | | | |
| This paper | ⓾ | ⓾ | ⓾ | ⓾ | ⓾ | ⓾ | ⓾ | ⓾ | ⓾ |

⓾ Covered in this survey   Not covered in this survey   Partially covered in this survey field and described the anatomy of a programmable switch. The proposed taxonomy categorizes the literature as state-based, abstraction-based, implementation-based, and layer-based. The layer-based consists of control/intent layer and data plane layer; the implementation-based encompasses software and hardware switches; the abstraction-based includes data flow graph and match-action pipelines; and the state-based differentiates between stateful and stateless data planes. This short survey lacks critical discussions on the existing P4 applications.

Kaljic et al. [40] presented a survey on data plane flexibility and programmability in SDN networks. The authors evaluated data plane architectures through several definitions of flexibility and programmability. In general, flexibility in SDN refers to the ability of the network to adapt its resources (e.g., changes in the topology or the network requirements). Afterwards, the authors identified key factors that influence the deviation from the original data plane given with OpenFlow. The survey concludes with future research directions.

Kannan et al. [41] presented a short survey related to the evolution of programmable networks. This work described the pre-SDN model and the evolution to SDN and programmable data plane. The authors highlighted some features of programmable switches such as stateful processing, accurate timing information, and flexible packet cloning and recirculation. The survey categorized data plane applications into two categories, namely, network monitoring and in-network computing. While this survey listed a considerable number of papers belonging to these categories, it barely explained the operation and main ideas of each paper. Also it lacks many other categories that are relevant in the programmable data plane context.

Tan et al. [42] presented a survey describing In-band Network Telemetry (INT). The survey explained the development stages and classifications of network measurement (traditional, SDN-based, and P4-based). It also outlined some existing applications that leverage INT such as congestion control, troubleshooting, etc. The survey concludes with discussions and potential future work related to INT.

Zhang et al. [43] presented a survey that focuses on stateful data plane. The survey starts with an overview of stateless and stateful data planes, then overviews and compares some stateful platforms (e.g., OpenState, FAST, FlowBlaze, etc.). The paper reviews a handful of stateful data plane applications and discusses challenges and future perspectives.

Table 1 summarizes the topics and the features described in the related surveys. It also highlights how this paper differs from the existing surveys. All previous surveys lack a microscopic comparison between the intra-category works. Also, none of them compare switch-based schemes against legacy server-based schemes. To the best of the authors' knowledge, this work is the first to exhaustively explore the whole programmable data plane ecosystem. Specifically, the paper describes P4 switches and provides a detailed taxonomy of applications using P4 switches. It categorizes and compares the applications within each category as well as with legacy approaches, and provides challenges and future perspectives.

## III. TRADITIONAL CONTROL PLANE AND SDN

### A. TRADITIONAL AND SDN DEVICES

With traditional devices, networks are connected using protocols such as Open Shortest Path First (OSPF) and Border Gateway Protocol (BGP) [47]) running in the control plane at each device. Both control and data planes are under full control of vendors. On the other hand, SDN delineates a clear separation between the control plane and the data plane, and consolidates the control plane so that a single centralized con-troller can control multiple remote data planes. The controller is implemented in software, under the control of the network owner. The controller computes the tables used by each switch and distributes them via a well-defined Application Programming Interface (API), such as Openflow [48]. While SDN allows for the customization of the control plane, it is limited to the OpenFlow specifications and the fixed-function data plane.

**TABLE 2:** Features, traditional, SDN, and P4 programmable devices.

| Feature | Traditional | SDN | P4 programmable |
|---|---|---|---|
| Control - data plane separation | No clear separation | Well-defined separation | Well-defined separation |
| Control and data plane interface | Proprietary | Standardized APIs (e.g. OpenFlow) | Standardized (e.g., OpenFlow, P4Runtine) and program-dependent APIs |
| Control and data plane program-dependent APIs | NA/Proprietary | NA/Proprietary | Target independent |
| Functionality separation at control plane | No modular separation of functions | Modular separation: (1) functions to build topology view (state) and (2) algorithms to operate on network state | Same as SDN networks |
| Customization of control plane | No | Yes | Yes |
| Visibility of events at data plane | Low | Low | High |
| Flexibility to define and parse new fields and protocols | No flexible, fixed | Subject to OpenFlow extensions | Easy, programmable by user |
| Customization of data plane | No | No | Yes |
| ASIC packet processing complexity | High, hard-coded | High, hard-coded | Low, defined by user's source code |
| Data plane match-action stages | Proprietary | OpenFlow assumes in series match-action stages | In series and/or in parallel |
| Data plane actions | Protocol-dependent primitives | Protocol-dependent primitives | Protocol-independent primitives |
| Infield runtime reprogrammability | No | No | Yes |
| Customer support | High | Medium | Low |
| Technology maturity | High | Medium | Low |

### B. COMPARISON OF TRADITIONAL, SDN, AND PROGRAMMABLE DATA PLANE DEVICES

Table 2 contrasts the main characteristics of traditional, SDN, and P4 programmable devices. In the latter, the forwarding behavior is defined by the user's code. Other advantages include the program-dependent APIs, where the same P4 program running on different targets requires no modifications in the runtime applications (i.e., the control plane and the interface between control and data planes are target agnostic); the protocol-independent primitives used to process packets; the more powerful computation model where the match-action stages can not only be in series but also in parallel; and the infield reprogrammability at runtime. On the other hand, the technology maturity and support for P4 devices can still be considered low in contrast to traditional and SDN devices.

### C. NETWORK EVOLUTION AND ANALOGY WITH OTHER DOMAIN SPECIFIC PROCESSORS

The introduction of the general-purpose computers in the early 1970s enabled programmers to develop applications running on CPUs. The use of high-level languages accelerated innovation by hiding the target hardware (e.g., x86). In signal processing, Digital Signal Processors (DSPs) were developed in the late 1970s and early 1980s with instruction sets optimized for digital signal processing. Matlab is used for developing DSP applications. In graphics, Graphics Processing Units (GPUs) were developed in the late 1990s and early 2000s with instruction sets for graphics. Open Computing Language (OpenCL) is one of the main languages for developing graphic applications. In machine learning, Tensor Processor Units (TPUs) and TensorFlow were developed in mid 2010s with instruction sets optimized for machine learning.

The programmable forwarding is part of the larger information technology evolution observed above. Specifically, over the last few years, a group of researchers developed a machine model for networking, namely the Protocol Independent Switch Architecture (PISA) [49]. PISA was designed with instruction sets optimized for network operations. The high-level language for programming PISA devices is P4.

## IV. PROGRAMMABLE SWITCHES

### A. PISA ARCHITECTURE

PISA is a packet processing model that includes the following elements: programmable parser, programmable match-action pipeline, and programmable deparser, see Fig. 3.

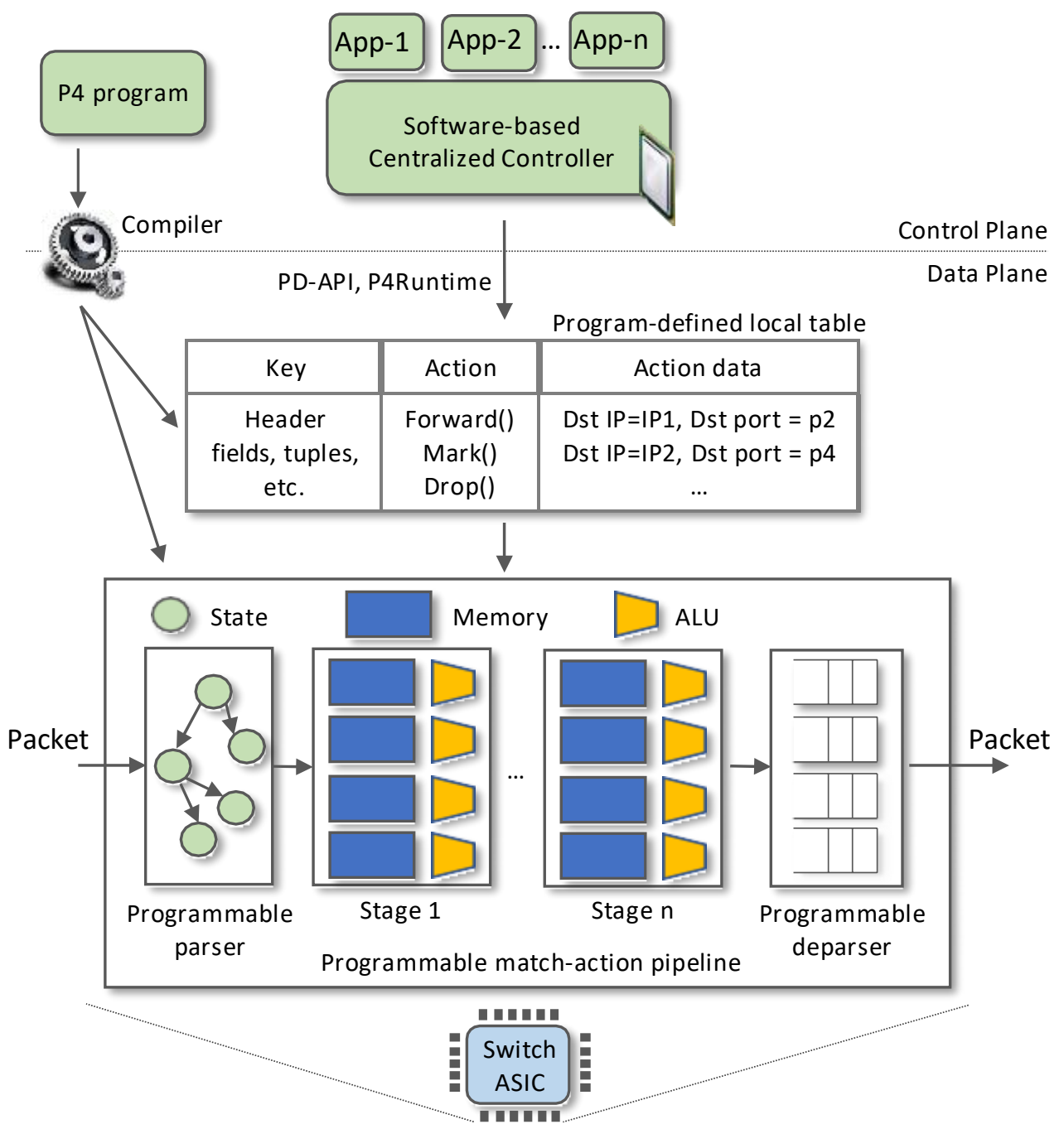

FIGURE 3: A PISA-based data plane and its interaction with the control plane.

The programmable parser permits the programmer to define the headers (according to custom or standard protocols) and to parse them. The parser can be represented as a state machine. The programmable match-action pipeline executes the operations over the packet headers and intermediate results. A single match-action stage has multiple memory blocks (tables, registers) and Arithmetic Logic Units (ALUs), which allow for simultaneous lookups and actions. Since some action results may be needed for further processing (e.g., data dependencies), stages are arranged sequentially. The programmable deparser assembles the packet headers back and serializes them for transmission. A PISA device is protocol-independent.

In Fig. 3, the P4 program defines the format of the keys used for lookup operations. Keys can be formed using packet header's information. The control plane populates table entries with keys and action data. Keys are used for matching packet information (e.g., destination IP address) and action data is used for operations (e.g., output port).

### B. PROGRAMMABLE SWITCH FEATURES

The main features of programmable switches are [51]:

- Agility: the programmer can design, test, and adopt new protocols and features in significantly shorter times (i.e., weeks or months rather than years).
- Top-down design: for decades, the networking industry operated in a bottom-up approach. Fixed-function ASICs are at the bottom and enforce available protocols and features to the programmer at the top. With programmable switches, the programmer describes protocols and features in the ASICs. Note that the physical layer and parts of the MAC layer may not be programmable.

**TABLE 3:** Comparison between a P4 programmable switch and a fixed-function switch [50].

| Characteristic | Programmable | Fixed-function |
|---|---|---|
| Throughput | 6.4Tb/s | 6.4Tb/s |
| Number of 100G ports | 64 | 64 |
| Max forwarding rate | 4.8B pps | 4.2B pps |
| Max 25G/10G ports | 256/258 | 128/130 |
| Programmable | Yes (P4) | No |
| Power draw | 4.2W per port | 4.9W per port |
| Large scale NAT | Yes (100k) | No |
| Large scale stateful ACL | Yes (100k) | No |
| Large scale tunnels | Yes (192k) | No |
| Packet buffers | Unified | Segmented |
| LAG/ECMP | Full entropy, programmable | Hash seed, reduced entropy |
| ECMP | 256-way | 128-way |
| Telemetry | Line-rate per flow stats | SFlow (sampled) |
| Latency | Under 400 ns | Under 450ns |

- Visibility: programmable switches provide greater visibility into the behavior of the network. INT is an example of a framework to collect and retrieve information from the data plane, without intervention of the control plane.
- Reduced complexity: fixed-function switches incorporate a large superset of protocols. These protocols consume resources and add complexity to the processing logic, which is hard-coded in silicon. With programmable switches, the programmer has the option to implement only those protocols that are needed.
- Differentiation: the customized protocol or feature implemented by the programmer needs not to be shared with the chip manufacturer.
- Enhanced performance: programmable switches do not introduce performance penalty. On the contrary, they may produce better performance than fixed-function switches. Table 3 shows a comparison between a programmable switch and a fixed-function switch, reproduced from [50]. Note the enhanced performance of the former (e.g., maximum forwarding rate, latency, power draw). When compared with general purpose CPUs, ASICs remain faster at switching, and the gap is only increasing as shown in Fig. 4.

The performance gain of switches relies on the multiple dimensions of parallelism, as described next.

- Parallelism on different stages: each stage of the pipeline processes one packet at a time [49]. In Fig. 3, the number of stages is [redacted]. Implementations may have more than 10 stages on the ingress and egress pipelines. While adding more stages increases parallelism, they consume more area on the chip and increase power consumption and latency.
- Parallelism within a stage: the ASIC contains multiple match-action units per stage. During the match phase, tables can be used for parallel lookups. In Fig. 3, there are four matches (in blue) on each stage that can occur at the same time. An ALU executes one operation over the header field, enabling parallel actions on all fields. Hundreds of match-action units exist per stage and thousands in an entire pipeline [49]. Since ALUs execute simple operations and use a simple Reduced Instruction Set Computer (RISC)-type instruction set, they can be implemented in the silicon at a minimal cost.
- Very Long Instruction Words: the set of instructions issued in a given clock cycle can be seen as one large instruction with multiple operations, referred to as Very Long Instruction Word (VLIW). A VLIW is formed from the output of the match tables. A stage executes one VLIW per packet, and each action unit within the stage executes one operation. Thus, for a given packet, one operation per field per stage is applied [52].
- Parallelism on pipelines: the switch chip may contain multiple pipelines per chip, also referred to as pipes. Pipes on a PISA device are analogous to cores on a general purpose CPU. Examples include chips containing two and four pipes [20, 49]. Each pipe is isolated from the other and processes packets independently. Pipes may implement the same functionality or different functionalities.

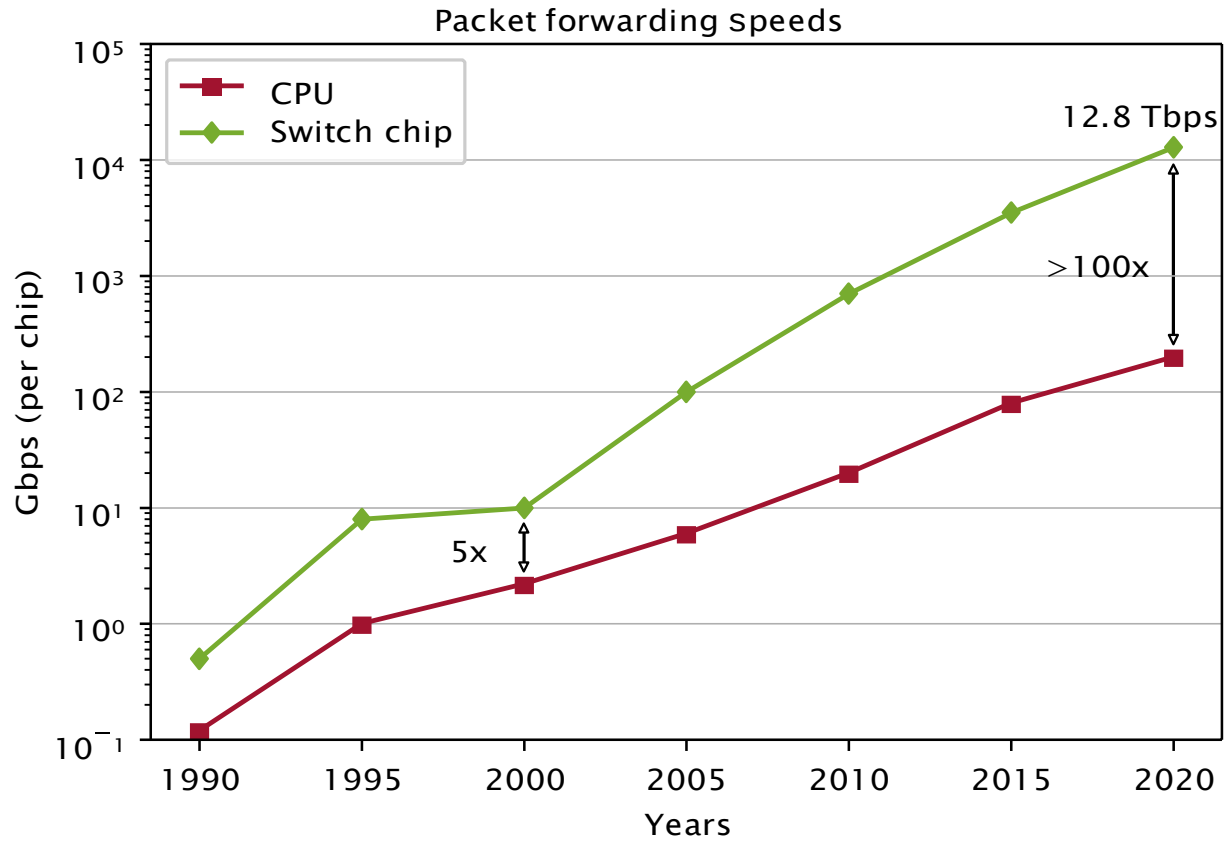

**FIGURE 4:** Evolution of the packet forwarding speeds of the general-purpose CPU and the switch chip (reproduced from [53]).

### C. P4 LANGUAGE

P4 has a reduced instruction set and has the following goals:

- Reconfigurability: the parser and the processing logic can be redefined in the field.

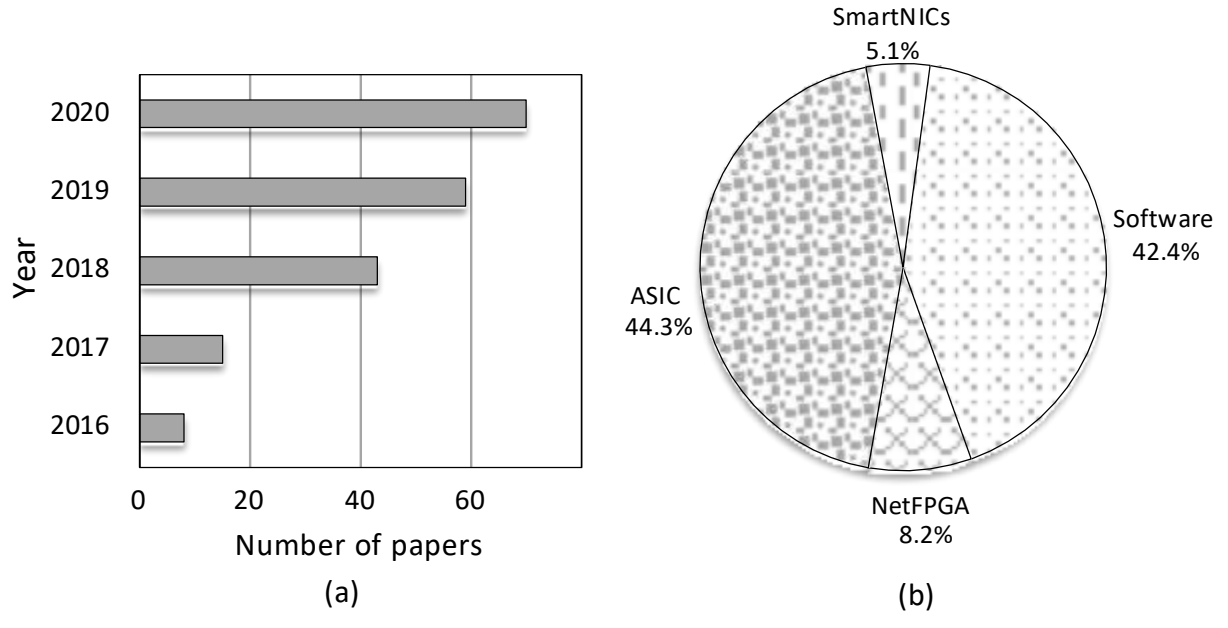

**FIGURE 5:** (a) Distribution of surveyed data plane research works per year. (b) Implementation platform distribution. The shares are calculated based on the studied papers in this survey.

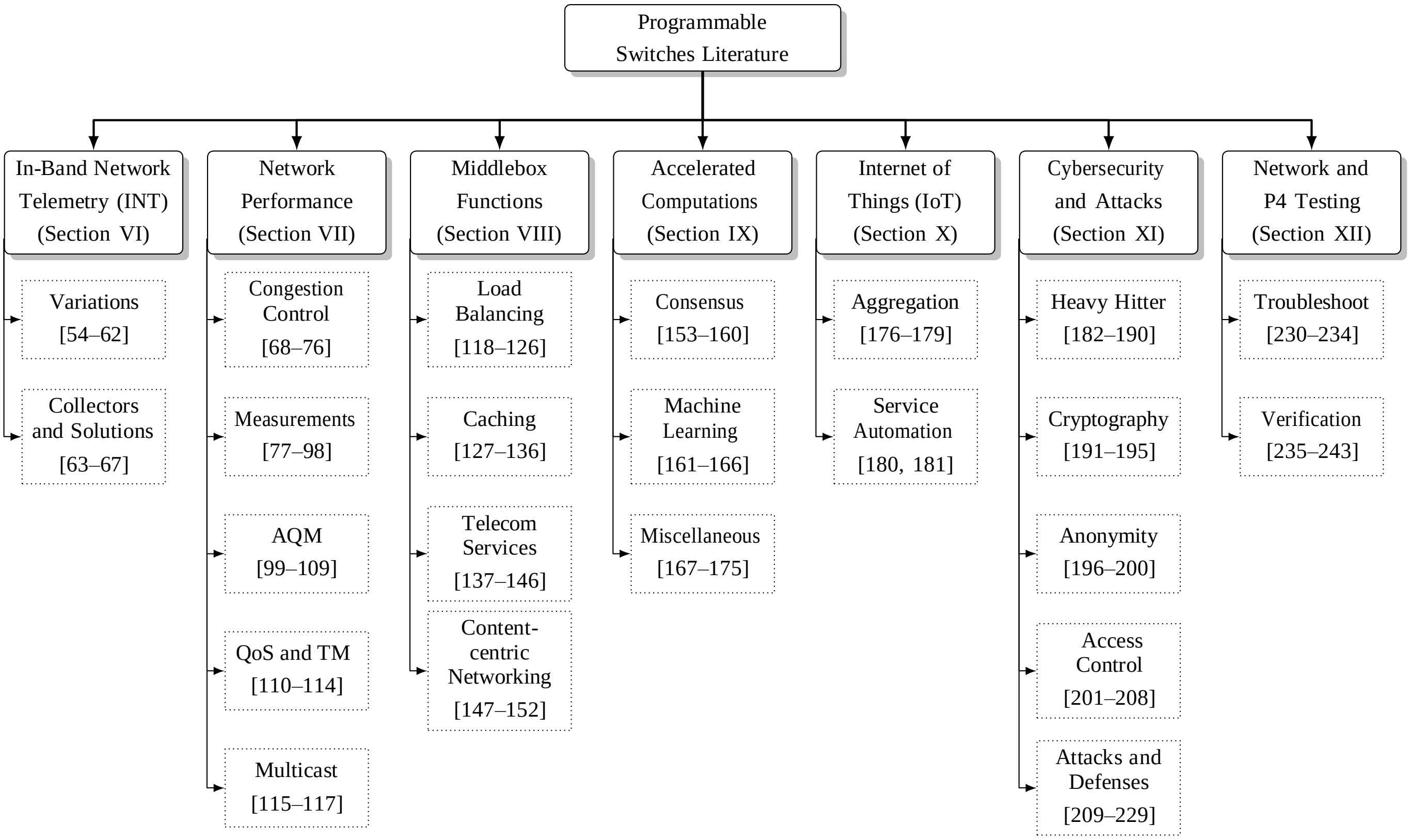

**FIGURE 6:** Taxonomy of programmable switches literature based upon relevant, explored research areas.

- Protocol independence: the switch is protocol-agnostic. The programmer defines the protocols, the parser, and the operations to process the headers.
- Target independence: the underlying ASIC is hidden from the programmer. The compiler takes the switch's capabilities into account when turning a target-independent P4 program into a target-dependent binary.

The original specification of the P4 language was released in 2014, and is referred to as $P4_{14}$. In 2016, a new version of the language was drafted, which is referred to as $P4_{16}$. $P4_{16}$ is a more mature language which extended the P4 language to broader underlying targets: ASICs, Field-Programmable Gate Arrays (FPGAs), Network Interface Cards (NICs), etc.

## V. METHODOLOGY AND TAXONOMY

This section describes the systematic methodology that was adopted to generate the proposed taxonomy. The results of this literature survey represent derived findings by thoroughly exploring more than 200 data plane-related research works starting from 2016 up to 2020. The distribution of which is summarized in Fig. 5 (a). Note that the survey additionally includes the important works of the first quarter of 2021.

Fig. 5 (b) depicts the share of each implementation platform used in the surveyed papers, grouped by software (e.g., BMv2, PISCES), ASIC (e.g., Tofino, Cavium), NetFPGA (e.g., NetFPGA SUME), and SmartNICs (e.g., Netronome NFP). The graph shows that the vast majority of the works were implemented on software and hardware switches. Note that behavioral software switches (e.g., BMv2 [244]) are not suitable indicators of whether the program could run on a hardware target; they are typically used for prototyping ideas and to foster innovation. On the other hand, non-behavioral software switches (e.g., PICSES [245], derived from Open vSwitch (OVS) [246]) are production-grade and can be de- ployed in data centers.

It is worth noting that the majority of works implemented on hardware switches are recent; this demonstrates the in- crease in the adoption of programmable switches by the in- dustry and academia. Currently, to acquire a switch equipped with Tofino chip (e.g., Edgecore Wedge100BF-32 [20]), and to get the development environment and the customer support, a Non-Disclosure Agreement (NDA) with Barefoot Networks (Intel) should be signed. Additionally, the client should attend a training course (e.g., [247]) to understand the architecture and the specifics of the platform. This process is somewhat lengthy and costly, and not every institution is capable of affording it.

The proposed taxonomy is demonstrated in Fig. 6. The taxonomy was meticulously designed to cover the most sig- nificant works related to data plane programmability and P4. The aim is to categorize the surveyed works based on various high-level disciplines. The taxonomy provides a clear separa- tion of categories so that a reader interested in a specific disci- pline can only read the works pertaining to the said discipline.

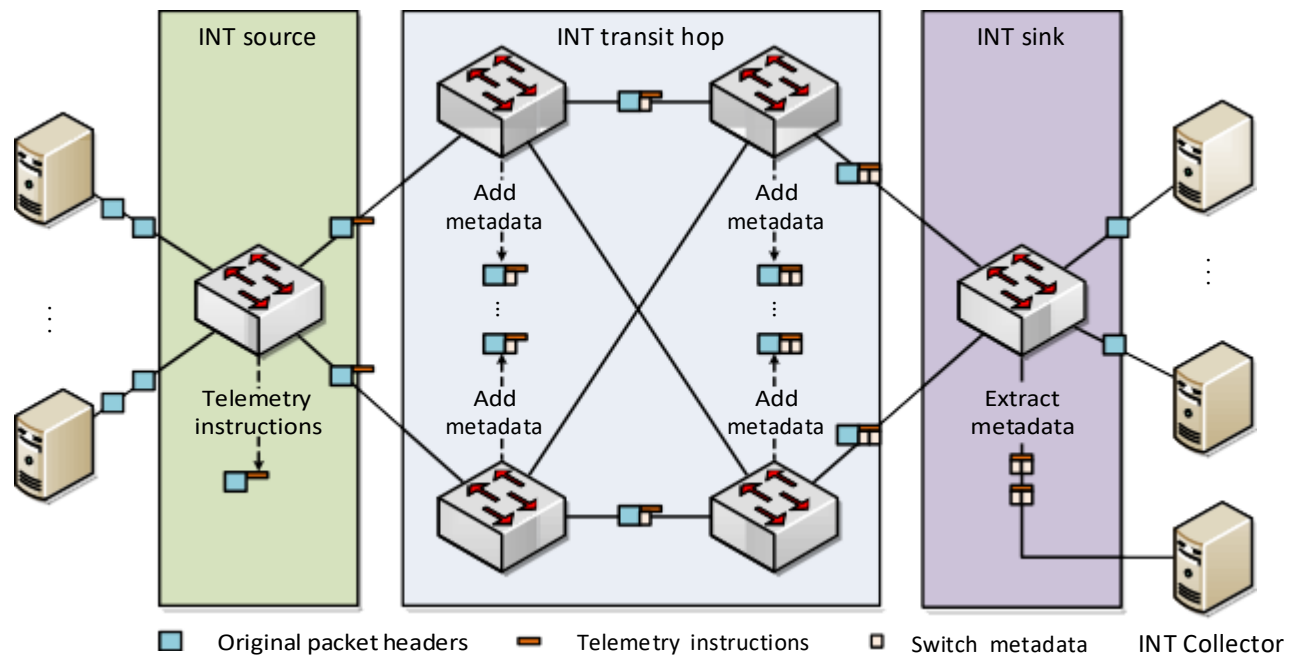

FIGURE 7: In-band Network Telemetry (INT).

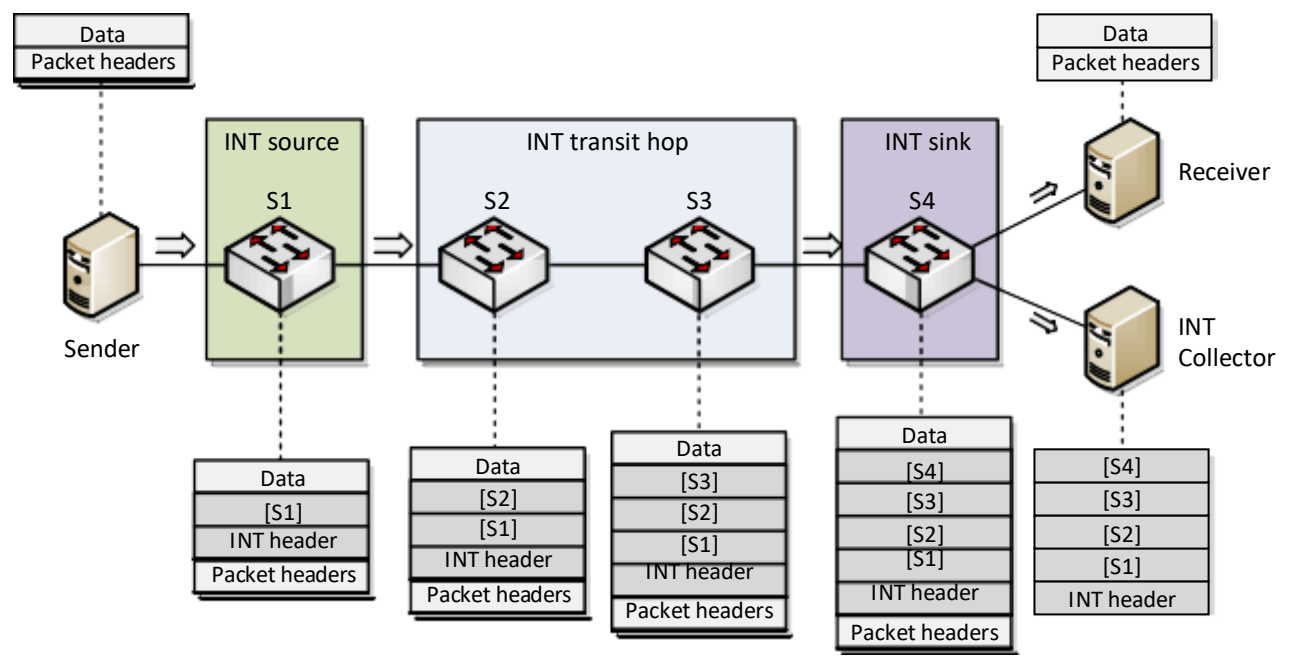

FIGURE 8: Example of how INT can be used to provide the path traversed by a packet in the network. The INT source inserts its label (S1) as well as the INT headers to instruct subsequent switches about the required operations (i.e., push their labels). Finally, switch S4 strips the INT headers from the packet and forwards them to a collector, while forwarding the original packet to the receiver.

The correctness of the taxonomy was verified by carefully examining the related work of each paper to correlate them into high-level categories. Each high-level category is further divided into sub-categories. For instance, various measurements works belong to the sub-category "Measurements" under the high-level category "Network Performance".

Further, the survey compares the results and the features offered by programmable data plane approaches (intra-category), as well as with those of the contemporary and legacy ones. This detailed comparison is elaborated upon for each sub-category, giving the interested reader a comprehensive view of the state-of-the-art findings of that sub-category. Additionally, the survey presents various challenges and considerations, as well as some current and future trends that could be explored as future work.

## VI. IN-BAND NETWORK TELEMETRY (INT)

Conventional monitoring and collecting tools and protocols (e.g., ping, traceroute, Simple Network Management Protocol (SNMP), NetFlow, sFlow) are by no means sufficiently accurate to troubleshoot the network, especially with the presence of congestion. These methods provide milliseconds accuracy at best and cannot capture events that happen on microseconds magnitude. Moreover, they cannot provide per-packet visibility across the network.

In-band Network Telemetry (INT) [248] is one of the earliest key applications of programmable data plane switches. It enables querying the internal state of the switch and provides fine-grained and precise telemetry measurements (e.g., queue occupancy, link utilization, queuing latency, etc.). INT handles events that occur on microseconds scale, also known as *microbursts*. Collecting and reporting the network state is performed entirely by the data plane, without any intervention from the control plane. Due to the increased visibility achieved with INT, network operators are able to troubleshoot problems more efficiently. Additionally, it is possible to perform instant processing in the data plane after measuring telemetry data (e.g., reroute flows when a link is congested), without having to interact with the control plane. Fig. 7 shows an INT-enabled network. INT enables network administrators to determine the following:

- The path a packet took when traversing the network (see Fig. 8). Such information is difficult to learn using existing technologies when multi-path routing strategies (e.g., Equal-cost Multi-Path Routing (ECMP) [249], flowlet switching [250]) are used.
- The matched rules that forwarded the packets (e.g., ACL entry, routing lookup).
- The time a packet spent in the queue of each switch.
- The flows that shared the queue with a certain packet.

The P4 Applications Working Group developed the INT telemetry specifications [251] with contributions from key enablers of the P4 language such as Barefoot Networks, VMware, Alibaba, and others. INT allows instrumenting the metadata to be monitored without modifying the application layer. The metadata to be inserted depends on the use case; for example, if congestion was the main concern to monitor, the programmer inserts queue metadata and transit latency. An INT-enabled network has the following entities: 1) INT source: a trusted entity that instruments with the initial instruction set what metadata should be added into the packet by other INT-capable devices; 2) INT transit hop: a device adding its own metadata to an INT packet after examining the INT instructions inserted by the INT source; 3) INT sink: a trusted entity that extracts the INT headers in order to keep the INT operation transparent for upper-layer applications; and 4) INT collector: a device that receives and processes INT packets.

The location of an INT header in the packet is intentionally not enforced in the specifications document. For example, it can be inserted as a payload on top of TCP, UDP, and NSH, as a Geneve option on top of Geneve, and as a VXLAN payload on top of VXLAN.

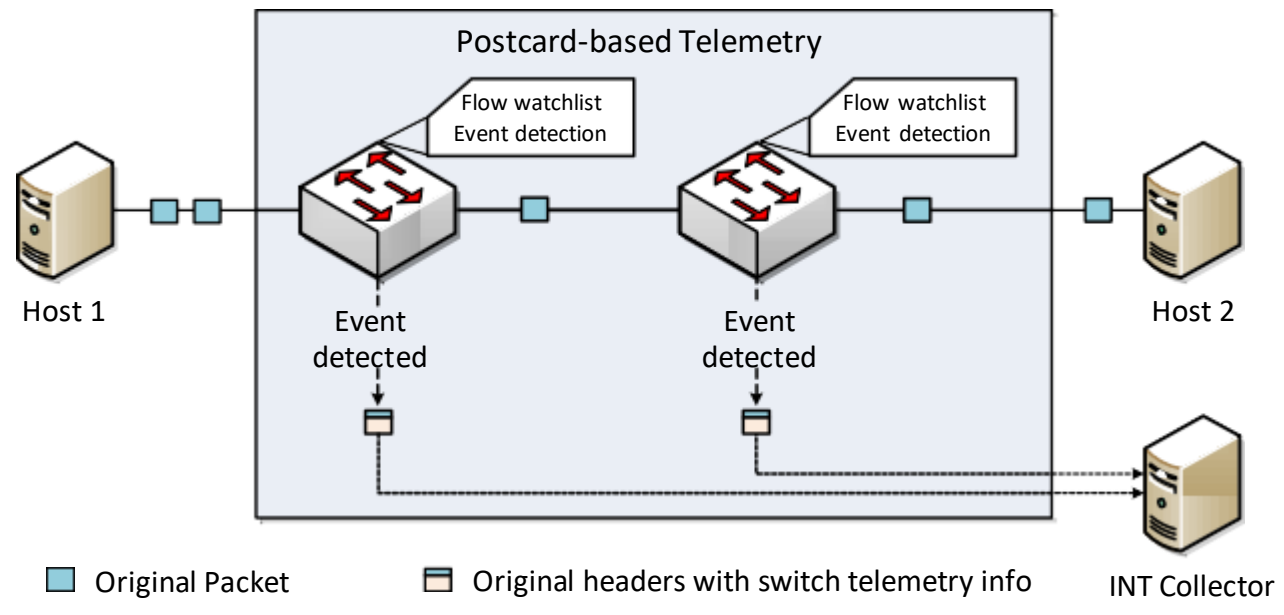

FIGURE 9: Postcard-based Telemetry (PBT).

### A. POSTCARD-BASED TELEMETRY (PBT)

INT provides the exact forwarding path, the timestamp and latency at each network node, and other information. Such detailed information is derived by augmenting user packets with data collected by each switch. Postcard-based Telemetry (PBT) is an alternative to INT which does not modify user packets. Fig. 9 shows an example of PBT. As a user packet traverses the network, each switch generates a postcard and sends it to the monitor. The event that triggers the generation of the postcard is defined by the programmer, according to the application's need. Examples include start and/or end of a flow, sampling (e.g., one report per second), packet dropped by the switch, queue congestion, etc.

### B. INT VARIATIONS

#### 1) Background

Despite the improvements that INT brings compared to legacy monitoring schemes, it introduces bandwidth overhead when enabled unconditionally by network operators. In such scenarios, INT headers are added to *every* packet traversing the switch, increasing bandwidth overhead which decreases the overall network throughput. To mitigate such limitation, conditional statements are included in the P4 program to send reports only when certain events occur (e.g., queue utilization exceeds a threshold). Such solution requires network operators to adjust thresholds and parameters manually based on the usual network traffic patterns. Consequently, several variations of INT have been developed, aiming at customizing its functionalities and addressing its limitations. Mainly, recent works focus on minimizing the bandwidth overhead of INT by adjusting thresholds and parameters automatically, based on measured traffic patterns and the desired application type.

#### 2) Active Network Telemetry

Network telemetry can be actively collected by generating and sending probes to a selected network path. Probes are typically used for minimizing the traffic overhead imposed by regular INT. Liu et al. [54] proposed NetVision, a probing-based telemetry system that actively sends the rightful amount and format of probe packets depending on the telemetry application (e.g., traffic engineering, network visualization). INT-path [57] is another probing-based approach that was the first to achieve network-wide telemetry. Network-wide telemetry provides a global view of the network, which simplifies the management and the control decisions. INT-Path uses Euler trail-based path planning policy to generate probe paths. This mechanism allows achieving non-overlapped probe paths. The idea is to transform network troubleshooting into pattern recognition problems after encoding the traffic status into a series of bitmap images. A subsequent work by Lin et al. [62] that extends NetVision, referred to as NetView [62], was proposed. The objective of NetView is to achieve on-demand network-wide telemetry. NetView considers various telemetry applications, has full coverage, and achieves scalable telemetry.

#### 3) Passive Network Telemetry

Instead of actively sending probes through the network, INT can determine telemetry information passively [252]. The standardized INT [251], which writes telemetry information along the path in packets, is an example of passive network telemetry.

Kim et al. [56] proposed selective INT (sINT), a scheme that dynamically adjusts the insertion frequency of INT headers. A monitoring engine observes changes in consecutive INT metadata and applies a heuristic algorithm to compute the insertion ratio. Marques et al. [58] described the orchestration problem in INT, which is associated with the optimal use of network resources for collecting the state and behavior of forwarding devices through INT. Niu at al. [59] proposed multilayer INT (ML-INT), a system that visualizes

**TABLE 4:** INT variations comparison.

| | Ref | Name | Overhead reduction strategy | Network-wide | Operator Intervention | Implementation |
|---|---|---|---|---|---|---|
| Active | [54] | NetVision | On-demand probing | ✓ | High; telemetry through queries | Mininet |
| | [57] | INT-Path | Non-overlapped probe paths generation | | High; telemetry through queries | Software (BMv2) |
| | [62] | NetView | Application-dependent probing frequency tuning | | High; telemetry through queries | ASIC (Tofino) |
| Passive | [55] | N/A | Flow subset selection by the knowledge plan | ✓ | Low; closed-loop network | Software (BMv2) w/ ONOS controller |
| | [56] | sINT | Monitoring ratio adjustment based on network change | ✓ | Low; telemetry based on network behavior | Software (BMv2) |
| | [58] | INTO | Telemetry orchestration based on heuristics | | High; telemetry specified by operators | N/A |
| | [59] | ML-INT | Per-flow packet subset selection through sampling | ✓ | High; telemetry specified by operators | ASIC (Tofino) and SmartNIC (NFP-4000) |
| | [61] | PINT | Telemetry encoding on multiple packets | ✓ | High; telemetry through queries | ASIC (Tofino) |
| | [60] | N/A | N/A | | High; telemetry through queries | Wireless Sensor Networks |

IP-over-optical networks in realtime. The proposed system encodes INT headers in a subset of packets pertaining to an IP flow. The encoded headers contain metadata that describes statistics of electrical and optical network elements on the flow's routing path. Ben et al. [61] proposed Probabilistic INT (PINT), an approach that probabilistically adds telemetry information into a collection of packets to minimize the per-packet overhead associated with regular INT. Hyun et al. [55] proposed an architecture for self-driving networks that uses INT to collect packet-level network telemetry, and Knowledge-Defined Networking (KDN) to create intelligence to the network management, considering the collected telemetry data. KDN accepts the network information as input and generates policies to improve the network performance. Karaagac et al. [60] extended INT from wired network to wireless network.

#### 4) INT Variations, Comparison, and Discussions

Table 4 compares the aforementioned INT variations solutions. The main motivation behind these solutions is that the majority of applications that leverage INT (e.g., congestion control, fast reroute) only require approximations of the telemetry data and therefore, do not need to gather per-packet per-hop INT information. NetVision, NetView, and INT-Path use probing to reduce the overhead of INT. The main limitation of such approaches is that probing might result in poor accuracy and timeliness as the probes might experience different network conditions than actual packets. All other works collect INT information passively. [55] and sINT select flows based on current network conditions, ML-INT uses a fixed sampling scheme to select a small portion of packets in a flow, and PINT uses a probabilistic approach to encode telemetry on multiple packets. Note that sampling and anomaly-based monitoring might lead to information loss since not all packets are being reported.

Some solutions require manual intervention from the operators to configure the telemetry process. The simplicity of the configuration interface is vital to make the solution easily deployable. Furthermore, some solutions (e.g., NetView, INT-Path) achieve network-wide telemetry. Note that network-wide traffic monitoring incurs additional overhead since multiple switches are being monitored at the same time. Finally, some solutions were implemented on software switches, while other were implemented on hardware. It is important to note that not all software implementations can fit into the pipeline of the hardware.

#### 5) INT, PBT, and Traditional Telemetry Comparison

Table 5 compares INT, PBT, and traditional telemetry. INT has higher potential vulnerabilities than PBT, such as eavesdropping and tampering. Adding extra protective measures (e.g., encryption) is difficult on the fast data path. On the other hand, PBT packets tolerate additional processing to enhance security. The flow tracking process is simpler with INT than with PBT. The latter requires the server receiving INT reports (i.e., INT collector, explained in Section VI-C) to correlate multiple postcards of a single flow packet passing through the network, to form the packet history at the monitor. This process also adds delay in reporting and tracking. Legacy schemes that rely on sampling and polling suffer from accuracy issues, especially when links are congested. INT on the other hand is push-based, has better accuracy, and is more granular (microseconds scale). Reports sent by an INT-capable device contain rich information (e.g., the path a packet took) that can aid in troubleshooting the network. Such visibility is minimal in legacy monitoring schemes. Programmable switches permit reporting telemetry after the occurrence of specific events (e.g., congestion). Moreover, they provide flexibility in programming reactive logic that executes promptly in the data plane. One drawback of INT is that it imposes bandwidth overhead if configured to report for every packet; however, when event-based reports are considered, the bandwidth overhead significantly decreases.

**TABLE 5:** In-band, postcard-based, and traditional network telemetry.

| Feature | INT | PBT | Traditional |
|---|---|---|---|
| User packet modification | Yes | No | No |
| User packet overhead | Yes | No | No |
| Potential vulnerabilities | Higher | Lower | Lower |
| Flow tracking process | Simpler | More complex | More complex |
| Delay in reporting, tracking | Lowest | Low | High |
| Microbursts detection | Yes | Yes | No |
| Accuracy | Higher | Higher | Lower; especially with congested links |
| Reporting type | Push-based, initiated by the data plane | Push-based | Polling (e.g., SNMP), initiated by the control plane; sampling (e.g., NetFlow), initiated by the data plane |
| Troubleshoot problems | Easier and cheaper | Easier and cheaper | Harder and more expensive |
| Granularity | Higher; microseconds scale | Higher | Lower; milliseconds scale at best |
| Event-based monitoring | Customizable based on conditions and thresholds | Customizable | Not possible |
| Reactive processing | Faster; reactive processing is executed in the data plane | Faster | Slower; reactive processing is executed in the control plane |
| Bandwidth overhead | High when all packets are reported, low when reported based on events | Higher than INT | Lowest |

**TABLE 6:** INT collectors comparison.

| | Ref | Name | Rate | Event detection | Processing acceleration | Historical data availability | Analytics | Implementation notes |
|---|---|---|---|---|---|---|---|---|
| Open-source | [63] | IntMon | 0.1Kpps | ✓ | ✓ | ✓ | Low | ONOS-BMv2 subsystem (ONOS 1.6) |
| | [64] | Prometheus INT exporter | 6.4Kpps | ✓ | ✓ | ✓ | Low | ONOS P4 Brigade project |
| | [65] | IntCollector | 154.8Kpps | | Yes; fast path with XDP | | Medium | C language, XDP for in-kernel processing |
| Closed-source | [66] | DeepInsight | N/A | | N/A | | High | SPRINT data plane telemetry (INT.p4) |
| | [67] | BroadView Analytics | N/A | ✓ | N/A | | High | Trident 3's in-band telemetry |

## C. INT COLLECTORS

### 1) Background

An INT collector is a component in the network that processes telemetry reports produced by INT devices. It parses and filters metrics from the collected reports, then optionally stores the results persistently into a database. Since a large number of reports is typically produced in INT, having a high-performance collector is essential to avoid missing important network events. To this end, a number of research works focus on developing and enhancing the performance of INT collectors running on commodity servers. Both open source and closed source INT collectors are proposed in the literature.

### 2) Open-source

IntMon [63] is an ONOS-based collector application for INT reports. It includes a web-based interface that allows controlling which flows to monitor and the specific metadata to collect. Another INT collector is the Prometheus INT exporter [64], which extracts information from every INT packet and pushes them to a gateway. A database server then periodically pulls information from the gateway. INTCollector [65] is a collector that extracts *events*, which are important network information, from INT raw data. It uses in-kernel processing to further improve the performance. INTCollector has two processing flows; the *fast path*, which processes INT reports and needs to execute quickly, and the *normal path* which processes events sent from the fast path, and stores information in the database.

### 3) Closed-source

Deep Insight [66] is a proprietary solution provided by Barefoot Networks that leverages INT capabilities to provide services such as real-time anomaly detection, congestion analysis, packet-drop analysis, etc. It follows a pay-as-you-grow business model, where customers pay based on the volume of collected telemetry. Another proprietary solution is BroadView Analytics used on Broadcom Trident 3 devices by Broadcom [67]. This solution enables real-time network latency analysis and facilitates Service Level Agreement (SLA) compliance.

### 4) INT Collectors Comparison, Discussions, and Limitations

Table 6 compare the aforementioned INT collectors. IntMon and Prometheus INT exporter were among the earliest collectors. Both have low processing rates since they are implemented without kernel nor hardware acceleration. Also, they are very limited with respect to the features they provide (e.g., lack of event detection, limited analytics, historical data unavailability, etc.). Prometheus INT exporter also suffers from increased overhead of sending the data for every INT packet to the gateway, and the potential loss of network events as the database only stores the latest data pulled from the gateway. INTCollector on the other hand has higher rate and uses the eXpress Data Path (XDP) [253] to accelerate the packet processing in the kernel space. It filters the data to be published based on significant changes in the network through its event detection mechanism. DeepInsight Analytics has a modular architecture and runs on commodity servers. It executes the Barefoot SPRINT data plane telemetry which consists of a P4 program (INT.p4) encompassing intelligent triggers. It also provides open northbound RESTful APIs that allow customers to integrate their third-party network management solutions. DeepInsight Analytics is advanced with respect to the features it provides (real-time anomaly detection, congestion analysis, packet-drop analysis, etc.). However, it is a closed-source solution and lacks reports of performance benchmarks.

Fig. 10 demonstrates the CPU efficiency of three INT collectors (IntMon, Prometheus INT exporter, and INTCollector) [65]. IntMon has the lowest throughput, and is 57 times slower than Prometheus INT. INTCollector on the other hand has the highest throughput and is 27 times faster than Prometheus INT exporter.

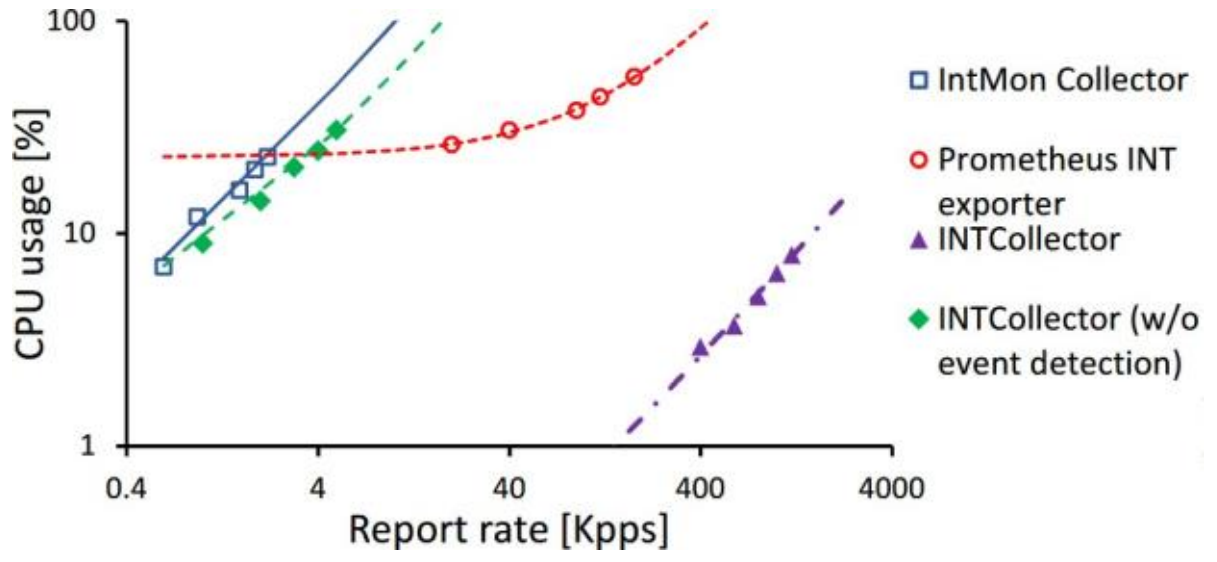

**FIGURE 10:** CPU efficiency with the three INT collectors. Source: INTCollector paper [65].

#### 5) Collectors in INT and Legacy Monitoring Schemes Comparison

Generally, collectors used with both INT and legacy monitoring schemes run on general purpose CPUs, and hence, have comparable performance. INT produces excessive amounts of reports when compared with legacy monitoring schemes (e.g., NetFlow), and therefore, requires having a collector with high processing capability. INT-based collectors are typically accelerated with in-kernel fast packet processing technologies (e.g., XDP) and hardware-based accelerators (e.g., Data Plane Development Kit (DPDK)).

### D. SUMMARY AND LESSONS LEARNED

Legacy telemetry tools and protocols are not capable of capturing *microbursts* nor providing fine-grained telemetry measurements. INT was developed to address these challenges; it enables the data plane developer to query with high-precision the internal state of switches. Telemetry data are then embedded into packets and forwarded to a high-performance collector. The collector typically performs analysis and applies actions accordingly (e.g., informs the control plane to update table entries). Current research efforts mainly focus on developing variations of INT to decrease its telemetry traffic overhead, considering the *overhead-accuracy* tradeoff. Other works aim at accelerating INT collectors to handle large volumes of traffic (in the scale of Kpps). Future work could possibly investigate further improvements for INT such as compressing packets' headers, broadening coverage and visibility, enriching the telemetry information, and simplifying the deployment.

## VII. NETWORK PERFORMANCE

Measuring and improving network performance is critical in nowadays' infrastructures. Low latency and high bandwidth are key requirements to operate modern applications that continuously generate enormous amounts of data [254]. Congestion control (CC), which aims at avoiding network overload, is critical to meet these requirements. Another important concept for expediting these applications is managing the queues that form in routers and switches through Active Queuing Management (AQM) algorithms. This section explores the literature related to measuring and improving the performance of programmable networks.

### A. CONGESTION CONTROL (CC)

#### 1) Background

One of the most challenging tasks in the Internet today is congestion control and collapse avoidance [255]. The difficulty in controlling the congestion is increasing due to factors such as high-speed links, traffic diversity and burstiness, and buffer sizes [68]. Today's CC algorithms aim at shortening delays, maximizing throughput, and improving the fairness and utilization of network resources.

Tremendous amount of research work has been done on congestion control, including end hosts algorithms such as loss-based CC algorithms (e.g., CUBIC [256], Hamilton TCP (HTCP) [257], etc.), model-based algorithms (e.g., Bottleneck Bandwidth and Round-trip Time (BBR) [258, 259]), congestion-signalling mechanisms (e.g., Explicit Congestion Notification (ECN) [260]), data-center specific schemes (e.g., TIMELY [261], Data Center Quantized Congestion Notification (DCQCN) [262], Data Center TCP (DCTCP) [263], pFabric [264], Performance-oriented Congestion Control (PCC) [265], etc.), and application-specific schemes (e.g., QUIC [266]).

With the advent of programmable data plane switches, researchers are investigating new methods for managing congestion. Such methods can be classified as 1) hybrid CC, where network-assisted congestion feedback is provided for end-hosts; and 2) in-network CC, where the switch performs traffic rerouting, steering, or other congestion control techniques, without modifications on end hosts.

#### 2) Hybrid CC

Handley et al. [68] proposed NDP, a novel protocol architecture for datacenters that aims at achieving low completion latency for short flows and high throughput for longer flows. NDP avoids core network congestion by applying per-packet multipath load balancing, which comes at the cost of reordering. It also trims the payloads of packets, similar to what is done in Cut Payload (CP) [267], whenever the queues of the switches become saturated. Once the payload is trimmed, the headers are forwarded using high-priority queues. Consequently, a Negative ACK (NACK) is generated and sent through high-priority queues so that a retransmission is sent before draining the low priority queue. Similarly, Feldmann et al. [69] proposed a method that uses network-assisted congestion feedback (NCF) in the form of NACKs generated entirely in the data plane. NACKs are sent to throttle elephant-flow senders in case of congestion. The method maintains three separate queues for mice flows, elephant flows, and control packets to ensure fair sharing of resources.

Li et al. [70] proposed High Precision Congestion Control (HPCC), a new CC mechanism that leverages INT-based data added by P4 switches to obtain precise link load information. HPCC computes accurate flow rate by using only one rate update, as opposed to legacy approaches that require a large number of iterations to determine the rate. HPCC provides near-zero queueing, while being almost parameterless. Fig. 11 shows the mechanism of HPCC. The switches add INT headers to every packet, and then the INT information is piggybacked into the TCP/RDMA Acknowledgement (ACK) packet. The end-hosts then use this information to adjust the sending rate through their smart Network Interface Controllers (NICs).

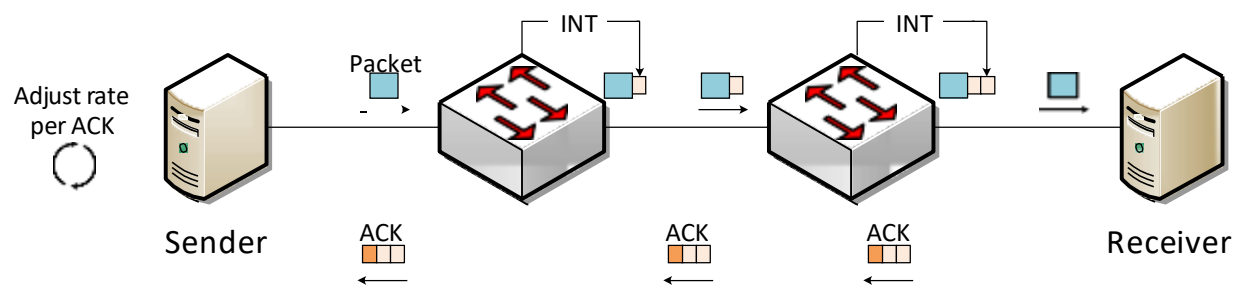

**FIGURE 11:** HPCC: INT-based high precision congestion control.

**TABLE 7:** Congestion control schemes comparison.

| | Ref | Name | Strategy | Feedback information | Rerouting | Traffic separation | End-device modification | Implementation |
|---|---|---|---|---|---|---|---|---|
| Hybrid CC | [68] | NDP | Trim packets to headers and priority forward | NACKs | | | | NetFPGA SUME |
| | [70] | HPCC | Use INT data to compute sending rate | INT | ✓ | ✓ | | Tofino |
| | [69] | NCF | Throttle elephant flows with NACKs | NACKs | ✓ | | ✓ | N/A |
| | [71] | N/A | Pace TCP traffic of elephant flows to safe targets | Flow count and BW | ✓ | ✓ | | BMv2 |
| | [72] | EECN | Remove the receiver from the regular ECN mechanism | ECN | ✓ | ✓ | ECN must be activated | BMv2 |
| In-network CC | [74] | P4Air | Separate flows according to their congestion control group | N/A | ✓ | | ✓ | Tofino |
| | [73] | N/A | Monitor queue latency to reroute traffic on congestion | N/A | | ✓ | ✓ | BMv2 |
| | [76] | P4QCN | Implementation and extension of QCN using P4 | QCN Fb value | ✓ | ✓ | | BMv2 |

Kfoury et al. [71] proposed a P4-based method to automate end-hosts' TCP pacing. It supplies the bottleneck bandwidths and the number of elephants flows to senders so that they can pace their rates to safe targets, avoiding filling routers' buffers. Shahzad et al. [72] proposed EECN, a system that uses ECN to signal the occurrence of congestion to the sender without involving the receiver. This is especially useful for networks with high bandwidth-delay product (BDP).

#### 3) In-network CC

Turkovic et al. [73] proposed a P4-based method that reroutes flows to backup paths during congestion. The system detects congestion by continuously monitoring the queueing delays of latency-critical flows. The same authors [74] proposed a method that separates the senders based on their congestion control algorithm. Each congestion control uses a separate queue in order to enforce the fairness among its competing flows. Apostolaki et al. [75] proposed FAB, a flow-aware and device-wide buffer sharing scheme. FAB prioritizes flows from port-level to the device-level. The goal of FAB is to minimize the flow completion time for short flows in specific workloads. Geng et al. [76] proposed P4QCN, a flow-level, rate-based congestion control mechanism that improves the Quantized Congestion Notification (QCN). P4QCN improves QCN by alleviating the problems of PFC within a lossless network. Furthermore, P4QCN extends the QCN protocol to IP-routed networks.

#### 4) CC Schemes Comparison, Discussions, and Limitations

Table 7 compares the aforementioned CC schemes. NDP and NCF are similar in the sense that both use NACKs as congestion feedback. NDP avoids congestion by applying per-packet multihop load balancing. This approach works adequately with symmetric topologies, but fails when topologies are asymmetric (e.g., BCube, Jellyfish), especially during heavy network load. Another limitation of NDP is the excessive retransmissions produced by the server. NCF adopted the idea of packet trimming from NDP, but generates NACKs from the trimmed packet and sends it directly to the sender. Such approach removes the receiver from the feedback loop, improving the sender's reaction time. One limitation of NCF is that it requires operators to manually tune some of the predefined parameters (e.g., threshold, queue size, etc.). Additionally, NCF might disclose network congestion information, making it less attractive to operators. Finally, the authors of NCF claim that the approach works with both datacenters and Internet-wide scenarios. However, no implementation results were presented to evaluate the effectiveness of the solution.

HPCC leverages INT data to control network congestion. It enhances the convergence time by using a Multiplicative-Increase Multiplicative-Decrease (MIMD) scheme. Note that previous TCP variants use the Additive-Increase Multiplicative-Decrease (AIMD), which is conservative when increasing the rate, and hence has a slow convergence time. The reason AIMD schemes are slow is that they use a single-bit congestion information (packet loss, ECN). With HPCC, end-hosts can perform aggressive increase as INT metadata encompasses precise link utilization and timely queue statistics. HPCC demonstrated promising results with respect to latency, bandwidth, and convergence time. The authors however did not evaluate the performance of HPCC with conventional congestion control algorithms in the Internet (e.g., CUBIC, BBR). Note that achieving inter-protocol fairness is essential so that the solution is adopted by operators.

The method in [71] uses TCP pacing. Pacing decreases throughput variations and traffic burstiness, and hence, minimizes queuing delays. However, this method works well only in networks where the number of large flows senders is small (e.g., in science Demilitarized Zone (DMZ) [254]).

**TABLE 8:** Congestion control schemes. 1) Programmable Switches (HPCC); 2) end-hosts; and 3) legacy network-assisted (ECN).

| Characteristic | Programmable switch | End-hosts | Legacy network-assisted (ECN) |
|---|---|---|---|
| Accuracy | Higher, INT-based, microbursts are detected and reported | Low, packet loss (e.g., CUBIC); Medium, estimated RTT and btlbw (e.g., BBR) | Lower with classic ECN; High with L4S |
| Required modifications | Switches, end-hosts | None; distributed nature of AIMD does not require storing state of flows | Minimal if ECN is used (most equipment have classic ECN implemented); High if L4S is used |
| Convergence | Faster (MIMD) | Slower (AIMD) | Adequate with ECN; Fast with L4S ECN |
| Queue utilization | Near-zero | High; possibility of Bufferbloat (e.g., CUBIC) | Low |
| Parameterization | Few | None | Few (e.g., thresholds) |
| Congestion information | Several fields (e.g., queue occupancy, link utilization, flow share, etc.) | Packets drop | 1-bit ECN mark |

Further, it is worth mentioning that methods which provide congestion feedback to end hosts must implement some security mechanisms to prevent packets from being modified.

As for the full in-network CC schemes, P4Air, which applies traffic separation, demonstrated significant improvements in fairness compared to contemporary solutions. However, it requires allocating a queue for each congestion control algorithm group (e.g., loss-based (Cubic), delay-based (TCP Vegas), etc.). Note that the number of queues is limited in switches, and production networks often reserve them for other applications' QoS [70]. P4QCN is not evaluated on hardware targets, and therefore their results (which are extracted based on software switches) are not that indicative.

#### 5) End-hosts, Programmable Switches, and Legacy Devices' CC Schemes

Table 8 compares the CC schemes assisted by programmable switches (e.g., HPCC) with end-hosts CC algorithms (e.g., CUBIC) and legacy congestion signalling schemes (e.g., ECN). End-hosts CC infer congestion through packet drops and estimations (e.g., btlbw and Round-trip Time (RTT) estimation with BBR), which is not always sufficient to infer the existence of congestion. Legacy devices use classic ECN to signal congestion so that end-hosts slow down their transmission rates. Classic ECN is limited as it only marks a single bit to signal congestion, and is not aggressive nor immediate. Programmable switches on the other hand use fine-grained prompt measurements to signal congestion (e.g., INT metadata), which results in higher detection accuracy, near-zero queueing delays, and faster convergence time. The distributed nature of end-hosts CC schemes allows them to operate without modifying the network infrastructure and without tweaking parameters. ECN-enabled devices and programmable switches on the other hand require few parameters (e.g., marking threshold) to adapt to different network conditions.

### B. MEASUREMENTS

#### 1) Background

Gaining an overall understanding of the network behavior is an increasingly complex task, especially when the size of the network is large and the bandwidth is high. Legacy measurements schemes have accuracy limitations since they rely on polling and sampling-based methods to gather traffic statistics. Typically, sampling methods have high sampling rates (e.g., one every 30,000 packets) and polling methods have large polling intervals. The literature [268] has shown that such methods are only suitable for coarse-grained visibility. The accuracy limitation of sampling and polling techniques hampers the development of measurement applications. For instance, it is not possible to accurately measure frequently changing TCP-specific fields such as congestion window, receive window, and sending rate.

Data streaming or sketching algorithms [269–272] were proposed to answer the limitation of sampling and polling. They address the following problem: *an algorithm is allowed to perform a constant number of passes over a data stream (input sequence of items) while using sub-linear space compared to the dataset and the dictionary sizes; desired statistical properties (e.g., median) on the data stream are then estimated by the algorithm*. The main problem with such algorithms is that they are tightly coupled to the metrics of interest. This means that switch vendors should build specialized algorithms, data structures, and hardware for specific monitoring tasks. With the constraints of CPU and memory in networking devices, it is challenging to support a wide spectrum of monitoring tasks that satisfy all customers. Legacy devices also lack the capability of customizing the processing behavior so that switches co-operate in the measurement process.

With the emergence of programmable switches, it is now possible to perform fine-grained measurements in the data plane at line rate. Moreover, data structures such as sketches and bloom filters can be easily implemented and customized for specific metrics of interest. Programmable switches pave the way for new areas of research in measurements since not only they provide flexibility in inspecting with high accuracy the traffic statistics, but also allow programmers to express reactive processing in real time (e.g., dropping a packet when a threshold is bypassed as done in Random Early Detection (RED) [273]).

INT provides path-level metrics, with data similar to that of polling-based techniques. Note that the metrics themselves are fixed; for instance, it is possible to determine the flow-level latency, but not the latency variation (jitter) [79]. The fixed metrics of INT also prevent performing network-wide measurements; note that the INT standard specification document does not mention methods to aggregate metadata and perform complex analytics in the data plane.

This section focuses on techniques that provide measurements that go beyond the fixed metrics extracted from the internal state of the switch.

#### 2) Generic Query-based Monitoring.

Operators constantly change their monitoring specifications. Adding new monitoring requirements on the fixed-function switching ASIC is expensive. Recent work explored the idea of providing a query-driven interface that allows operators to express their monitoring requirements. The queries can then be converted into switch programs (e.g., P4) to be deployed in the network. Alternatively, the queries can be executed on the control plane considering the measured information extracted from the data plane.

A simplistic attempt is FlowRadar [77], a system that stores counters for all flows in the data plane with low memory footprint, then exports periodically (every 10ms) to a remote collector. Liu et al. [78] proposed Universal Monitoring (UnivMon), an application-agnostic monitoring framework that provides accuracy and generality across a wide range of monitoring tasks. UnivMon benefits from the granularity of the data plane to improve accuracy and runs different estimation algorithms on the control plane. Narayana et al. [79] presented Marple, a query language based on common query constructs (i.e., map, filter, group by). Marple allows performing advanced aggregation (e.g., moving average of latencies) at line rate in the data plane. Similarly, Sonata [87] provides a unified query interface that uses common dataflow operators, and partitions each query across the stream processor and the data plane. PacketScope [93] also uses dataflow constructs but allows to query the internal switch processing, both in the ingress and the egress pipelines.

Many of the previous works use the sketch data structure. The work in [96] extended the sketching approach used in previous works to support the notion of time. The motivation of this work is that recently captured traffic trends are the most relevant in network monitoring. Huang et al. [97] proposed OmniMon, an architectural design that coordinates flow-level network telemetry operations between programmable switches, end-hosts, and controllers. Such coordination aims at achieving high accuracy while maintaining low resource overhead. Chen et al. [98] proposed BeauCoup, a P4-based measurement system that handles multiple heterogeneous queries in the data plane. It offers a general query abstraction that counts the *attributes* across related packets identified by *keys*, and flags packets that surpass a defined threshold.

Other approaches such as Elastic sketch [81] performs measurement that are adaptive to changes in network conditions (e.g., bandwidth, packet rate and flow size distribution). *Flow [85] supports concurrent measurements and dynamic queries. Such approach aims at minimizing the concurrency problems and the network disruption resulting from compiling excessive queries into the data plane. TurboFlow [86] aims at achieving high coverage without sacrificing information richness. Bai et al. [94] proposed FastFE, a system that performs traffic features extraction by leveraging programmable data planes. Extracted features are then used by traffic analysis and behavior detector ML techniques.

#### 3) Performance Diagnosis Systems.

Recent works are leveraging programmable data planes to diagnose network performance. The main motivation here is that fine-grained information can be monitored at line rate, mitigating the slow reaction to “gray failures” experienced by diagnosing end-hosts in legacy approaches.

Ghasemi et al. [80] proposed Dapper, an in-network TCP performance diagnosis system. Dapper analyzes packets in real time, and identifies and pinpoints the root cause of the bottleneck (sender, network, or receiver). Blink [90] also diagnoses TCP-related issues. In particular, it detects failures in the data plane based on retransmissions, and consequently, reroutes traffic. Other approaches attempt to diagnose performance degradation manifested by an increase of latency. Wang et al. [92] proposed SpiderMon, a system that performs network-wide performance degradation diagnosis. The key idea is to have every switch maintain fine-grained telemetry data for a short period of time, and upon detecting performance degradation (e.g., increased delay), the information is offloaded to a collector. Liu et al. [89] proposed a memory-efficient approach for network performance monitoring. This solution only monitors the top-*k* problematic flows.

#### 4) Queue and Other Metrics Measurement.

Programmable data planes allows querying the internal state of the queue with fine-grained visibility. Recent works leveraged this feature to provide better queueing information which can be used by various applications (e.g., AQMs, congestion control, etc.).

Chen et al. [88] proposed ConQuest, a P4-based queue measurement solution that determines the size of flows occupying the queue in real time, and identifies flows that are grabbing a significant portion of the queue. Joshi et al. [83] proposed BurstRadar, a system that uses programmable switches to monitor microbursts in the data plane. Mircorbursts are events of sporadic congestion that last for tens or hundreds of microseconds. Microbursts increase latency, jitter, and packet loss, especially when links' speeds are high and switch buffers are small.

Other works enabled measuring further metric. For instance, Ding et al. [91] proposed P4Entropy, an algorithm to estimate network traffic entropy (Shannon entropy) in the data plane. Tracking entropy is useful for calculating traffic distribution in order to understand the network behavior. Another example is the system proposed by Chen et al.

**TABLE 9:** Measurements schemes comparison.

| | Ref | Name | Core idea | Approx. | External computation | Data structure | Network wide | Platform HW | Platform SW |
|---|---|---|---|---|---|---|---|---|---|
| Generic query-based monitoring | [97] | OmniMon | Coordinates flow-level telemetry among devices | ✓ | | Slots (bloom filter) | | | |
| | [87] | Sonata | Uses scalable stream processor | | | Sketch | ✓ | | |
| | [77] | FlowRadar | Stores flow counters and periodically exports results | | | Bloom filter | | | |
| | [81] | Elastic Sketch | Adapts to network changing conditions | | | Sketch | | | |
| | [79] | Marple | Aggregates based on “map, filter, group by” constructs | ✓ | | Key-value store | | | |
| | [98] | BeauCoup | Enables simultaneous multiple distinct counting queries | | ✓ | Coupon collect (bloom filter) | ✓ | | |
| | [78] | UnivMon | Provides application-agnostic monitoring | | | Universal sketches | | | |
| | [85] | *Flow | Groups traffic in the switch and computes statistics in servers | ✓ | | GPV (register arrays) | ✓ | | |
| | [86] | TurboFlow | Produces fine-grained and unsampled flow records | ✓ | | Hash table | ✓ | | |
| | [96] | N/A | Enables time-aware monitoring | | | Time-aware sketch | ✓ | | |
| | [93] | PacketScope | Monitors packets’ lifecycle inside the switch | | | Key-value store (hash table) | ✓ | | |
| | [94] | FastFE | Extracts traffic features for ML models | ✓ | | key-value store | ✓ | | |

| | Ref | Name | Core idea | Scope | Reactive processing | Measured information | Network wide | Platform HW | Platform SW |
|---|---|---|---|---|---|---|---|---|---|
| Performance diagnosis systems | [80] | Dapper | Diagnoses TCP performance issues in the data plane | Identifies TCP bottleneck | N/A | Flight size, MSS, sender’s reaction time, loss, RTT, CWND, RWND | ✓ | | |
| | [92] | SpiderMon | Diagnoses latency with small memory footprint | Identifies flows affecting latency | Limits rate | Queue latency | | | |
| | [90] | Blink | Detects failures based on the predictable behavior of TCP | Identifies retransmitters | Reroutes traffic | RTO-induced retransmissions | ✓ | | |
| | [89] | N/A | Improves monitoring scalability by measuring subset of flows | Identifies top-k influential flows | N/A | Retransmissions, latency, packet loss, out-of-order | ✓ | | |

| | Ref | Name | Core idea | Passive measurement | Analysis | Measured information | Data structure | Platform HW | Platform SW |
|---|---|---|---|---|---|---|---|---|---|
| Queue/other measurement | [88] | ConQuest | Identifies flows contributing heavily to the queue | | Data plane | Queue occupancy | Count-min sketch | | |
| | [95] | N/A | Measures the RTT of TCP traffic in ISP networks | | Data plane | RTT from an ISP vantage point | Hash table | | |
| | [83] | BurstRadar | Monitors microbursts and captures telemetry for the contributing packets | ✓ | Control plane | Queue occupancy | Ring buffer | | |
| | [91] | P4Entropy | Estimates network traffic entropy | ✓ | Data plane | Shannon entropy | Count-min sketch | | |

[95] which passively measures the RTT of TCP traffic in ISP networks. RTT measurement is important for detecting spoofing and routing attacks, ensuring Service Level Agreements (SLAs) compliance, measuring the Quality of Experience (QoE), improving congestion control, and many others.

#### 5) Measurements Schemes Comparison, Discussions, and Limitations

Table 9 compares the measurements schemes.

**Generic Query-based Monitoring.** Some schemes (e.g., Sonata, FlowRadar, UnivMon) performed approximations of the metrics by using probabilistic data structures (e.g., sketch, bloom filter, etc), sampling methods, and top-*k* counting. In addition, some focused on a subset of traffic by leveraging event matching techniques. Such techniques are primarily used to achieve high resource efficiency (i.e., low memory footprint), but cannot achieve full accuracy. On the other hand, systems like OmniMon carefully coordinates the collaboration among different types of entities in the network. Such coordination will result in efficient resource utilization and fully accuracy. OmniMon follows a *split-merge* strategy where the *split* operation decomposes telemetry operations into partial operations and schedules them among the entities (switches, end-hosts, and controller), and the *merge* operation coordinates the collaboration among these entities. The idea is to leverage the strength of the data plane in the switches and end-hosts (i.e., per-flow measurements with high accuracy) and the control plane (i.e., network-wide collaboration). OmniFlow also ensures *consistency* through a synchronization mechanism and *accountability* through a system of linear equation considering packet loss and other data center characteristics. Results show that OmniMon reduces the memory by 33%-96% and the number of actions by 66%-90% when compared to state-of-the-art solutions.

Another criterion that differentiates the measurements schemes is whether there are computations being performed outside the data plane. Most of the systems use the control plane or external servers to perform complex computations since the data plane has limited support to complex arithmetic functions. While some systems (e.g., BeauCoup) do not require an external computation device, they often support less measurement operations.

The selection of the data structure to be used in the data plane strongly affects the measurements features supported by a certain scheme. For instance, the goal of BeauCoup is to enable simultaneous distinct counting queries; for such task, the authors based their design on the coupon-collection problem [274], which computes the number of random draws from $n$ coupons such that all coupons are drawn at least once. For example, if the threshold of distinct destination IPs for detecting superspreaders is 130, instead of recording all distinct destination IPs, 32 coupons are defined. Consequently, the destination IPs of incoming packets are mapped to those 32 coupons. While this data structure uses less memory than the other state-of-the-art measurement sketches, it is limited to specific objectives (distinct counting). Other works (e.g., UnivMon) focused on generalizing the measurement scenarios, and hence, used universal sketches as data structures.

Qiu et al. [96] focused on capturing traffic trends that are the most relevant in network monitoring and attacks' detection. The notion of time is not supported by native streaming algorithms. For instance, *count-min sketch*, which is a data structure that uses constant memory amount to record data, is oblivious to the passage of time. Existing solutions that consider recency are easily implemented on software, but not on programmable ASICs. For example, resetting a sketch after a timer expires requires iterating over the elements in the sketch, an operation that cannot be implemented in the data plane due to the lack of loops. Likewise, creating multiple sketches require additional stages which is limited in the hardware. Time-adaptive sketches utilize the idea of Dolby noise reduction [275, 276]; a *pre-emphasis* function inflates the update when a new key is inserted and a *de-emphasis* function restores the original value. This mechanism ages the old events over time, and therefore, improves the accuracy of recent events. The authors implemented the pre-emphasis function in the data plane using simple bit shifts, and the de-emphasis function in the control plane.

Finally, some systems considered network-wide monitoring, while others only restricted their capabilities to local per-switch measurements. Network-wide measurement is essential and can significantly improve the visibility of traffic, as discussed in Section XIII-D.

**Performance Diagnosis Systems.** Some performance diagnosis schemes restricted their scope to troubleshooting TCP. For instance, Dapper infers sending rate, Maximum Segment Size (MSS), sender's reaction time (time between received ACK and new transmission), loss rate, latency, congestion window (CWND), receiver window (RWND), and delayed ACKs. Based on the inferred variables, Dapper can identify the root cause of the bottleneck. Similarly, the authors in [89] monitored conditions such as retransmissions, packet loss, round-trip-time, out-of-order packets to identify the top-k problematic flows. Furthermore, Blink detects failures based on the predictable behavior of TCP, which retransmits packets at epochs exponentially spaced in time, in the presence of failure. Other schemes (i.e., SpiderMon) identify failures based on the increase of latency.

Some schemes use reactive processing to mitigate the network performance issue. For instance, Blink promptly reroutes traffic whenever failures signals are generated by the data plane, while SpiderMon limits the sending rate of the root cause hosts.

Finally, it is worth mentioning that some systems (e.g., Blink, Dapper) considered traces from real-world captures such as the ones provided by CAIDA for evaluation. Using real-world traces gives more credibility to the proposed solution.

**Queue and other Metrics Measurement.** Understanding the occupancy of the queue is useful for use cases such as mitigating congestion-based attacks, avoiding conflicting workloads, implementing new AQMs, optimizing switch configurations, debugging switch implementation, off-path monitoring of queues in legacy devices, etc. ConQuest performs queue measurements and identifies flows depending on the purpose (e.g., detecting bursty connections). It maintains compact snapshots of the queue, updated on each incoming packet. The snapshots are then aggregated in a round-robin fashion to approximate the queue occupancy. Afterwards, it cleans the previous snapshots to reuse it for further packets. Similarly, BurstRadar detects microbursts, which can increase latency, jitter, and packet loss, especially when links' speeds are high and switch buffers are small. It is almost impossible to detect microbursts in legacy switches which use sampling and polling-based techniques. BurstRadar detects microbursts, and captures a snapshot of the telemetry information of all the involved packets. Afterwards, an analysis is conducted on the snapshot to identify the microburst-contributing flow and the burst characteristics. Note that BurstRadar does not support measuring the queues of legacy devices passively, but ConQuest does. In addition, BurstRadar performs the analysis on the control plane, while ConQuest uses the data plane for analysis.

### 6) In-Network versus Legacy Measurements

Fig. 12 compares the legacy measurements to those conducted on programmable switches. There are two main classes of legacy measurements techniques. First, there are techniques that rely on polling and sampling (e.g., NetFlow). The differences between in-network measurements and polling/sampling-based schemes are closely related to the differences between legacy measurements and INT (see Table 5). For instance, the granularity of the measurements conducted in the data plane is much higher than those collected in traditional measurements (e.g., NetFlow). Further, it is not possible to conduct event-based monitoring in legacy approaches, whereas with in-network measurements, the programmer has the flexibility of customizing the monitoring based on conditions and thresholds. Second, there are techniques that rely on sketching or streaming algorithms to estimate the metric of interest. Such methods are tightly coupled with the metric, which forces hardware vendors to invest time and effort in building customized algorithms and data structures that might not be used by various customers. Moreover, with the constraints of routers and switches, it is not possible to implement a variety of monitoring tasks while still supporting the standard routing/switching functionalities. Therefore, such approaches are not scalable for the long run.

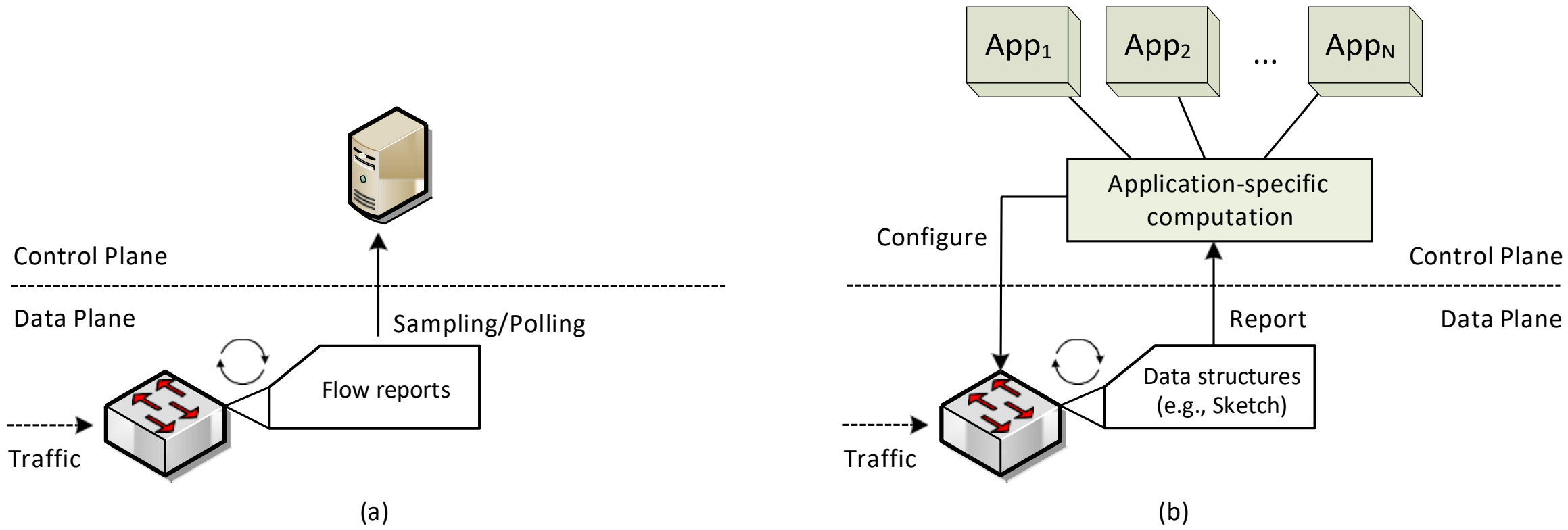

FIGURE 12: (a) Traditional measurements with sampling/polling. The switch uses sampling and polling protocols (e.g., NetFlow, SNMP) to generate fixed network flow records. Instead of collecting every packet, sampling collects only one every *N* number of packets. Records are then exported to an external server for further analysis. (b) Measurements with programmable switches (e.g., UnivMon [78]). The switch runs a universal algorithm over a universal data structure (e.g., universal sketch). The control plane then estimates a wide range of metrics for various applications. Note that this is not the only design possible for measurement tasks with programmable switches. The programmer has the flexibility to use customized algorithms than run at line rate in the data plane. Such algorithms can leverage various data structures in the P4 program (e.g., sketch, bloom filter) to store flow statistics. The switch then push statistics reports to the control plane for further analysis and reactive processing.

With programmable switches, it is possible to customize the monitoring tasks by implementing customized sketching/streaming algorithms as P4 programs. This advantage improves scalability as the operator can always modify the algorithms whenever needed.

## C. ACTIVE QUEUE MANAGEMENT (AQM)

### 1) Background

A fundamental component in network devices is the *queue* which temporarily buffers packets. As data traffic is inherently bursty, routers have been provisioned with large queues to absorb this burstiness and to maintain high link utilization. The majority of delays encountered in a communication session is a result of large backlogs formed in queues. Previous legacy devices are limited in the visibility of the queue as they provide little or no insight about which flows are occupying or sharing the queue [88]. Consequently, researchers have been investigating queue management algorithms to shorten the delay and mitigate packet losses, while providing fairness among flows. AQM is a set of algorithms designed to shorten the queueing delay by prohibiting buffers on devices from becoming full. The undesirable latency that results from a device buffering too much data is known as "Bufferbloat". Bufferbloat not only increases the end-to-end delay, but also decreases the throughput and increases the jitter of a communication session. Modern AQMs help in mitigating the bufferbloat problem [277–280]. Unfortunately, modern AQMs are typically not available in state-of-the-art network equipment; for instance, Controlled Delay (CoDel) AQM, which was proposed in 2013, and was proven in the literature to be effective in mitigating Bufferbloat [281], is still not available in most network equipment. With programmable switches, it is now possible to implement AQMs as P4 programs, which not only accelerates support for new AQMs, but also provides means to customize its parameters programmatically in response to network traffic. Moreover, programmable switches thrives for innovation on newer AQMs that can be easily implemented and rapidly tested.

### 2) Standardized AQMs Implementation

Kundel et al. [99] implemented the CoDel queueing discipline on a programmable switch. CoDel eliminates Bufferbloat, even in the presence of large buffers [100]. Sharma et al. [101] proposed Approximate Fair Queueing (AFQ), a mechanism built on top of programmable switches that approximates fair queuing on line rate. Fair Queueing (FQ) aims at fairly dividing the bandwidth allocation among active flows. Laki et al. [102] described an AQM evaluation testbed with P4 in a demo paper. The authors tested the framework with two AQMs: Proportional Integral Controller Enhanced (PIE) and RED. Papagianni et al. [103] implemented Proportional Integral $PI^2$ AQM on a programmable switch. $PI^2$ is an extension of PIE AQM to support coexistence between classic and scalable congestion controls in the public Internet. Kunze et al. [104] analyzed the implementation details of three AQMs, namely, RED, CoDel, and PIE on a hardware programmable switch (Tofino). Toresson [105] implemented a combination of PIE and Per-Packet Value (PPV) concept on a programmable switch.

**TABLE 10:** AQM schemes comparison.

| | Ref | Name | Idea | Params/thresh. | Multiple queues | Data structure | Implementation |
|---|---|---|---|---|---|---|---|
| Standardized AQMs | [99] | P4-CoDel | Implements CoDel on P4 | 2 | ✓ | Registers | BMv2 |
| | [101] | AFQ | Approximates fair queueing in the switch | 4 | | Count-min sketch | Cavium OCTEON |
| | [102] | N/A | Evaluation testbed for PIE and RED | RED 1, PIE 5 | ✓ | Registers | BMv2 |
| | [103] | PI2 for P4 | Implements of PI2 on P4 | 3 | ✓ | Registers | BMv2 |
| | [104] | N/A | Implementation challenges of standardized AQMs on Tofino | N/A | | Registers | Tofino |
| Custom AQMs | [106] | ADS | Approximates Shortest Remaining Processing Time (SRPT) | 2 | | N/A | BMv2 |
| | [107] | P4-ABC | Implements activity-based congestion management | 2 | ✓ | Registers | BMv2 |
| | [108] | SP-PIFO | Approximates Push-In First-Out (PIFO) queues | 2 | | Registers | Tofino |
| | [109] | MTQ/QTL | Implements MTQ/QTL on programmable switches | 3 | ✓ | Registers | BMv2 |

#### 3) Custom AQMs

Mushtaq et al. [106] approximated Shortest Remaining Processing Time (SRPT) on a programmable switch. Their method, which they refer to as Approximate and Deployable SRPT (ADS), was evaluated and it was shown that it can achieve performance close to SRPT. Menth et al. [107] implemented activity-based congestion management (ABC) on programmable switches. ABC aims at ensuring fair resource sharing as well as improving the completion times of short flows. Alcoz et al. [108] proposed SP-PIFO, a method that approximates Push-In First-Out (PIFO) queues on programmable data planes. The method consists of an adaptive scheduling algorithm that dynamically adapts mapping between packet ranks and Strict Policy (SP) queues. Kumazoe et al. [109] implemented MTQ/QTL scheme on P4.

#### 4) AQM Schemes Comparison, Discussions, and Limitations

Table 10 compares the aforementioned AQM schemes. Some schemes require tuning a number of parameters and thresholds so that they operate well in certain network conditions. It is worth mentioning that a scheme becomes hard to manage and less autonomous when the number of parameters and thresholds is high.

Some schemes are simple to implement in the data plane. CoDel's algorithm can be easily expressed in the data plane as it consists of comparisons, counting, basic arithmetic, and dropping packets. Similarly, $PI^2$ is simple to implement as it is mostly based on basic bit manipulations. FQ algorithms on the other hand are difficult to implement on hardware as they require complex flow classification, per-packet scheduling, and buffer allocation. Such requirements make FQ algorithms expensive to be implemented on high-speed devices. AFQ aims at *approximating* fair queueing by using programmable switches' features such as mutating switch state, performing basic calculations, and selecting the egress queue of a packet. AFQ's operations can be summarized as follows: 1) per-flow state, which includes the number and timing information of the previous packet pertaining to that flow, is approximated; 2) the position of each packet in the output schedule is determined; 3) the egress queue to use is selected; and 4) the packet is dequeued based on the approximate sorted order. Note that AFQ uses a probabilistic data structure (count-min sketch) since it only approximates the states, and uses multiple queues in its implementation.

#### 5) AQMs on Programmable Switches and Fixed-function Devices

Inventing novel AQMs that control queueing delay, mitigate bufferbloat, and achieve fairness with different network conditions (e.g., short/long RTTs, lossy networks, WANs) is an active research area. Typically, new AQMs are implemented and tested in software (e.g., as a Linux queueing discipline (*qdisc*) used with traffic control (*tc*)), which is limited when the objective is to deploy the AQMs on production networks. With programmable switches, AQMs are implemented in P4 programs, which foster innovation and enhance testing with production networks. Additionally, operators can create their own customized AQMs that perform efficiently with their typical network traffic.

Historically, deploying AQMs on network devices is a lengthy and costly process; once an effective AQM is published and thoroughly tested, equipment vendors start investigating whether it is feasible to implement it on future devices. Such process might take years to finish, and by then, new network conditions evolve, requiring new AQMs. With programmable switches, this process is cost-efficient and relatively fast (can be completed in weeks).

**TABLE 11:** AQMs on programmable and fixed-function switches.

| Feature | Programmable switches | Fixed-function devices |
|---|---|---|
| Innovation | Higher; new AQMs are expressed in P4 programs | Lower; only developed by equipment vendors |
| Exclusivity | Higher; operators can implement their own custom AQMs without disclosing technical information | Lower; most supported AQMs are standards |
| Readiness | Faster (weeks to months); once an AQM is expressed in P4, it can be immediately available | Slower (years) |
| Cost | Lower | Higher |
| Tweakable | Higher; even standard AQMs can be customized and tweaked based on network traffic | Lower; only through parameters |

Table 11 compares the features of AQMs on programmable switches versus fixed-function devices. While new AQMs can be devised on programmable switches, there are some constraints that should be taken into account. First, the traffic manager of a programmable switch is not programmable itself using P4; this is where AQM algorithms are typically implemented in legacy devices. Nevertheless, there are efforts that started investigating methods for emulating programmable traffic managers [282]. Second, current AQMs do not consider the constraints of high-speed ASICs, and thus, cannot be directly implemented as they are using P4. Researchers overcome such limitations through approximations or through rewriting the AQM logic in an high-speed ASIC-friendly way. Third, queue state information is not available in the ingress before packet enqueue. Consequently, AQM are usually implemented in the egress. Such limitations can be addressed in future research works pertaining to AQMs.

## D. QUALITY OF SERVICE AND TRAFFIC MANAGEMENT

### 1) Background

Meeting diverse Quality of Service (QoS) requirements is a fundamental challenge in today's networks. Traffic Management (TM) provides access control that guarantees that the traffic admitted to the network conforms to the defined QoS specifications. TM often regulates the rate of a flow by applying traffic policing. New generation of programmable switches facilitate traffic policing and differentiation by allowing network operators to express their logic in a programming language (P4). This section explores the works on programmable switches that involve QoS and TM.

### 2) Quality of Service

Bhat et al. [110] described a system where programmable switches intelligently route traffic by inspecting application headers (layer-5) to improve users' QoE. Chen et al. [111] proposed a bandwidth manager for end-to-end QoS provisioning using programmable switches. The system classifies packets into different categories based on their QoS demands and usages, and uses two-level queue when prioritizing. Chen et al. [112] proposed a system that uses the metering capabilities of a programmable switch to measure the flow rate. It then marks packets when the flow rate exceeds a certain threshold. The sender then adjusts its congestion window proportional to the marking packet ration. The goal of this approach is to avoid the frequent packet drops of TCP when rate limiting QoS scheme is present.

**TABLE 12:** QoS/TM schemes comparison.

| | Ref | Idea | Input | Many Queues | Target |
|---|---|---|---|---|---|
| QoS | [110] | App-layer headers inspect | Layer-5 headers | ✓ | SW |
| | [111] | BW manager for e2e QoS | Flow ID, rate | | HW |
| | [112] | Rate limiting to improve QoS | Flow rate | ✓ | SW |
| TM | [113] | Traffic mgmt. (RL-SP / DRR) | Traffic rate | | SW |
| | [114] | MCM-based traffic meter | Traffic rate, VN ID | ✓ | SW |

### 3) Traffic Management

Tokmakov et al. [113] proposed RL-SP-DRR, a traffic management system that combines Rate-limited Strict Priority (RL-SP) and Deficit round-robin (DRR) to achieve low latency and fair scheduling while improving link utilisation, prioritization and scalability. Lee et al. [114] implemented a traffic meter based on Multi-Color Markers (MCM) on programmable switches to support multi-tenancy environments.

### 4) QoS/TM Schemes Comparison, Discussions, and Limitations

Table 12 compares the QoS/TM schemes. The main idea in [110] is to translate application-layer header information into link-layer headers (Q-in-Q 802.1ad) for the core network in order to perform QoS routing and provisioning. The authors adopted the Adaptive Bit Rate (ABR) video streaming as a use case to showcase the QoS improvements and the flexibility of traffic management. Such approach is interesting since switches are inspecting higher layers in the protocol stack. This capability is not available in non-programmable devices. Note however that the solution was only implemented on a software switch (BMv2). When it comes to hardware switches, the solution might face challenges to run at line rate when processing L5 headers. Therefore, the authors left the hardware implementation as a future work.

The other approaches considered traffic rates as inputs rather than inspecting application-layer headers. [114] focused on isolating virtual networks (VN). A VN has to have its own dedicated bandwidth (i.e., other networks' traffic should not impact the bandwidth) and should be able to differentiate priorities in order to provide QoS for its flows. While the solution was not implemented on hardware (the authors left the hardware implementation as future work), it is worth noting that this system relies on metering primitives which are available in today's hardware targets (e.g., meters in Tofino). Similarly, [113] was only implemented on a software switch (BMv2) and was evaluated by comparison against standard priority-based and best-effort scheduling. This system uses multiple priority queues, a feature supported in hardware targets. Therefore, the system could be implemented on hardware switches. The approach in [111] aims at limiting the maximum allowed rate and at maximizing bandwidth utilization. This is the only work that was implemented on a hardware switch (Tofino), and its design was compared against approaches based on OpenFlow.

#### 5) Comparison of QoS/TM between Legacy and Programmable Networks

The ability to perform QoS-based traffic management in legacy networks is restricted to algorithms that consider standard header fields (e.g, differentiated services [283]). On the other hand, programmable switches can parse, modify, and process customized protocols. Hence, operators now have the ability to perform TM by inspecting custom headers fields. Moreover, it is possible to extract with high-granularity metadata pertaining to the state of the switch (e.g., queue occupancy, packet sojourn time, etc.) at line rate. Such information can significantly help switches take better decisions while performing traffic management.

### *E. MULTICAST*

#### 1) Background

Multicast routing enables a source node to send a copy of a packet to a group of nodes. Multicast uses in-network traffic replication to ensure that at most a single copy of a packet traverses each link of the multicast tree. Perhaps the most widely multicast routing protocol deployed in traditional networks is the Protocol-Independent Multicast (PIM) protocol [284]. PIM and other multicast routing protocols require a signaling protocol such as the Internet Group Management Protocol (IGMP) [285] to create, change, and tear-down the multicast tree. Traditional multicast presents some challenges. For example, it is not suitable for environments where multicast group members constantly move (e.g., virtual machine migration and allocation). In such cases, the multicast tree must be updated dynamically, which may require substantial time and overhead. Also, some routers support a limited number of group-table entries, which does not scale in environments such as datacenters. Additionally, the signaling protocol and multicast algorithm are hard coded in the router, which reduces flexibility in building and managing the tree. Finally, it is not possible to implement multicast based on non-standard header fields.

#### 2) Source-routed Multicast

Shahbaz et al. [115] presented ELMO, a multicast scheme based on programmable P4 switches for datacenter applications. ELMO encodes the multicast tree in the packet header, as opposed to maintaining group-table entries inside routers. Kadosh et al. [116] implemented ELMO using a hybrid dataplane with programmable and non-programmable elements. ELMO is intended for multi-tenant datacenter applications requiring high scalability. Braun et al. [117] presented an implementation of the Bit Index Explicit Replication (BIER) architecture [286] with extensions for traffic engineering. Similar to ELMO, BIER removes the per-multicast group state information from switches by adding a BIER header, which is used to forward packets. BIER does not require a signaling protocol for building, managing, and tearing down trees.

**TABLE 13:** Source-routed multicast schemes comparison (source: [115]).

| Ref | Name | Group size | Network size | Heavy processing | Platform | |
|---|---|---|---|---|---|---|
| | | | | | HW | SW |
| [115] | ELMO | None | None | ≺ | | |
| [117] | BIER | 2.6K | 2.6K | | | |

#### 3) Priority-based Decentralized Multicast

Cloud applications in data centers often require file transfers to be completed in a prioritized order. Luo et al. [287] proposed Priority-based Adaptive Multicast (PAM), a preemptive and decentralized rate control protocol for data center multicast. The switches explicitly and preemptively compute sending rates based on priorities encoded in scheduling headers, and the real-time link loads.

#### 4) Multicast Schemes Comparison, Discussions, and Limitations

Table 13 compares the source-routed multicast schemes. Both ELMO and BIER are source-routed multicast schemes. In BIER, group members are encoded as bit strings and are then inspected by switches to identify the output port. Such scheme requires heavy processing on the switch, hampering the execution at line rate. Consequently, the authors only implemented BIER on a software switch (BMv2). ELMO on the other hand has no restrictions on the group and network sizes, and was implemented on a hardware switch, running at line rate.

Other schemes like PAM addressed the challenges faced in file transfers by data center cloud applications. For instance, when sharing the link with other latency-sensitive flows, file transfers suffer from continuous changes in the link's bandwidth, affecting the flow completion times. To solve this problem, PAM adopted a scheduling scheme that performs adaptive rate allocations in RTT scales. Other aspects that were addressed by PAM include: fault tolerance and scalability of file transfers; limited number of priority queues; and the challenges of performing complex computations in data plane.

#### 5) Comparison P4-based and Traditional Multicast

Table 14 compares P4-based multicast and traditional multicast. The main advantages of implementing multicast routing with programmable P4 switches are: i) the group membership is encoded in the packet itself, which permits the creation of arbitrary multicast tree based on the application. For example, a multicast tree to update software devices may prioritize bandwidth over latency, while one for media traffic may prioritize latency; ii) switches do not need to store per-group state information, although tables can be customized and used in conjunction with the tree encoded in the packet header; iii) groups can be reconfigured easily by changing the information in the header of the packet; and iv) the elimination of the signaling protocol to build, manage, and tear-down the tree results in consider simplification and flexibility for the operator.

**TABLE 14:** Comparison between P4-based and traditional multicast.

| Feature | P4-based multicast | Traditional multicast |
|---|---|---|
| Scalability | High; no state information required in switches | Low; state information required in switchers per-group |
| Tree management | Flexible; custom multicast algorithm and features can be implemented | Inflexible; signaling protocol required and hard coded in the switch |
| Packet overhead | High; multicast tree is encoded in packet header | No packet overhead |
| Dynamic tree updates | Easy; packet header carries update information | Complex; topology challenges may trigger time-consuming tree changes |
| IP address constraint | Flexible; switch can multicast packets independently of the type of IP address | Fixed; switch is hard-coded to only multicast packets when the destination IP is a multicast address |

### F. SUMMARY AND LESSONS LEARNED

Performing network-wide monitoring and measurements is of utmost importance for network operators to diagnose performance degradation. A wide range of research efforts harness streaming methods that utilize various data structures (e.g., sketches, bloom filters, etc.) and approximation algorithms. Further, the majority of measurements work provide a query-based language to specify the monitoring tasks. Future measurement works should consider generalizing the monitoring jobs, reducing storage requirements, managing accuracy-memory trade-off, extending monitoring primitives, minimizing controller intervention, and optimizing the placement of switches in a legacy network. Another line of research aim at combating congestion and reducing packet losses by analyzing measurements collected in the data plane and by applying queue management policies. Congestion control is enhanced by adopting techniques such as throttling senders, cutting payloads, enforcing sending rates by leveraging telemetry data, and separating traffic into different queues. Furthermore, a handful of works are investigating methods to improve QoS by applying traffic policing and management. Techniques adopted include application-layer inspection, traffic metering, traffic separation, and bandwidth management. Finally, the scalability concerns of multicast in legacy networks are being mitigated with programmable switches. Recent efforts proposed encoding multicast trees into the headers of packets, and using programmable switches to parse these headers and to determine the multicast groups. Future endeavours should investigate incremental deployment (i.e., interworking with legacy multicast schemes), and reliability enhancement (e.g., by adopting layering protocols such as Pragmatic General Multicast (PGM) and Scalable Reliable Multicast (SRM)).

## VIII. MIDDLEBOX FUNCTIONS

RFC 3234 [288] defines *middlebox* as a device that performs functions other than the standard functions of an IP router between a source and a destination host. In legacy devices, middlebox functions are designed and implemented by manufacturers. Hence, they are limited in the functionalities they provide, and typically include standard well-known functions (e.g., NAT, protocol converters (6to4/4to6), etc.). To overcome this limitation, the trend moved towards implementing middleboxes in x86-based servers and in data centers as Network Function Virtualization (NFVs). While this shift accelerated innovation and introduced a wide range of new applications, there was some performance implications resulting from operating systems' scheduling delays, interrupt processing latency, pre-emptions, and other low-level OS functions. Since programmable switches offer the flexibility of inspecting and modifying packets' headers based on custom logic, they are excellent candidates for enabling middlebox functions, while operating at line rate without performance implications.

### A. LOAD BALANCING

#### 1) Background

A cloud data center, such as a Google or Facebook data center, provides many applications concurrently, such as email and video applications. To support requests from external clients, each application is associated with a publicly visible IP address to which clients send their requests and from which they receive responses. This IP address is referred to as Virtual IP (VIP) address. The external requests are then directed to a software load balancer whose task is to distribute requests to the servers, balancing the load across them. The load balancer is also referred to as layer-4 load balancer because it makes decisions based on the 5-tuple source IP address and port, destination IP address and port, and transport-layer protocol. This state information is stored in a connection table containing the 5-tuple and the Direct IP (DIP) address of the server serving that connection. State information is needed to avoid disruptions caused by changes in the DIP pool (e.g., server failures, addition of new servers). The load balancer also provides a translation functionality, translating the VIP to the internal DIP, and then translating back for packets traveling in the reverse direction back to the clients. The traditional software-based load balancer is illustrated in Fig. 13(a).

#### 2) Stateful Load Balancing

Recent works presented schemes where load balancing functionality is implemented in programmable P4 switches. The main idea consists of storing state information directly in the switch's dataplane. The connection table is managed by the software load balancer, which can be implemented either in the switch's control plane or as an external device, as shown in Fig. 13(b). The software load balancer adds new entries in the switch's table as they arrive, or removes old entries as flows end.

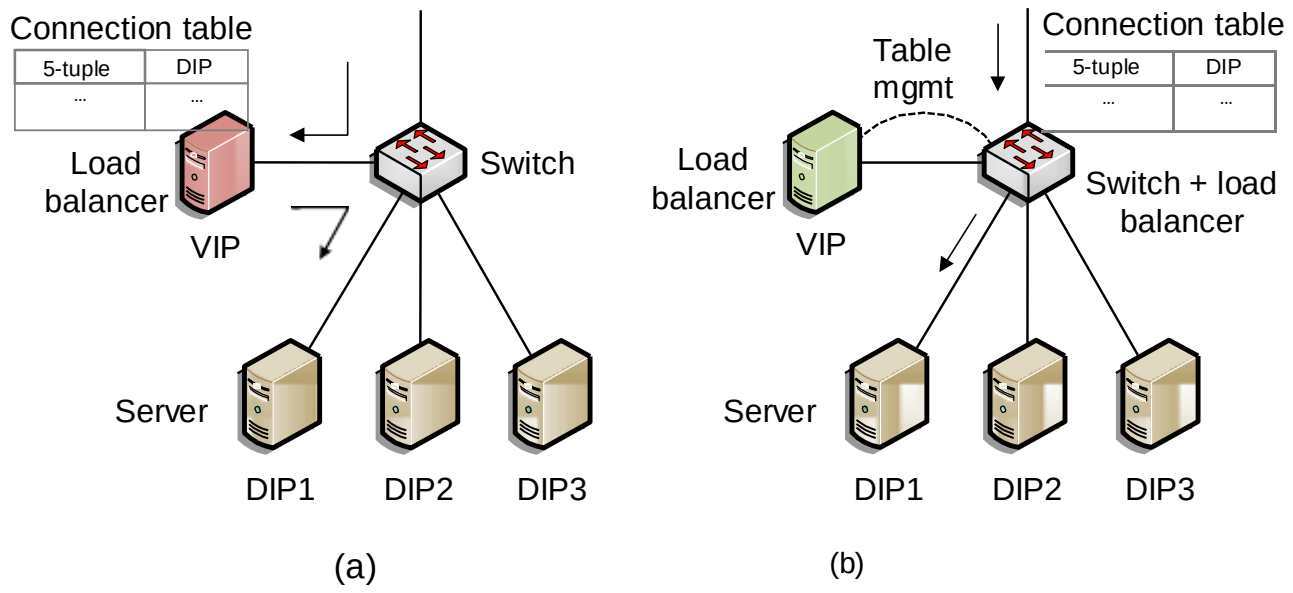

**FIGURE 13:** (a) Traditional software-based load balancing. (b) Load balancing system implemented by a programmable switch.

Katta et al. [118] proposed HULA, a load balancer scheme where switches store the best path to the destination via their neighboring switches. This strategy avoids storing the congestion status of all paths in leaf switches. Bennet et al. [119] extended this approach to support multi-path transport protocols (e.g., Multi-path TCP (MPTCP)). Another significant work is SilkRoad, [120], a load balancer that provides a direct path between application traffic and servers. Other mechanisms such as DistCache [121] enables load balancing for storage systems through a distributed caching method. DASH [122] proposed a data structure that leverages multiple pipeline stages and per-stage SALUs to dynamically balance data across multiple paths. The aforementioned approaches work under specific assumptions about the network topology, routing constraints, and performance. Contra [123] generalized load balancing to work with various topologies and under multiple constraints by using a performance-aware routing mechanism.

### 3) Stateless Load Balancing

Recent advances in customized and stateful packet processing in programmable switches not only forked a variety of stateful load balancing schemes, but also stateless ones. Stateless load balancing in this context avoids storing per-connection state in the switch.

Perhaps the first and most significant P4-based stateless load balancing scheme is Beamer [124]. Instead of storing the state in the switch, Beamer leverages the connection state already stored in backend servers to perform the forwarding. Another scheme is SHELL [125], which is an application-agnostic, application-load-aware approach that uses a power-of-choices scheme to dispatch flows to a suitable instance. Other approaches such as W-ECMP [126] were built to solve the issue of hash collision in the well-known Equal-Cost Multi-Path (ECMP) scheme. W-ECMP maintains a maximum utilization table, which is used as weights to determine the routing probability for each path. Note that W-ECMP is not storing *per-connection* state information in the data plane.

### 4) Load Balancing Schemes Comparison, Discussions, and Limitations

Table 15 compares the aforementioned load balancing schemes. The key idea of switch-based stateful load balancing is to eliminate the need for a software-layer while mapping a connection to the same server, ensuring Per-Connection Consistency (PCC) property. The majority of the proposed approaches are stateful, meaning that the switches store information locally to perform load balancing.

Some approaches (e.g., HULA, MP-HULA, Contra) use active probing to collect network performance metrics. Such metrics are then analyzed by the switches to make load balancing decisions. Note that probing increases the bandwidth overhead which might result in performance degradation.

In the presence of multi-path transport protocols (e.g., MPTCP), systems such as HULA provide sub-optimal forwarding decisions when several subflows pertaining to a single connection are pinned on the same bottleneck link. As a result, schemes such as MP-HULA, Contra, and Dash were proposed to support multi-path transport protocols. For instance, MP-HULA is a transport layer multi-path aware load-balancing scheme that uses the best-k paths to the destination through the neighbor switches.

Other approaches are stateless. Beamer relies on using the connection state already stored in backend servers to ensure that connections are never dropped under churn. On the other hand, SHELL, which assigns new connections to a set of pseudo-randomly chosen application instances, marks packets, allowing the load-balancer to direct them without storing state. W-ECMP makes forwarding decisions based on weights adjusted according to the link utilization. Finally, it is important for a load balancing scheme to be adaptive and

**TABLE 15:** Load balancing schemes comparison.

| | Scheme | Name | Centralized | Active probing | MP-TCP support | Failure handling | Platform | |
|---|---|---|---|---|---|---|---|---|
| | | | | | | | Hardware | Software |
| Stateful | [118] | HULA | ✓ | | ✓ | | ✓ | |
| | [120] | SilkRoad | ✓ | ✓ | ✓ | | | ✓ |
| | [119] | MP-HULA | ✓ | | | | ✓ | |
| | [121] | DistCache | ✓ | ✓ | ✓ | | | ✓ |
| | [122] | DASH | ✓ | | | ✓ | ✓ | |
| | [123] | Contra | ✓ | | ✓ | | ✓ | |
| Stateless | [124] | Beamer | | ✓ | | ✓ | | |
| | [125] | SHELL | ✓ | ✓ | ✓ | ✓ | | ✓ |
| | [126] | W-ECMP | ✓ | ✓ | ✓ | ✓ | ✓ | |

**TABLE 16:** Switch-based and server-based load balancers.

| Feature | Switch-based | Server-based |
|---|---|---|
| Throughput | Higher; (e.g., SilkRoad with 6.4Tbps ASIC can achieve about 10Gpps) | Lower (e.g., 9Mpps per core [289]) |
| Latency | Lower; sub-microseconds from ingress to egress | Higher; additional latency when processing new requests* |
| Scalability | Lower; connection is stored in limited SRAM | Higher |
| Policy flexibility | Limited; hash-based flow assignments may lead to imbalance | Flexible policies can be written in software |
| System complexity | Simpler; it requires a customized parser, match-action tables | More complex; it requires coordination with routers, tunneling (e.g., GRE encapsulation) |

*After the first packet is processed, no additional latency is observed [289].

handle network failures. Furthermore, it should mitigate load imbalance in asymmetric topologies.

#### 5) Comparison between Switch-based and Server-based Load Balancer

Table 16 shows a comparison between switch-based and server-based load balancers. There is a significant improvement in the throughput when load balancing is offloaded to the switches; for instance, SilkRoad [120], which is a load balancing scheme in the data plane, achieves 10 billion packets per second (pps) while operating at line rate. Software load balancers on the other hand achieve a much lower throughput, nine million PPS on average. Software-based load balancers also incur additional latency overhead when processing new requests. It is relatively easy to install additional software load balancers, which makes it more scalable than switch-based load balancing schemes. Moreover, software load balancers are more flexible in assigning flow identification policies. Finally, switch-based schemes are simpler as the whole logic is expressed in a program (customized parser and match-action tables), whereas server-based balancers might require additional coordination with routers (e.g., tunneling).

### B. CACHING

#### 1) Background

Modern applications (e.g., online banking, social networks) rely on key-value stores. For example, retrieving a single web page may require thousands of storage accesses. As the number of users increases to millions or billions, the need for higher throughput and lower latency is needed. A challenge of key-value stores is the non-uniform access of items. Instead, popular items, referred to as “hot items”, receive more queries than others. Furthermore, popular items may change rapidly due to popular posts, limited-time offers, and trending events [127]. Fig. 14(a) shows a typical skew key-value store system which presents load imbalance among servers storing key-value objects. The performance of such systems may present reduced throughput and long latencies.

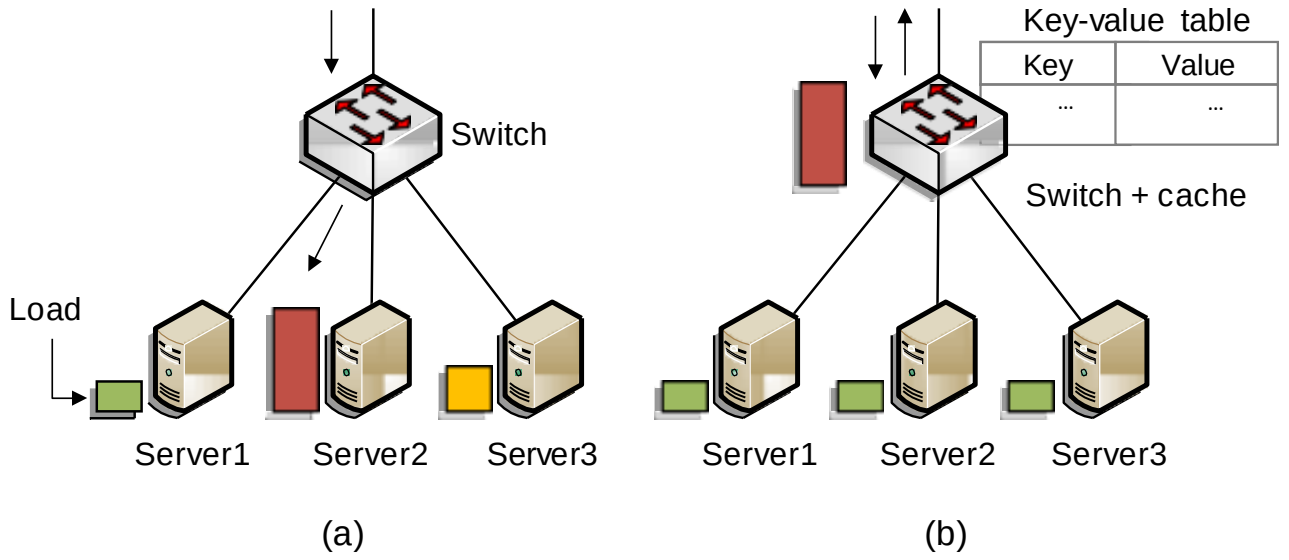

**FIGURE 14:** (a) Traditional software-based caching. (b) Switch-based caching.

For example, server 2 may add substantial latency as a result of storing a hot item and being over-utilized, while server 1 is under-utilized.

#### 2) Key-value caching

Fig. 14(b) illustrates a system where a programmable switch receives a query before forwarding them to the server storing the key. The switch is used as an “in-network cache”, where the hottest items are stored. When a read request for a hot key is received, the switch consults its local table and returns the value corresponding to that key. If the key is missed (i.e., the case for non-hot keys) then the switch forwards the request to the appropriate server. When a write request is received, the switch checks its local table and evicts the entry if the key is stored there. It then forwards the request to the appropriate backend server. A controller periodically collects statistics to update the cache with the current hot items.

A noteworthy approach is NetCache [127], an in-network architecture that uses programmable switches to store hot items and balance the load across storage nodes. Similarly, Liu et al. [128] proposed IncBricks, a hardware-software co-designed in-network caching fabric for key-value pairs with basic computing primitives in the data plane. Cidon et al. [129] proposed AppSwitch, a packet switch that performs load balancing for key-value storage systems, while exchanging only a single message from the key-value client to the server. Wang et al. [130] proposed CONCORDIA, a rack-scale Distributed Shared Memory (DSM) with in-network cache coherence. While the system targets cache coherence, the authors implemented a distributed key-value store to demonstrate the practical benefits of the system. Similarly, Li et al. [131] proposed Pegasus which acts as an in-network coherence directory tracking and managing the replication of objects.

#### 3) Application-specific caching

Other class of caching schemes target specific applications rather than caching arbitrary key-value pairs. Signorello et al. [132] developed a preliminary implementation of Named Data Networking (NDN) instance using P4 that caches requests to optimize its operations. Grigoryan et al. [133] proposed a system that caches Forwarding Information Base (FIB) entries (the most popular entries) in fast memory in order to minimize the TCAM consumption and to avoid the TCAM overflow problem. Zhang et al. [134] proposed B-Cache, a framework that bypasses the original processing pipeline to improve the performance of caching. Vestin et al. [135] proposed FastReact, a system that enables caching for industrial control networks. Finally, Woodruff et al. [136] proposed P4DNS, an in-network cache for Domain Name System (DNS) entries.

**TABLE 17:** Caching schemes comparison.

| | Scheme | Name | Cached data | Network accelerator | Automatic entry indexing | Custom protocol | Multi-level cache | Platform | |
|---|---|---|---|---|---|---|---|---|---|
| | | | | | | | | HW | SW |
| **Generic key-value caching** | [127] | NetCache | Key-value | ✓ | | | ✓ | | |
| | [128] | IncBricks | Key-value | | ✓ | | ✓ | | |
| | [129] | AppSwitch | Key-value | ✓ | ✓ | | ✓ | | |
| | [130] | CONCORDIA | Cache metadata | ✓ | | | ✓ | | |
| | [131] | Pegasus | Generic objects | ✓ | | ✓ | ✓ | | |
| **Application-specific caching** | [132] | NDN.p4 | NDN names | ✓ | | | ✓ | | |
| | [133] | PFCA | FIB entries | ✓ | | ✓ | | | |
| | [134] | B-Cache | FIB entries | ✓ | | ✓ | ✓ | | |
| | [135] | FastReact | Sensor readings | ✓ | ✓ | | ✓ | | |
| | [136] | P4DNS | DNS entries | ✓ | | ✓ | ✓ | | |

### 4) Caching Schemes Comparison, Discussions, and Limitations

Table 17 compares the aforementioned caching schemes. Schemes can be separated based on the type of data they aim to cache. For instance, NetCache, AppSwitch, and IncBricks cache arbitrary key-value pairs, while NDN.p4 caches only NDN names. Further, some schemes (e.g., NetCache, P4DNS, etc.) automatically index entries to be cached based on their access frequencies, while others require the operators to manually specify the entries. Another important distinction is whether the scheme uses a custom protocol or not. For instance, switches in NetCache parse a custom protocol that carries key-value pairs, while switches in P4DNS parse standard DNS headers.

The main motivation of switch-based caching schemes is to improve the performance issues of server-based schemes. For instance, NetCache, which efficiently detects hot key-value items and serves them in the data plane, was capable of handling two billion queries per second for 64,000 items with 16-bytes keys and 128-bytes values. Compared to commodity servers, NetCache improves the throughput by 3-10 times and reduces the latency of 40% of queries by 50%. In addition to the throughput, the latency of the queries is also a major metric to improve. In IncBricks, the latency of requests is reduced by over 30% compared to client-side caching systems.

Similarly, B-Cache aims at improving the performance by caching behaviors defined along the processing pipeline into a single cache match-action table. The motivation behind B-Cache is that the performance of the data plane decreases significantly as the complexity of the P4 program and the packet processing pipeline grows. When a match occurs, the packet bypasses the original pipeline, making the performance of caching independent of the pipeline length. Note however that this system was evaluated on a software switch (BMv2), and it is not certain whether this design is always feasible on hardware targets.

**TABLE 18:** Switch-based and server-based caching.

| Feature | Switch-based | Server-based |
|---|---|---|
| Throughput | Higher; (e.g., NetCache, 2BQPS🗁) | Lower; 0.2BQPS |
| Latency | Lower; (e.g., NetCache, 🖮 µ♦, mostly caused by the client) | Higher; 🗁🗄 µ♦ |
| Key size | Not flexible (limited by packet header length) | Arbitrary |
| Value size | Not flexible (limited by the amount of state accessed when processing a packet) | Arbitrary |
| Load imbalance | No | Yes |
| System complexity | Simpler; it requires a customized parser, match-action tables | More complex; it requires coordination with routers, tunneling (e.g., GRE encapsulation) |
| Table size | Limited by RAM | Arbitrary |
| Cache policies | Limited by table size | Arbitrary |

🗁BQPS: Billion Queries Per Second.

Other caching schemes are more targeted for specific applications. As examples, FastReact enables caching for industrial control networks, while P4DNS caches DNS entries. Further, some schemes offer multi-level caching (e.g., level-1 and level-2 caches).

Unlike the other approaches which store cached data in the data plane, CONCORDIA coordinates coherence among the cache of servers, and therefore only stores the cache's metadata in the switch.

### 5) Comparison between Switch-based and Server-based Caching

Table 18 compares the switch-based versus server-based caching schemes. The throughput when data is cached on the switch is order of magnitude larger than that of general purpose servers. The latency is also reduced by 50%, and most of it is induced by the client. The switched-based caching solves the load imbalance problem and is simpler as the whole logic is expressed in a program. Server-based caching on the other hand is more flexible regarding cache policies, as well as keys, values, and tables' sizes.

**TABLE 19:** Telecom schemes comparison.

| | Ref | Idea | Deployment | Latency scale | Concurrent users | Platform | |
|---|---|---|---|---|---|---|---|
| | | | | | | HW | SW |
| **5G functions** | [137] | Enhances the data path in 5G multi-tenants | Backhaul | Microseconds | N/A | | |
| | [138] | Implements a 5G firewall in the switch | Backhaul | Microseconds | 1K | | |
| | [139] | Performs signaling in the data plane | Core | Milliseconds | 65K | | |
| | [140] | Offloads MPC user plane functions to switch | Core | Microseconds | 65K-1M | | |
| | [141] | Hybrid next-generation NodeB (gNB) | 5G RAN | N/A | N/A | | |
| | [142] | Provides smart handover for mobile UE | Between CU and DU | N/A | N/A | | |
| | [143] | Potentials of programmability in SDN/NFV | Edge | Microseconds | N/A | | |
| | [144] | Implements UPF in the data plane | Core | Milliseconds | 300-10,000 | | |
| **Media offloading** | [145] | Offloads media traffic relay to switch | Edge | Nanoseconds | 65K-1M | | |
| | [146] | Offloads video processing to switch | Edge | N/A | N/A | | |

## C. TELECOMMUNICATION SERVICES

### 1) Background

The evolution of the current mobile network to the emerging Fifth-Generation (5G) technology implies significant improvements of the network infrastructure. Such improvements are necessary in order to meet the Key Performance Indicators (KPIs) and requirements of 5G [290]. 5G requires ultra-reliable low latency and jitter (microseconds-scale). As programmable switches fulfill these requirements, researchers are investigating the idea of offloading telecom-oriented VNFs running on x86 servers to programmable hardware.

### 2) 5G Functions

Ricart-Sanchez et al. [137] proposed a system that uses programmable data plane to enhance the performance of the data path from the edge to the core network, also known as the backhaul, in a 5G multi-tenant network. The same authors [138] proposed a 5G firewall that detects, differentiates and selectively blocks 5G network traffic in the backhaul network.

In parallel, attempts such as TurboEPC [139] proposed offloading a subset of user state in mobile packet core to programmable switches in order to perform signaling in the data plane. Similarly, Singh et al. [140] designed a P4-based element of 5G Mobile Packet Core (MPC) that merges the functions of both signaling gateway (SGW) and the Packet Data Network Gateway (PGW). Additionally, Voros et al. [141] proposed a a hybrid next-generation NodeB (gNB) that combines the capabilities of P4 switches and the external services built on top of NIC accelerators (DPDK). Another important function required in 5G is handover. Palagummi et al. [142] proposed SMARTHO, a system that uses programmable switches to perform handover efficiently in a wireless network.

Paolucci et al. [143] demonstrated the potential and the disruptiveness of data plane programmability as opposed to the SDN/NFV network with no programmable switches. Specifically, the authors focused on the problems of full softwarization in current SDN networks (high latency and jitter, low precision traffic and advanced monitoring, etc.) and how P4 is paving the way to novel orchestration frameworks enabling innovation at the edge. Lin et al. [144] integrated P4 switches in their 5G testbed to implement the User Plane Function (UPF) in the data plane.

### 3) Media offloading

Kfoury et al. [145] proposed a system for offloading conversational media traffic (e.g., Voice over IP (VoIP), Voice over LTE (VoLTE), WebRTC, media conferencing, etc.) from x86-based relay servers to programmable switches. While this system is not tailored for 5G network specifically, it provides significant performance improvements for Over-The-Top (OTT) VoIP systems.

Andrus et al. [146] offloaded video processing to the switch. Essentially, the switch dynamically filters and separate control traffic from video streams, and then redirect them to the desired destinations. The authors implemented this scheme due to processing constraints on the software when the number of devices is high (the authors noted that CCTV cameras in London, UK is estimated at roughly 500,000).

### 4) Telecom Schemes Comparison, Discussions, and Limitations

Table 19 compares the aforementioned telecom schemes on P4. In general, all schemes aim at offloading various functionalities originally executed on x86-based servers to the data plane. Such strategy improves the network performance (e.g., latency, throughput) significantly and aim at achieving the KPIs of 5G. For instance, the experiments conducted in [137] show that the attained QoS metrics meet the latency requirements of 5G. Similarly, the results reported in [138] demonstrate that the system meets the reliability KPI of 5G, which states that the network should be secured with zero downtime. Furthermore, the results reported in [142] show that there are 18% and 25% reductions in handover time with respect to legacy approaches, for two- and three-handover sequences, respectively.

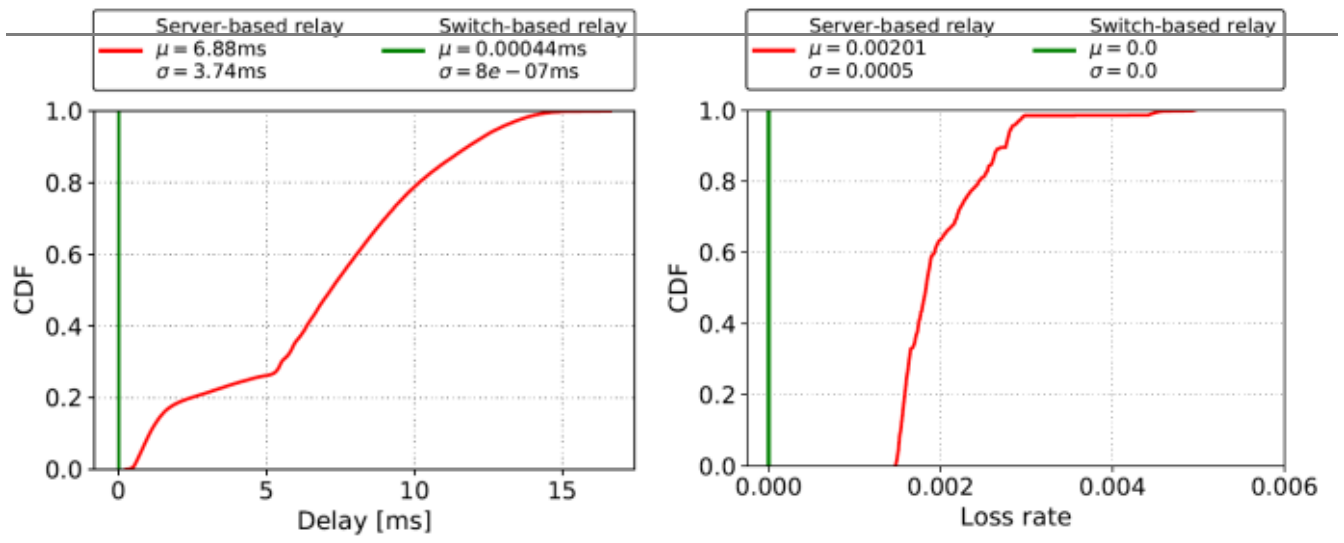

**FIGURE 15:** CDF of delay and packet loss rate of 900 offloaded VoIP calls [145].

TABLE 20: Switch-based and server-based media relaying.

| Metric | Switch-based relay [145] | Server-based relay |
|---|---|---|
| Relay server CPU | Lower; negligible with 900 active sessions | Higher; averages at 50% for 900 active sessions |
| Latency | Lower; almost constant at 440ns with 900 sessions | Higher; from 0.2ms to 17ms with 900 sessions |
| Jitter | Lower; negligible with 900 active sessions | Higher; ranges from 100us to 3ms |
| Packet loss | None contributed by the switch | High; increases as the number of sessions increases |
| Maximum number of sessions | Higher; more than one million with additional resources to spare | Lower; thousand sessions per core before QoS degrades |
| Mean opinion score (MOS) | Higher; maximum MOS (4.4) with 1800 concurrent sessions | Lower; for 1800 sessions, 50% of sessions have a MOS score below 3.7 |
| Table size | Limited by SRAM | Arbitrary |
| Additional functions | Limited to relaying | Arbitrary; e.g., media mix, lawful interception |

The system in [145] emulates the behavior of the relay server which is primarily used to solve the NAT problem. Results show that ultra-low latency and jitter (nanoseconds-scale) are achieved with programmable switches as opposed to x86-based relay servers where the latency and the jitter are in the milliseconds-scale (see Fig. 15). The solution also improves the packet loss rate, CPU usage of the server, Mean Opinion Score (MOS), and can scale to more than one million concurrent sessions, with additional resources to spare in the switch.

Other systems allow offloading the signaling part to the data plane. For instance, TurboEPC offloads messages that constitute a significant portion of the total signaling traffic in the packet core, aiming at improving throughput and latency of the control plane's processing.

#### 5) Switch-based and Server-based Media Relay

Offloading media traffic from general purpose servers to programmable switches greatly improves the quality of service. Table 20 shows the metrics achieved when media is relayed by a relay server versus when it is relayed by the switch, based on [145]. The results show that the latency, jitter and packet loss rates are significantly lower when media is being relayed by the switch. Not only the QoS metrics are improved, but also the maximum number of concurrent sessions. With Tofino 3.2Tbps, more than one million sessions were accommodated in the switch's SRAM, with additional resources to spare for other functionalities. On the other hand, only one thousand sessions per CPU core were handled in the server-based relay, before QoS starts to degrade. The drawback of offloading media traffic to the switch is that some functionalities are complex to be implemented in the data plane (e.g., media mixing for conference calls, noise reduction, etc.).

### D. CONTENT-CENTRIC NETWORKING

#### 1) Background

Emerging network architectures (e.g., [291]) promote content-centric networking, a model where the addressing scheme is based on named data rather than named hosts. In other words, users specify the data they are interested in instead of specifying where to get the data from. A branch of content-centric networking is the publish/subscribe (pub/sub) model. The goal of the model is to provide a scalable and robust communication channel between producers and consumers of information. A large fraction of today's Internet applications follow the publish/subscribe paradigm. With the IoT, this paradigm proliferated as sensors/actuators are often deployed in dynamic environments. Other applications that use pub/sub model include instant messaging, Really Simple Syndication (RSS) feeds, presence servers, telemetry and others. Current approaches to content-centric networking use software-based middleboxes, which limits the performance in terms of throughput and latency. Recent works are leveraging programmable switches to overcome the performance limitations of software-based middleboxes.

#### 2) Publish/subscribe

Jepsen et al. [147] presented "packet subscription", a new abstraction that generalizes the forwarding rules by evaluating stateful predicates on input packets. Wernecke et al. [148, 149] presented distribution strategies for content-based publish/subscribe systems using programmable switches. The authors described a system where the notification distribution tree (i.e., the subscribers that should receive the notification) is encoded in the packet headers, similar to multicast source routing. Similarly, Kundel et al. [150] implemented a publish/subscribe system on programmable switches. The system is flexible in encoding attributes/values in packet headers.

#### 3) Named Data Networking

Signorello et al. [132] developed NDN.p4, a preliminary implementation of a Named Data Networking (NDN) instance that caches requests to optimize its operations. Miguel et al. [151] extended NDN.p4 to include the content store and to solve the scalability issues of the previous FIB design. Karrakchou et al. [152] proposed ECDN, another CDN implementation on P4 where data plane configuration is generated according to application requirements and supports extensions to the regular CDN such as adaptive forwarding, customized monitoring, in-network caching control, and publish/subscribe forwarding.

#### 4) Content-centric Networking Schemes Comparison, Discussions, and Limitations

Table 21 compares the aforementioned pub/sub schemes. In [147], the authors described a compiler that generates P4 tables from logical predicates. It utilizes a novel algorithm based on Binary Decision Diagrams (BDD) to preserve switch resources (TCAM and SRAM). This feature simplifies the configuration as operators do not need to manually install tables entries switches, which is a cumbersome process when the topology is large. The prototype was evaluated on a hardware switch (Tofino), and the authors considered the Nasdaq's ITCH protocol as the pub/sub use case. Results show that the system was able to process messages at line rate while using the full switch capacity (6.5 Tbps). The other systems considered different encoding strategies. For example, in [148, 149], the authors described a system where the notification distribution tree (i.e., the subscribers that should receive the notification) is encoded in the packet headers, similar to multicast source routing. The advantage of storing the distribution tree in the packet header instead of storing it in the switch is that rules in the switches do not need to be updated when subscriptions change. Another distinction between the pub/sub systems is whether they require a dedicated language to describe the subscriptions, and the configuration complexity.

**TABLE 21:** Content-centric networking schemes comparison.

| | Ref. | Dedicated language | Conf. complex. | Encoding structure | Platform HW | Platform SW |
|---|---|---|---|---|---|---|
| Pub / sub | [147] | | Medium | Hierarchical (BDD) | | |
| Pub / sub | [148] [149] | ✓ | High | Distribution Tree | | |
| Pub / sub | [150] | ✓ | High | Attribute-value pair | | |
| NDN | [132] | N/A | High | Type-Length -Value | | |
| NDN | [151] | N/A | High | Type-Length -Value | | |
| NDN | [152] | N/A | Low | FCTrees | | |

Regarding the NDN schemes, ENDN focused on making the data plane adaptive and easily programmable to meet the application needs. This flexibility is lacking in the other P4-based CDN schemes. It is worth mentioning that P4 has its shortcomings when it comes to supporting a stateful variable length protocol. This is an important aspect that should be tackled when implementing NDN on the data plane.

#### 5) Comparison between Switch-based and Server-based Pub/Sub Systems

Fig. 16 illustrates the operations of traditional software-based pub/sub systems (a) and switch-based pub/sub systems (b). Latency and its variations are significantly reduced when the switch acts as a pub/sub broker. However, the size of memory in the switch limits the amount of data to be distributed. Moreover, implementing features provided by software-based pub/sub systems such as QoS levels, session persistence, message retaining, last will and testament (notify users after a device disconnects) in hardware is challenging.

### E. SUMMARY AND LESSONS LEARNED

Programmable switches offer the flexibility of customizing the data plane to enable middlebox functions. A middlebox can be defined as a device that performs functions that are beyond the standard capabilities of routers and switches. A number of works demonstrated the implementation of middlebox functions such as caching, load balancing, offloading services, and others on programmable switches. The majority of load balancing schemes took advantage of the stateful nature of the data plane to store the load balancing connection table. Future work should consider minimizing the storage requirement to improve the scalability, supporting flow priority, and developing further variations for novel multipath transport protocols such as multipath QUIC.

The switch can also act as an "in-network cache" that serves hot items at line rate. Some schemes indexes entries automatically, while others require operator's intervention. Future endeavours could investigate items compression, communication minimization, priority-based caching, and aggregated computations caching (e.g., cache the average of hot items).

An additional middlebox application is offloading telecom functions. The switch is capable of relaying media traffic and user plane functions. Future work could investigate scalability improvement (i.e., to accommodate more concurrent sessions), offloading signalling traffic, and in-network media mixing.

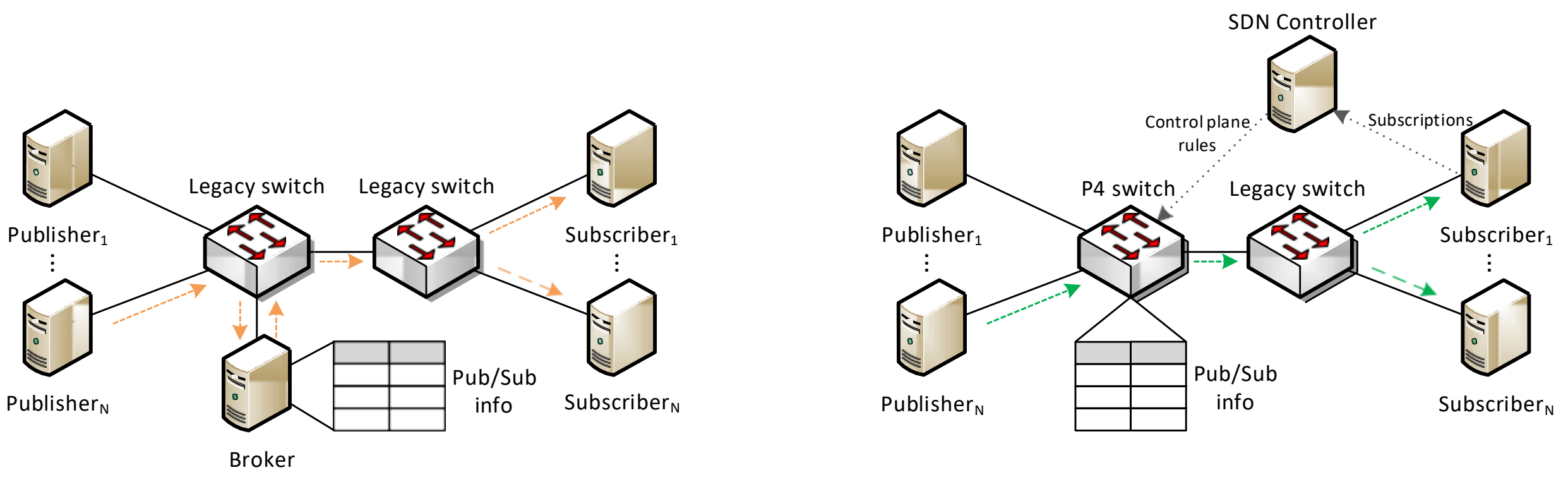

(a) (b)

**FIGURE 16:** (a) Traditional software-based pub/sub architecture. (b) Pub/sub implemented on a programmable switch.

Finally, the switch can also act as a broker to distribute packets in a publish/subscribe system. Future work could investigate reliability insurance (e.g., packet deliver guarantee), message retaining, and QoS differentiation (e.g., QoS features of MQTT).

## IX. NETWORK-ACCELERATED COMPUTATIONS

Programmable switches offer the flexibility of offloading some upper-layer logic to the ASIC, referred also as in-network computation. Since switch ASICs are designed to process packets at terabits per second rates, in-network computation can result in an order of magnitude or more of improvement in throughput when compared to applications implemented in software. The potential performance improvement has motivated programmers to built in-network computation for different purposes, including consensus, machine learning acceleration, stream processing, and others.

The idea of delegating computations to networking devices was perceived with Active Networks [292], where packets are replaced with small programs ("capsules") that are executed in each traversed device along the path. However, traditional network devices were not capable of performing computations. With the recent advancements in programmable switches, performing computations is now a possibility.

### A. CONSENSUS

#### 1) Background

Consensus algorithms are common in distributed systems where machines collectively achieve agreement on a single data value, or on the current state of a distributed system. Reliability is achieved with consensus algorithms, even in the presence of some malicious or faulty processes. Consensus algorithms are used in applications such as blockchain [293], load balancing, clock synchronization, and others [294].

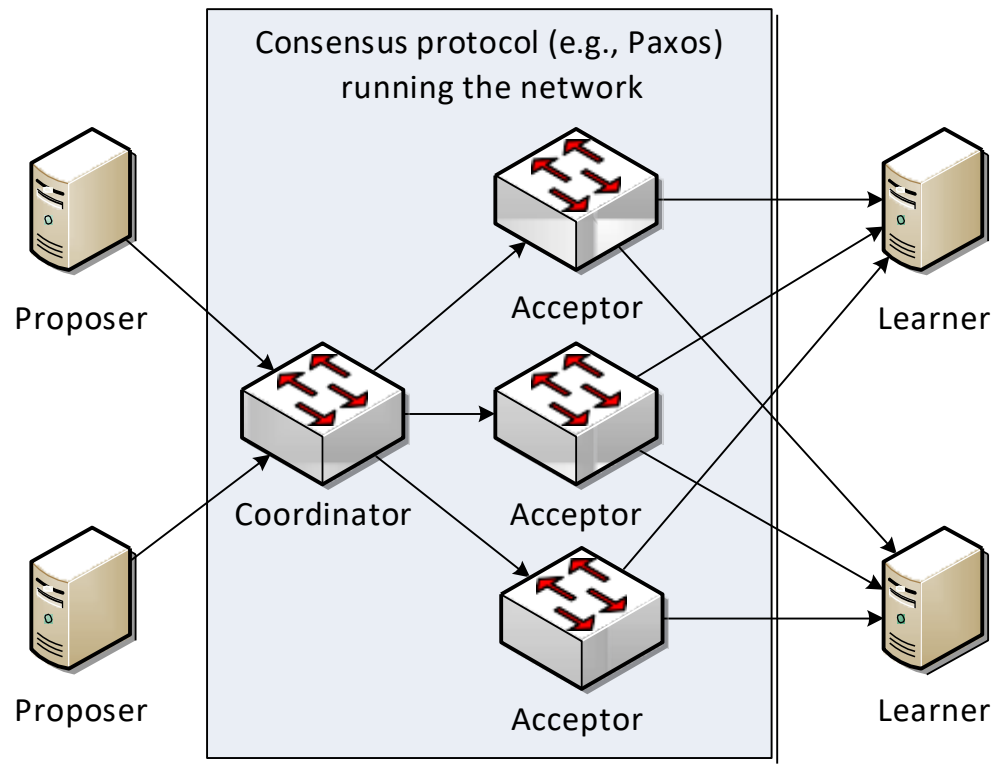

FIGURE 17: Consensus protocol in the data plane model [154]. An application sends a request to the proposer which resides on a commodity server. The proposer then creates a Paxos message and sends it to the coordinator, running in the data plane. The role of the coordinator is be the broker of requests on behalf of proposers. Afterwards, the acceptor, which also runs on the data plane, receives the messages from the coordinator, and ensures consistency through the system by deciding whether to accept/reject proposals. Finally, learners provide replication by learning the result of consensus.

Latency has always been a bottleneck with consensus algorithms as protocols require expensive coordination on every request. Lately, researchers have started investigating how programmable switches can be leveraged to operate consensus protocols in order to increase throughput and decrease latency. Fig. 17 shows a consensus model in the data plane.

#### 2) Paxos Implementations

Li et al. [153] proposed Network-Ordered Paxos (NOPaxos), a P4-based Paxos [295] system that applies replication in the data center to reduce the latency imposed from communication overhead. Similarly, Dang et al. [154] presented an implementation of Paxos using P4 on the data plane. Jin et al. [157] proposed NetChain, a variant of the Paxos protocol that provides scale-free sub-RTT coordination in data centers. It is strongly-consistent, fault-tolerant, and presents an in-network key-value store. Dang et al. [158] proposed Partitioned Paxos, a P4-based system that separates the two aspects of Paxos, namely, agreement and execution, and optimizes them separately. Furthermore, The same authors also proposed P4xos [160], a P4-based solution that executes Paxos logic directly in switch ASICs, without strengthening assumptions about the network (e.g., ordered delivery, packet loss, etc.).

#### 3) Other Implementations

Another line of research focused on consensus algorithms other than Paxos. Li et al. [155] proposed Eris, a P4-based solution that avoids replication and transaction coordination overhead. It processes a large class of distributed transactions in a single round trip, without any additional coordination between shards and replicas. Sakic et al. [159] proposed P4 Byzantine Fault Tolerance (P4BFT), a system that is based on BFT-enabled SDN, where controllers act as replicated state machines. The system offloads the comparison of controllers' outputs required for correct BFT operations to programmable switches. Finally, Han et al. [156] offloaded part of the Raft consensus algorithm [296] to programmable switches in order to improve its performance. The authors selected Raft due to the fact that it has been formally proven to be more safe than Paxos, and it has been implemented on popular SDN controllers.

#### 4) Consensus Schemes Comparison, Discussions, and Limitations

Table 22 compares the aforementioned consensus schemes. In general, consensus algorithms such as Paxos are complex and cannot be easily implemented with the constraints of the data plane. For instance, [154] only implemented phase-2 logic of Paxos leaders and acceptors. Similarly, NetChain uses a variant of the Paxos protocol that divides it into two parts: steady state and reconfiguration. This variant is known as Vertical Paxos, and is relatively simple to implement in the network as the division's parts can be mapped to the control plane and the data plane.

**TABLE 22:** Consensus schemes comparison.

| | Name & ref | Algo. | Weak assumpt. | Full proto. | Platform | |
|---|---|---|---|---|---|---|
| | | | | | HW | SW |
| Paxos | NOPaxos [153] | Paxos | ✓ | ✓ | | |
| | N/A [154] | Paxos | | ✓ | | |
| | NetChain [157] | Novel | ✓ | | | |
| | Partitioned Paxos [158] | Paxos | | | | |
| | P4xos [160] | Paxos | | | | |
| Others | Eris [155] | Novel | | | | |
| | P4BFT [159] | BFT | | | | |
| | N/A [156] | Raft | | ✓ | | |

Unordered and completely asynchronous networks require the full implementation and complexity of Paxos. NOPaxos suggests that the communication layer should provide a new Ordered Unreliable Multicast (OUM) primitive; that is, there is a guarantee that receivers will process the multicast messages in the same order, though messages can be lost. NOPaxos relies on the network to deliver ordered messages in order to avoid entirely the coordination. Dropped packets on the other hand are handled through coordination with the application. Other systems like Eris avoid replication and transaction coordination overhead. The main contribution of Eris compared to NOPaxos is that it establishes a consistent ordering across messages delivered to many destination shards. Eris also allows receivers to detect dropped messages.

Partitioned Paxos [158] improved the existing systems. The motivation behind Partitioned Paxos is that existing network-accelerated approaches do not address the problem of how replicated application can cope with the high rate of consensus messages; NOPaxos only processes 13,000 transactions per second since it presents a new bottleneck at the host side. Other systems (e.g. NetChain) are specialized replication services and can not be used by any off-the-shelf application.

Finally, P4xos improves both the latency and the tail-latency. The throughput is also improved compared to hardware servers which require additional memory management and safety features (e.g., user and kernel separation). P4xos was implemented on a hardware switch (Tofino), and results show that it reduces the latency by three times compared to traditional approaches, and it can process over 2.5 billion consensus messages per second (four orders of magnitude improvement).

#### 5) Network-Assisted and Legacy Consensus Comparison

Consensus algorithms have been traditionally implemented as applications on general purpose CPUs. Such architecture inherently induces latency overhead (e.g., Paxos coordinator has a minimum latency of 96us [297]).

There are numerous performance benefits gained when consensus algorithms are implemented in programmable devices. When consensus messages are processed on the wire, the latency significantly decreases (Paxos coordinator had a minimum latency of 340ns [297]). Moreover, when compared to legacy consensus deployments, network-assisted consensus require fewer hops traversal.

### B. MACHINE LEARNING

#### 1) Background

The remarkable success of Machine Learning (ML) today has been enabled by a synergy between development in hardware and advancements in machine learning techniques. Increasingly complex ML models are being developed to handle the large size of datasets and to accelerate the training process. Hardware accelerators (e.g., GPU, TPU) were introduced to speedup the training. These accelerators are installed in large clusters and collaborate through distributed training to exploit parallelism. Nevertheless, training ML models is time consuming and can last for weeks depending on the complexity and the size of the datasets. Researchers have traditionally investigated methods to accelerate the computation process, but not the communication in distributed learning. With the advancements in programmable switches, it is now possible to accelerate the ML training process through the network.

#### 2) In-network Training

Sapio et al. [161] proposed DAIET, a system that performs in-network data aggregation to accelerate applications that follow a partition/aggregate workload pattern. Similarly, Yang et al. [162] proposed SwitchAgg, a system that performs similar functions as DAIET, but with a higher data reduction rate. Perhaps the most significant work in the training acceleration literature is SwitchML [163], a system that performs in-network aggregation for ML model updates sent from workers on external servers.

#### 3) In-network Inference

Other schemes have shown interest in speeding the inference process by leveraging programmable switches. Siracusano et al. [164] proposed N2Net, a system that runs simplified neural networks (NN) on programmable switches. Sanvito et al. [165] proposed BaNaNa Split, a solution that evaluates the conditions under which programmable switches can act as CPUs' co-processors for the processing of Neural Networks (e.g., CNN). Finally, Xiong et al. [166] proposed IIsy, a system that enables programmable switches to perform in-network classification. The system maps trained ML classification models to match-action pipelines.

#### 4) ML Schemes Comparison, Discussions, and Limitations

Table 23 compares the aforementioned ML schemes. While the goal of DAIET is to discuss what computations the network can perform, the authors did not design a complete system, nor did they address the major challenges of supporting ML applications. Moreover, their proof-of-concept presented a simple MapReduce application on a software switch, and it is not certain whether the system can be implemented on a hardware switch. Compared to DAIET, SwitchAgg does not require modifying the network architecture, and offers better processing abilities with a significant data reduction rate. Moreover, SwitchAgg was implemented on an FPGA, and the results show that the job completion time can be reduced as much as 50%.

**TABLE 23:** Machine learning schemes comparison.

| | Scheme | Name | Core idea | Evaluated model/algorithm | Quantization | Platform | |
|---|---|---|---|---|---|---|---|
| | | | | | | HW | SW |
| **In-network Training** | [161] | DAIET | In-network computation for partition/aggregate work pattern | SGD, Adam | N/A | | |
| | [162] | SwitchAgg | In-network aggregation without modifying the network | MapReduce-like system | | | |
| | [163] | SwitchML | Accelerates distributed parallel training in ML | Synchronous SGD | | | |
| **In-network Inference** | [164] | N2Net | In-network classification using BNN | Binary neural networks | N/A | | |
| | [165] | BaNaNa Split | NN processing division between switches and CPUs | Binary neural networks | | | |
| | [166] | IIsy | Maps trained ML classification models to match-action pipeline | Decision tree, SVM, naïve bayes, k-means | ◄ | | |

SwitchML extended the literature on accelerating ML models training by providing a complete implementation and evaluation on a hardware switch. A commonly used training technique for deep neural networks is synchronous stochastic gradient descent [299]. In this technique, each worker has a copy of the model that is being trained. The training is an iterative process where each iteration consists of: 1) reading the sample of the dataset and locally perform some computation-intensive learning using the worker's accelerators. This yields to a gradient vector; and 2) updating the model by computing the mean of all gradient vectors. The main motivation of this idea is that the aggregation is computationally cheap (takes 100ms), but is communication-intensive (transfer hundreds of megabytes each iteration). SwitchML uses computation on the switch to aggregate model update in the network as the workers are sending them (see Fig. 18). An advantage is that there is minimal communication; each worker sends its update vector and receives back the aggregated updates. The design challenges of this system include: 1) the limitation of storage available on the switch, addressed by using a streaming approach; 2) switches cannot perform much computations per packet, addressed by partitioning the work between the switch and the workers; 3) ML systems use floating point numbers, addressed by quantization approaches; and 4) failure recovery is needed to ensure correctness. The system is implemented on a hardware switch (Tofino), and results show that the system speeds up training by up to 300% compared to existing distributed learning approaches.

With respect to in-network inference, it is challenging to implement full-fledged models as they require extensive computations (e.g., multiplications and activation functions). Simple variation such as the Binary Neural Network (BNN) only requires bitwise logic functions (e.g., XNOR, POPCNT, SIGN). N2Net provides a compiler that translates a given BNN model to switching chip's configuration (P4 program). The authors did not mention on which platform N2Net was evaluated; however, based on their evaluations, they concluded that a BNN can be implemented on most current switching chips, and with small additions to the chip design, more complex models can be implemented. IIsy studied other ML models. The authors of IIsy acknowledged that the work is limited in scope as it does not address popular ML algorithms such as neural networks. Furthermore, it is bounded to the type of features it can extract (i.e., packet headers), and has accuracy limitations. IIsy tries to find a balance between the limited resources on the switch and the classification accuracy. Finally, BaNaNa Split took a different approach by partitioning the processing of NN to offload a subset of layers from the CPU to a different processor. Note that the solution is far from complete, and the authors evaluated a single binary fully connected layer with 4096 neurons using a network processor-based SmartNIC.

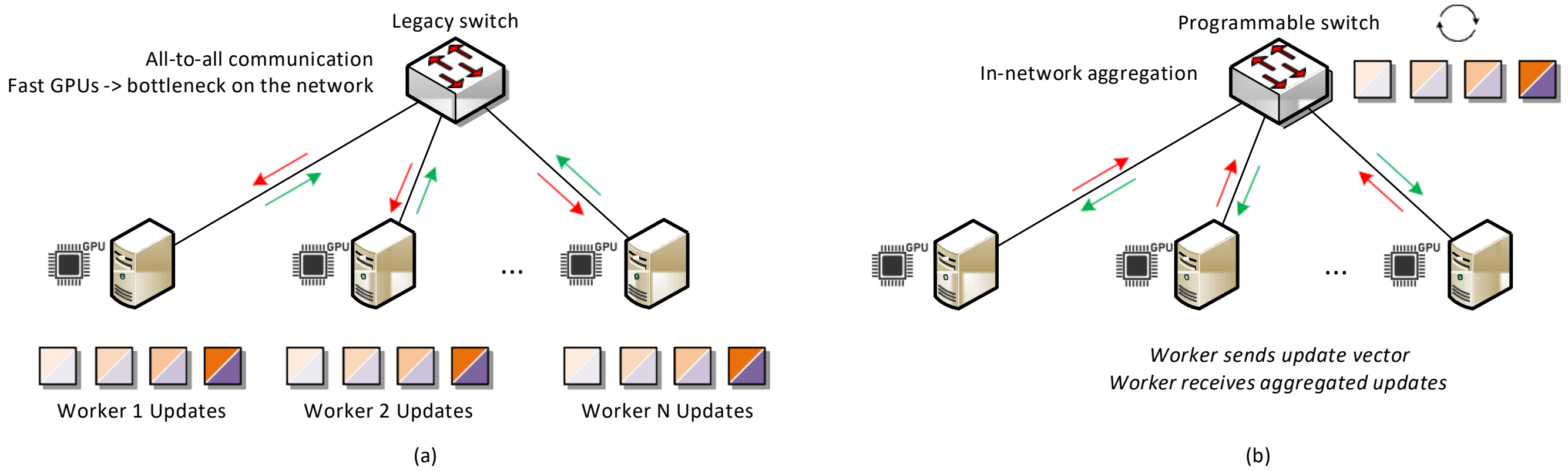

**FIGURE 18:** (a) ML model updates in legacy networks. The aggregation process is communication-intensive and follows an all-to-all communication pattern. This means that the workers should receive all the other workers' updates. Since accelerators on end-hosts are becoming faster, the network should speed up so that it does not become the bottleneck. Therefore, it is expensive to deploy additional accelerators since it requires re-architecting the network. The red arrow in (a) shows that the bottleneck source is the network. (b) ML model updates accelerated by the network. Aggregation is performed in the network by the programmable switches while the workers are sending them. The workers do not need to obtain the updates of all other workers, hence there is minimal communication. They only obtain the aggregated model from the switch. The red arrow in (b) shows that the bottleneck source is the worker rather than the network [163, 298]

TABLE 24: Switch-based and server-based ML approaches.

| Feature | Inference | | Training | |
|---|---|---|---|---|
| | Switch-based | Server-based | Switch-based | Server-based |
| Speed | Faster, inference at line rate | Slower | Faster, aggregations at line rate | Slower; aggregations on an x86 server |
| Complex computations support | Lower, basic arithmetic and bitwise logic function | Higher | Lower | Higher |
| Communication overhead | Low | Low | Lower, switch is the centralized aggregator | Higher, updates are exchanged with a remote aggregator |
| Storage | Lower | Higher | Lower, update is not stored entirely at once | Higher |
| Encrypted traffic | Difficult | Easy | Difficult | Easy |

### C. COMPARISON BETWEEN SWITCH-BASED AND SERVER-BASED ML

Table 24 shows a comparison between switch-based and server-based ML approaches. ML works that were extracted from the literature can be divided into two main categories: 1) expedited inference in the data plane, and 2) accelerated training in the network. The main advantage of switch-based over server-based inference is the ability to execute at line rate, and hence provides faster results to the clients. Performing complex computations in the switch is achieved through estimations, and hence is limited. Moreover, the SRAM capacity of the switch is small, impeding the storage of large models. Such limitations are not problematic with server-based inference approaches.

Distributed training can be significantly faster when aggregations are offloaded to a centralized switch. However, due to the small capacity of the switch memory, it is not possible to store the whole model update at once. Additionally, encrypted traffic remains a challenge when inference or training is handled by the switch.

### D. SUMMARY AND LESSONS LEARNED

Accelerating computations by leveraging programmable switches is becoming a trend in data centers and backbone networks. Although switches only support basic and limited operations, it was shown in the literature that the performance of various tasks (e.g., consensus, training models in machine learning), could significantly improve if computations are delegated to the network.

The majority of the in-network consensus works aim at implementing common consensus protocols such as Paxos and Raft in the data plane. Due to the hardware constraints, current schemes implement only simplified variations of the protocols. Future work could investigate implementing novel consensus algorithms that diverge from the existing complex ones. Further, such schemes should encompass failure recovery mechanisms.

Another interesting in-network application is ML training/inference acceleration. The literature has shown that significant performance improvements are attained when the switch aggregates model updates or classifies new samples. Future work could explore developing ML models for various tasks such as classification, regression, clustering, etc.

In addition to the aforementioned categories, data plane programming is being used for stream processing [167, 168], parallel processing [169], string searching [170], erasure coding [171], in-network lock managers [172], database queries acceleration [173], in-network compression [174], and computer vision offloading [175].

## X. INTERNET OF THINGS (IOT)

The Internet of Things (IoT) is a novel paradigm in which pervasive devices equipped with sensors and actuators collect physical environment information and control the outside world. IoT applications include smart water utilities, smart grid, smart manufacturing, smart gas, smart metering, and many others. Typical IoT scenarios entail a large number of devices periodically transmitting their sensors' readings to remote servers. Data received on those collectors is then processed and analyzed to assist organizations in taking data-driven intelligence decisions.

### A. AGGREGATION

#### 1) Background

Since IoT devices are constrained in size and processing capabilities, they typically generate packets that carry small payloads (e.g., temperature sensor readings). While such packets are small in size, their headers occupy a significant portion of the total packet size. For instance, Sigfox Low-Power Wide Area Network (LPWAN) [300] can support a maximum of 12-bytes payload size per packet. The overhead of headers is 42-bytes (Ethernet 14-bytes + IP 20-bytes + UDP 8-bytes), which represent approximately 78% of the packet total size. When numerous devices continuously transmit IoT packets, a significant percentage of network bandwidth is wasted on transmitting these headers. Packet aggregation is a mechanism in which the payloads of small packets are aggregated into a single larger packet in order to mitigate the bandwidth overhead caused by transmitting multiple headers.

**TABLE 25:** IoT aggregation schemes comparison.

| Scheme | Evaluation | | Constraints | | | Line rate | | Platform | |
|---|---|---|---|---|---|---|---|---|---|
| | Theoretical | Implementation | Same payload size | Payload <= 16 bytes | Number of packets | Aggregation | Disaggregation | HW | SW |
| [176] | | | | | 8 | | ✓ | | |
| [177] | | | ✓ | ✓ | Up to MTU | | | | |
| [178] | | | | | 8 | ✓ | ✓ | | |

Legacy packet aggregation mechanisms operate on the CPUs of servers or on the control plane of switches [301–306]. While legacy mechanisms reduce the overhead of packet headers, they unquestionably increase the end-to-end latency and decrease the throughput. As a result, some studies have suggested aggregating only packets that are not real-time.

#### 2) IoT Bandwidth Overhead Reduction

Wang et al. [176] presented an approach where small IoT packets are aggregated into a larger packet in the switch data plane (see Fig. 19). The goal of performing this aggregation is to minimize the bandwidth overhead of packets' headers. The same authors [177] extended this work to solve some constraints related to the payload size and the number of aggregated packets. Similarly, Madureira et al. [179] proposed IoTP, a layer-2 communication protocol that enables the aggregation of IoT data in programmable switches. The solution gathers network information that includes the Maximum Transmission Unit (MTU), link bandwidths, underlying protocol, and delays. These properties are used to empower the aggregation algorithm.

#### 3) Aggregation Comparison, Discussions, and Limitations

Table 25 compares the aforementioned IoT aggregations schemes. [176] and [177] operate in the same way. Upon receiving a packet, the P4 switch parses its headers and identifies whether the packet is an IoT packet. If the packet was identified as an IoT packet, the switch parses and extracts the payload. Afterwards, the payload is stored in switch registers along with some other metadata, and the packet is dropped. Once packets are aggregated, the resulting packet is sent across the WAN to reach the remote server. Before the packet reaches the server, it is disaggregated by another P4 switch situated close to the server and several packets identical to the original ones are generated. An important observation is that the aggregation/disaggregation processes are transparent to both the IoT devices and the servers; hence, no modifications are required on either end. The main advantages of [177] over [176] are: 1) packets can have different payload sizes; 2) the payload size is no longer limited to 16 bytes; 3) the number of packets is dynamic and only limited by the packet MTU; and 4) both the disaggregation and the aggregation run at line rate.

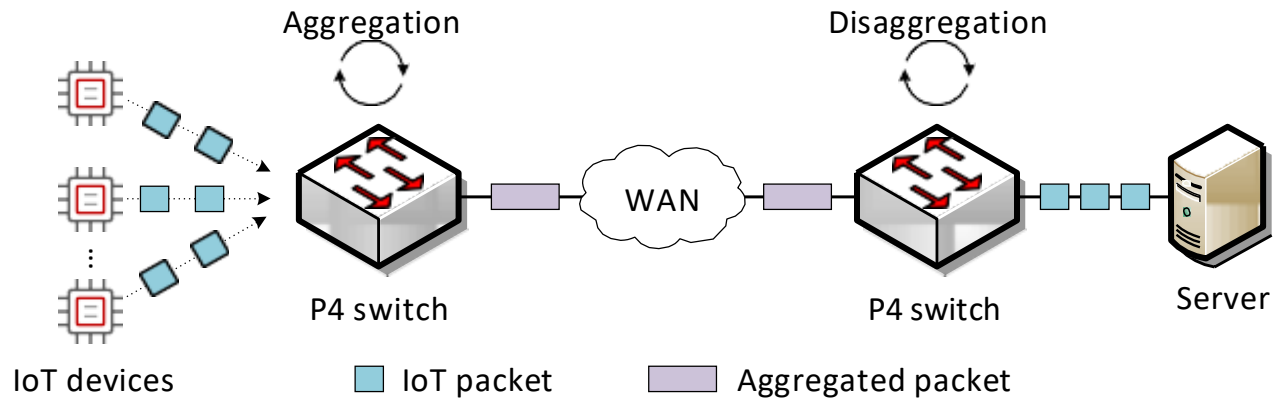

**FIGURE 19:** IoT packets aggregation [176]. Frequent small IoT packets are aggregated by a P4 switch and encapsulated in a larger packet. Another switch across the WAN disaggregates the large packet to restore the original IoT packets. Such mechanism prevents the frequent transmissions of headers, and thus, minimizes the bandwidth overhead.

#### 4) Comparison between Server-based and Switch-based Aggregation

Table 26 shows a comparison between switch-based and server-based packet aggregation. When aggregation is performed on the switch (ASIC), the throughput is higher and the latency and jitter are lower than that of the server-based approaches (e.g., switch CPU, x86-based server). On the other hand, the server-based aggregation has more flexibility in defining the number of packets and the amount of data that can be aggregated. Note that if aggregation and disaggregation are executed on the IoT device itself, the session would suffer from long delays and low throughput. More importantly, the IoT device is limited in computational and energy resources.

### B. SERVICE AUTOMATION

#### 1) Background

Low-power low-range IoT communication technologies (e.g., Bluetooth Low Energy (BLE) [307], Zigbee [308], Z-wave [309]) typically follow a peer-to-peer model. IoT devices in such technologies can be divided into two distinct types, *peripheral* and *central*. Peripheral devices, which consist of sensors and actuators, receive commands and execute subsequent actions. Central devices on the other hand run applications that analyze information collected from peripheral devices and subsequently issue commands.

**TABLE 26:** Switch-based and server-based packet aggregation.

| Feature | Switch-based (ASIC) | Server-based (CPU) |
|---|---|---|
| Throughput | Higher; (e.g., [176], 100Gbps, i.e., max capacity) | Lower; (e.g., [176], 2.58Gbps) |
| Latency and Jitter | Lower; | Higher; |
| Count of packets to be aggregated | Not flexible (limited by the switch SRAM) | Arbitrary |
| Amount of data to be aggregated | Not flexible (limited by the switch SRAM, parsing capacity) | Arbitrary |

The interconnection of devices and services can follow a Peer-to-Peer (P2P) model or a cloud-centric approach. In the P2P model, the automation service runs on the central device which processes and analyzes sensor data published by peripheral devices in order to issue commands. The main advantages of the P2P include the low end-to-end latency and the subtle power consumption as devices are physically close to each other. The drawbacks of the P2P model include poor scalability, short reachability, and inflexibility of policy enforcement. The cloud-centric model addresses the limitations of the P2P model by adding a gateway node that connects peripheral devices to a middleware hosted on the cloud (Internet). While this approach solves the poor scalability and the policy enforcement flexibility issues, it incurs additional delays and jitters in collecting and reacting to data. Moreover, the middleware represents a single point of failure which can shutdown the whole service in the event of an outage. With programmable switches, researchers are investigating in-network approaches to manage transactional relationships between low-power, low-range IoT devices.

#### 2) Service Management and Multi-protocol Processing

Uddin et al. [180] proposed Bluetooth Low Energy Service Switch (BLESS), a programmable switch that automates IoT applications services by encoding their transactions in the data plane. It maintains link-layer connections to the devices to support P2P connectivity. The same authors proposed Muppet [181], an extension to BLESS to support multiple non-IP protocols.

#### 3) Service Automation Comparison, Discussions, and Limitations

In BLESS, the data plane operations are performed at the Attribute Protocol (ATT) service layer which consists of three operations: read attributes, write attributes, and attributes' notification. BLESS parses ATT packets, then processes and forwards them to the devices. The control plane on the other hand is responsible for address assignment, device and service discovery, policy enforcement, and subscription management. The switch was implemented on a software switch (PISCES), and the results show that BLESS combines the advantages of P2P and the cloud-center approaches. Specifically, it achieves small communication latency, low device power consumption, high scalability, and flexible policy enforcement. Muppet extended this approach to support multiple IoT protocols. The system studied two popular IoT protocols, namely BLE and Zigbee. Being in the middle, Muppet switch is responsible for translating actions (e.g., on/off switch of a light bulb) between Zigbee and BLE protocols, as well as logging important events to a database which resides on the Internet via the Hypertext Transfer Protocol (HTTP). Note that parsers and actions policies have to be implemented for each supported protocol. Another difference from BLESS is that the implementation of Muppet's control plane leverages ONOS controller with Protocol Independent (PI) framework.

**TABLE 27:** Switch-based, P2P, and cloud service automation.

| Feature | Switch-based | Peer-to-peer | Cloud-based |
|---|---|---|---|
| Latency | Low | Low | High |
| IoT energy | Low | Low | High |
| Scalability | High | Low | High |
| Reachability | High | Low | High |

#### 4) Comparison between Server-based and Switch-based Service Automation

Table 27 shows a comparison between switch-based, P2P, and cloud-based service automation. Generally, the switch-based approach overcomes the limitations of both approaches. It achieves the low energy and latency characteristics of P2P while increasing scalability and reachability.

### C. SUMMARY AND LESSONS LEARNED

In the context of IoT, there exist broadly two categories, namely, packets aggregation and service automation. The goal of packet aggregation is to minimize the overhead of IoT packets' headers. Typically, headers in IoT packets represent a significant portion of the whole packet size. By aggregating several packets into a single packet, the bandwidth overhead is reduced. Future work should study the performance side-effects (e.g., delay, jitter, loss rate, retransmission) that aggregation causes to packets. Furthermore, timers should be implemented to avoid excessive delays resulting from waiting for enough packets to be aggregated.

With respect to service automation, the goal is to automate IoT applications services by encoding their transactions in the data plane while improving scalability, reachability, energy consumption, and latency. Future work should design and develop translators for non-IP IoT protocols so that applications on various devices that run different protocols can exchange data. Additionally, production-grade software switches should be leveraged to support non-Ethernet IoT protocols.

Other works that involve IoT include flowlet-based stateful multipath forwarding [310] and SDN/NFV-based architecture for IoT networks [311].

## XI. CYBERSECURITY

Extensive research efforts have been devoted on deploying programmable switches to perform various security-related functions in the data plane. Such functions include heavy hitter detection, traffic engineering, DDoS attacks detection and mitigation, anonymity, and cryptography. Fig. 20 demonstrates the difference between contemporary security appliances and programmable switches with respect to layers inspection in the OSI model. Although programmable switches are limited in the computation power, they are capable of inspecting upper layers (e.g., application layer) at line rate. Such functionality is not available in any of the existing solutions.

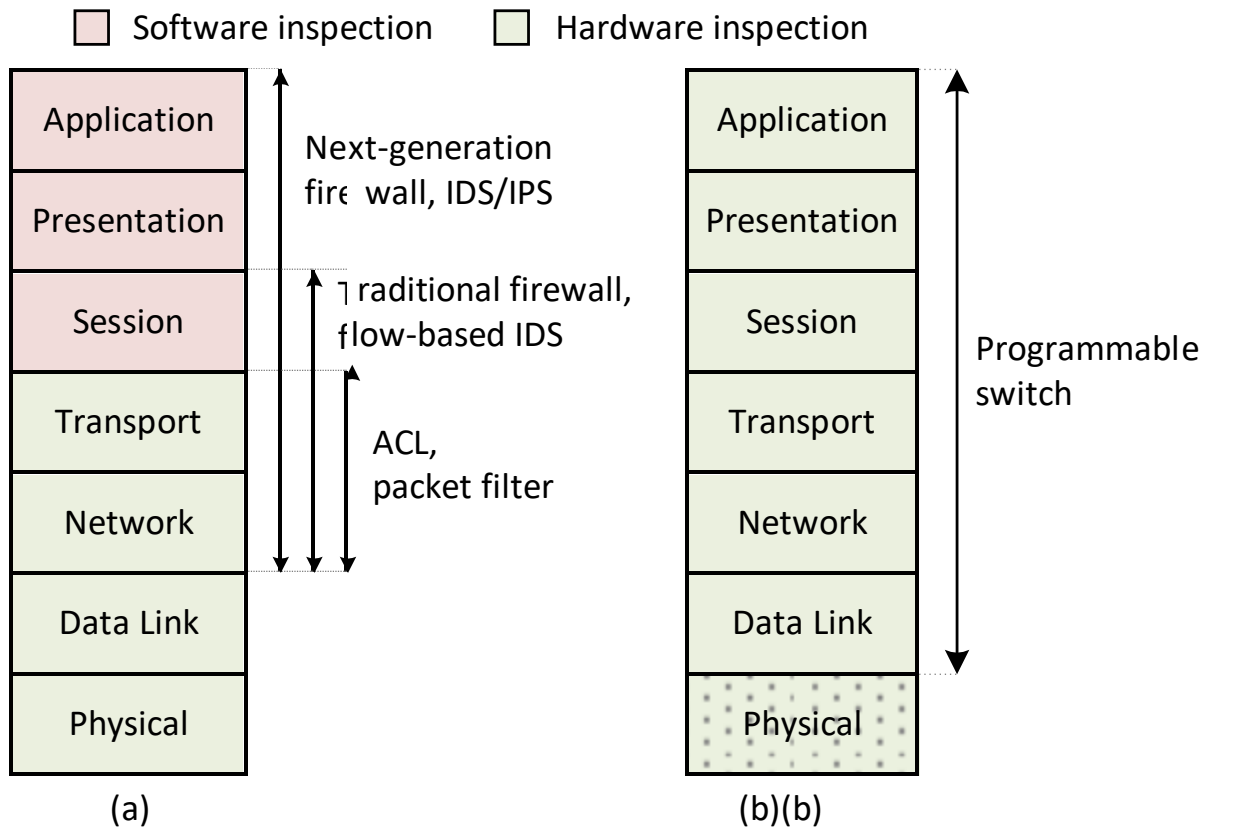

**FIGURE 20:** Layers inspection in the OSI model. (a) Contemporary security appliances. (b) Programmable switch.

## A. HEAVY HITTER

### 1) Background

Heavy hitters are a small number of flows that constitute most of the network traffic over a certain amount of time. They are identified based on the port speed, network RTT, traffic distribution, application policy, and others. Heavy hitters increase the flow completion time for delay-sensitive mice flows, and represent the major source of congestion. It is important to promptly detect heavy hitters in order to react to them; for instance, redirect them to a low priority queue, perform rate control and traffic engineering, block volumetric DDoS attacks, and diagnose congestion. Traditionally, packet sampling technique (e.g., NetFlow) was used to detect heavy hitters. The main problem with such technique is the limited accuracy due to the CPU and bandwidth overheads of processing samples in the software. Advancements in programmable switches paved the way to detect heavy hitters in the data plane, which is not only orders of magnitude faster than sampling, but also enables additional applications (e.g., flow-size aware routing). The detection schemes can be classified as local and network-wide. In the former, the detection occurs on a single switch; in the latter, the detection covers the whole network.

### 2) Local Detection

Sivaraman et al. [182] proposed HashPipe, a heavy hitter detection algorithm that operates entirely in the data plane. It detects the *k*-th heavy hitter flows within the constraints of programmable switches while achieving high accuracy. Furthermore, Kučera et al. [183] proposed Elastic Trie, a solution that detects hierarchical heavy hitters, in-network traffic changes, and superspreaders in the data plane. Hierarchical heavy hitters include the total activity of all traffic matching relevant IP prefixes. Basat et al. [184] proposed PRECISION, a heavy hitter detection algorithm that probabilistically recirculates a fraction of packets for a second pipeline traversal. The recirculation idea greatly simplifies the access pattern of memory without significantly degrading throughput. Tang et al. [185] proposed MV-Sketch, a solution that exploits the idea of majority voting to track the candidate heavy flows inside the sketch data structure. Finally, Silva et al. [186] proposed a solution that identifies elephant flows in Internet eXchange Points (IXP) networks.

### 3) Network-wide Detection

A work proposed by Harrison et al. [187] considers a network-wide distributed heavy-hitter detection. The approach reports heavy hitters deterministically and without errors; however, it incurs significant communication costs that scale with the number of switches. Accordingly, the same authors proposed another scheme (Carpe [188]) which reports probabilistically with negligible communication costs. Ding et al. [189] proposed an approach for incrementally deploying programmable switches in a network consisting of legacy devices with the goal of monitoring as many distinct network flows as possible. The same authors of MV-Sketch proposed SpreadSketch [190], an extension to Count-min sketch where each bucket is associated with a distinct counter to track the distinct items of a stream. SpreadSketch aims at mitigating the high processing overhead of MV-Sketch.

### 4) Heavy Hitter Detection Comparison, Limitations, and Discussions

Table 28 compares the aforementioned heavy hitter schemes. A major criterion that differentiates the solutions is the selection and the implementation of the data structure. Hash tables and sketches are frequently used to store counters for heavy flows. Note that several variations of such data structures are being used in the literature, mainly to tackle the memory-accuracy tradeoff; the choice of data structure reflects on the accuracy of the performed measurements. For example, with probabilistic data structures, only approximations are performed.

In HashPipe, the programmable switch stores the flows identifiers and their byte counts in a pipeline of hash tables. HashPipe adapts the space saving algorithm which is described in [312]. The system was evaluated using an ISP trace provided by CAIDA (400,000 flows), and the results show that HashPipe needed only 80KB of memory to identify the 300 heaviest flows, with an accuracy of 95%. Another hashtable-based solution is Elastic Trie, which consists of a prefix tree that expands or collapses to focus only on the prefixes that grabs a large share of the network. The data plane informs the control plane about high-volume traffic clusters in an event-based push approach only when some conditions are met. Other systems explored different data structures for the task. For instance, in [189] the authors used the HyperLogLog algorithm [313] which approximates the number of distinct elements in a multi-set. The solution is capable of detecting heavy hitters by only using partial input from the data plane.

**TABLE 28:** Heavy hitter schemes comparison.

| | Ref | Name | Core idea | Data structure | Adaptive thresholds | Approximations | Platform | |
|---|---|---|---|---|---|---|---|---|
| | | | | | | | HW | SW |
| **Local Detection** | [182] | HashPipe | Maintains counts of heavy flows in a pipeline of hash tables | Hash tables | ✓ | ✓ | | |
| | [183] | Elastic Trie | Detects hierarchical heavy hitters using hashtable prefix tree | Prefix tree | | | | |
| | [184] | PRECISION | Recirculates a small fraction of packets to simplify memory access | Hash tables | ✓ | | | |
| | [185] | MV-Sketch | Supports the queries of recovering all heavy flows in a sketch | Invertible sketches | ✓ | | | |
| | [186] | N/A | Identifies elephant flows using dynamic thresholds in IXPs | Hash tables | | ✓ | | |
| **Network-wide Detection** | [187] | N/A | Switch store locally the counts & coordinator aggregates the results | Hash tables | | | | |
| | [188] | Carpe | Probabilistic counting on the switch with probabilistic reporting | Sample and hold | | | — | — |
| | [189] | N/A | Monitors distinct flows using HyperLogLog algorithm | HyperLogLog | | | | |
| | [190] | SpreadSketch | Tracks superspreaders using probabilistic counting | Invertible sketches | ✓ | | | |

Another important criteria is whether the scheme tracks heavy hitters across the whole network. For example, unlike HashPipe which considers a single switch, [187] tracks network-wide heavy hitters. Tracking network-wide heavy hitter is important as some applications (e.g., port scanners, superspreaders, etc.) cannot go undetected within a single location. Moreover, aggregating the results of switches separately for detecting heavy hitter is not sufficient; flows might not exceed a threshold locally, but when the total volume is considered, the threshold might be crossed.

#### 5) Comparison between P4-based and Traditional Heavy Hitter Detection

The main advantage of heavy hitters detection schemes in the data plane over sampling-based approaches is the ability to operate at line rate. This means that every packet is considered in the detection algorithm, which improves accuracy and the speed of detection. Moreover, additional applications that exploit reactive processing can be implemented. For instance, switches can perform a flow-size aware routing method to redirect traffic upon detecting a heavy hitter.

### B. CRYPTOGRAPHY

#### 1) Background

Performing cryptographic functions in the data plane is useful for a variety of applications (e.g., protecting the layer-2 with cryptographic integrity checks and encryption, mitigating hash collisions, etc.). Computations in cryptographic operations (e.g., hashing, encryption, decryption) are known to be complex and resource-intensive. The supported operations in switch targets and in the P4 language are limited to basic arithmetic (e.g., additions, subtractions, bit concatenation, etc.). Recently, a handful of works have started studying the possibility of performing cryptographic functions in the data plane. Generally, cryptographic functions are executed externally (e.g., on a CPU) and invoked from the data plane.

#### 2) External Cryptography

The authors in [191] argue on the need to implement cryptographic hash functions in the data plane to mitigate potential attacks targeting hash collisions. Consequently, they presented prototype implementations of cryptographic hash functions in three different P4 target platforms (CPU, SmartNIC, NetFPGA SUME). Another work by Hauser et al. [192] attempted to implement host-to-site IPsec in P4 switches. For simplification, only Encapsulating Security Payload (ESP) in tunnel mode with different cipher suites is implemented. The same authors also proposed P4-MACsec [314], an implementation of MACsec on P4 switches. MACsec is an IEEE standard for securing Layer 2 infrastructure by encrypting, decrypting, and performing integrity checks on packets.

Malina et al. [193] presented a solution where P4 programs invoke cryptographic functions (externs) written in VHDL on FPGAs. The goal of this work is to avoid coding cryptographic functions on hardware (VHDL), and thus enables rapid prototyping of in-network applications with security functions. Another work that relies on externs for cryptographic functions is P4NIS [194].

#### 3) Data Plane Cryptography

The previous works delegated the complex computations to the control plane. Chen et al. [195] implemented the Advanced Encryption Standard (AES) protocol in the data plane using scrambled lookup tables. AES is one of the most widely used symmetric cryptography algorithms that applies several encryption rounds on 128-bit input data blocks.

#### 4) Cryptography Schemes Comparison, Discussions and Limitations

Table 29 compares the aforementioned cryptography schemes. With respect to hashing, P4 currently implements hash functions that do not have the characteristics of cryptographic hashing. For example, Cyclic Redundancy Check (CRC), which is commonly used in P4 targets, is originally developed for error detection. CRC can be easily implemented in embedded hardware, and is computationally much less complex than cryptographic hash functions (e.g., Secure Hash Algorithm (SHA)-256); however, it is not secure and has a high collision rate. Evaluation results in [191] show that 1) implementing cryptographic hash functions on CPU is easy, but has high latency (several milliseconds); 2) SmartNICs has the highest throughput, but can only process packets up to 900 bytes; and 3) NetFPGA has the lowest latency, but cannot be integrated using native P4 features. The authors found that the performance of hashing is highly dependent on the application, the input type, and the hashing algorithm, and therefore there is no single solution that fits all requirements. However, P4 targets should benefit from the characteristics of each solution (CPU, SmartNICs, FPGA, and ASICs) to implement cryptographic hashing.

**TABLE 29:** Cryptography schemes comparison.

| | Ref | Name | Core idea | Security goal | | | Algorithms | Platform | |
|---|---|---|---|---|---|---|---|---|---|
| | | | | Conf. | Integ. | Auth. | | HW | SW |
| **External** | [191] | N/A | Implementations of cryptographic hash function | ✓ | ✓ | | SipHash-2-4, Poly1305-AES, BLAKE2b, HMAC-SHA256-51 | | |
| | [192] | P4-IPsec | Implementation of host-to-site IPsec in P4 switches | | | | AES-CTRHMAC-MD | | |
| | [192] | P4-MACsec | Implementation of MACsec on P4 switches | | | ✓ | AES-GCM | | |
| | [193] | N/A | VHDL-based cryptographic functions on FPGAs | | | | AES-GCM-256, SHA-3, EdDSA | | |
| | [194] | P4NIS | Splits and encrypts traffic to protect from eavesdropping | | | ✓ | RSA (512-1152) | | |
| **Data plane** | [195] | N/A | AES implementation using scrambled lookup tab | | ✓ | ✓ | AES-128, AES-192, AES-256 | | |

As for more complex protocol suites (e.g., IPsec), Hauser et al. [192] only implemented Encapsulating Security Payload (ESP) in tunnel mode for simplification. The Security Policy Database (SPD) and the Security Association Database (SAD) are represented as match-action tables in the P4 switch. To avoid complex key exchange protocols such as the Internet Key Exchange (IKE), this work delegates runtime management operations to the control plane. Moreover, since encryption and decryption are not supported by P4, the authors relied on user-defined P4 externs to perform complex computations. Note that implementing user-defined externs is not applicable for ASIC (e.g., Tofino), and consequently, the main CPU module of the switch is used for performing encryption/decryption computations, at the cost of increased latency and degraded throughput. Same ideas are applied to P4-MACsec by the same authors. Other works that rely on externs include [193, 194].

The system proposed by Chen et al. [195] has significant performance advantages as it is fully implemented in the data plane. The idea of the proposed system is to apply permuted lookup tables by using an encryption key. The authors found that a single switch pipeline is capable of performing two AES rounds. Consequently, the system leverages *packet recirculation* technique which re-injects the packet into the pipeline. By doing so, it is possible to complete the 10 rounds of encryption required by the AES-128 algorithm by using five pipeline passes. Note that recirculation uses loopback ports and hence is limited by their bandwidth. The implementation on Tofino chip shows that ≈10Gbps throughput was attained. The authors argued that this throughput is sufficient to support various in-network security applications. Nevertheless, it is possible to enhance the throughput by configuring additional physical ports as loopback ports.

Note that there are other schemes that implements some cryptographic primitives in the data plane but are in the Privacy and Anonymity category (Section XI-C).

#### 5) Comparison between In-network and Contemporary Cryptography

Cryptographic primitives often require performing complex arithmetic operations on data. Implementing such computations on general purpose servers is simple; memory and processing units are not constrained. The literature has shown that there is a need to implement cryptographic functions in the data plane. For instance, cryptographic hash functions can significantly improve existing data plane applications with respect to collisions; encryption can protect confidential information from being exposed to the public. However, switches have limitations when it comes to computing. Supported hash functions in P4 are non-cryptographic (e.g., CRC), and therefore, produce collisions when the table is not large. Consequently, researchers are continuously investigating techniques to perform such operations in the data plane.

### C. PRIVACY AND ANONYMITY

#### 1) Background

Packets in a network carry information that can potentially identify users and their online behavior. Therefore, user privacy and anonymity have been extensively studied in the past (e.g., ToR and onion routing [315]). However, existing solutions have several limitations: 1) poor performance since overlay proxy servers are maintained by volunteers and have no performance guarantees; 2) deployability challenges; some solutions require modifying the whole Internet architecture, which is highly unlikely; 3) no clear partial de-ployment pathway; and 4) most solutions are software-based. Consequently, recent works started investigating methods that exploit programmable switches to develop partially-deployable, low-latency, and light-weight anonymity systems. With respect to anonymity and privacy in the network, new class of attacks which target the topology, requires the attacker to know the topology and understand it's forwarding behavior. Such attacks can be mitigated by obfuscating (hiding) the topology from external users. P4-based schemes are also being developed to achieve this goal.

**TABLE 30:** Privacy and anonymity schemes comparison.

| Name/Scheme | Goal | Strategy | Platform | |
|---|---|---|---|---|
| | | | HW | SW |
| ONTAS [196] | Protect PII in packet traces | Headers fields hashing | | |
| PANEL [197] | Protect Internet users' identities | Source info rewriting | | |
| SPINE [198] | Protect Internet users' identities | Header fields concealing | | |
| PINOT [199] | Anonymization when using DNS | Header fields concealing | | |
| NetHide [200] | Mitigate topology attacks | Topology obfuscation | N/A | N/A |

### 2) Users Privacy Protection

Kim et al. [196] proposed Online Network Traffic Anonymization System (ONTAS), a system that anonymizes traffic *online* using P4 switches. Moghaddam et al. [197] proposed Practical Anonymity at the NEtwork Level (PANEL), a lightweight and low overhead in-network solution that provides anonymity into the Internet forwarding infrastructure. Likewise, Datta et al. [198] proposed Surveillance Protection in the Network Elements (SPINE), a system that anonymizes traffic by concealing IP addresses and relevant TCP fields (e.g., sequence number) from adversarial Autonomous Systems (ASes) on the data plane. Wang et al. [199] proposed Programmable In-Network Obfuscation of DNS Traffic (PINOT), a system where the packet headers are obfuscated in the data plane to protect the identity of users sending DNS requests.

On the other hand, Meier et al. [200] proposed NetHide, a P4-based solution that obfuscates network topologies to mitigate against topology-centric attacks such as Link-Flooding Attacks (LFAs).

### 3) Privacy and Anonymity Schemes Discussions

Table 30 compares the privacy and anonymity schemes. NetHide aims at mitigating the attacks targeting the network topology. The solution formulates network obfuscation as a multi-objective optimization problem, and uses accuracy (hard constraints) and utility (soft constraints) as metrics. The system then uses ILP solver and heuristics. The P4 switches in this system capture and modify tracing traffic at line rate. The specifics of the implementation were not disclosed, but the authors claim that the system was evaluated on realistic topologies (more than 150 nodes), and more than 90% of link failures were detected by operators, despite obfuscation.

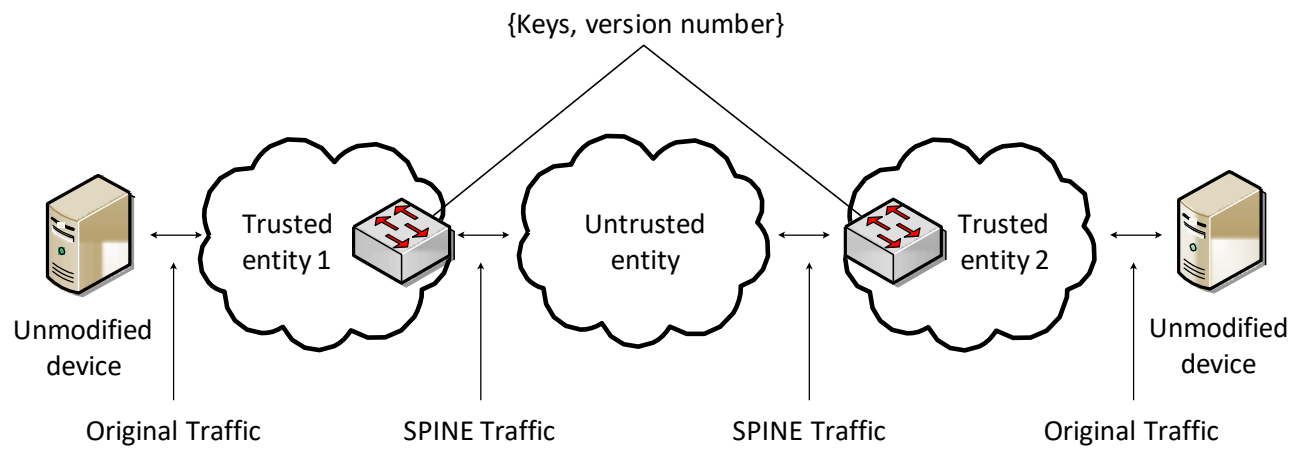

**FIGURE 21:** SPINE architecture [198].

ONTAS had a slightly different goal; it aims at protecting the personally identifiable information (PII) from online traces. The system overcomes the limitations of existing systems which either requires network operators to anonymize packet traces before sharing them with other researchers and analysts, or anonymize traffic online but with significant overhead. ONTAS provides a policy language used by operators for expressing anonymization tasks, which makes the system flexible and scalable. The system was implemented and tested on a hardware switch, and results show that ONTAS entails 0% packet processing overhead and requires half storage compared to existing offline tools. A limitation of this system is that it does not anonymize TCP/UDP field values. Another limitation is that it does not support applying multiple privacy policies concurrently.

Other line of research (i.e., PANEL, SPINE) focused on protecting the identities of Internet user. PANEL overcomes the performance limitations of popular anonymity systems (e.g., Tor), and does not require modifying entirely the Internet routing and forwarding protocols as proposed in [316] and [317]. Partial deployment is possible as PANEL can co-exist with legacy devices. The solution involves: 1) source address rewriting to hide the origin of the packet; 2) source information normalization (IP identification and TCP sequence randomization) to mitigate against fingerprinting attacks; and 3) path information hiding (TTL randomization) to hide the distance to the original sender at any given vantage point.

As for SPINE, it does not require cooperation between switches and end-hosts, but assumes that at least two entities (typically two ASes or two ISPs) are trusted. Fig. 21 shows the SPINE architecture. The solution encrypts the IP addresses before the packets enter the intermediary ASes. Therefore, adversarial devices only see the encrypted addresses in the headers. It also encrypts the TCP sequence and ACK numbers to mitigate against attributing packets to flows. SPINE transforms IPv4 headers into IPv6 headers when packets leave the trusted entity and restore the IPv4 headers upon entering the trusted entity. These operations enable routing to be performed in intermediary networks. The encrypted IPv4 address is inserted in the last 32-bits of the IPv6 destination address. The encryption works by XORing the IP address with the hash of a pre-shared key and a nonce. The system uses SipHash since it is easily implemented in the data plane.

Note that SPINE was implemented on software. If ASIC implementation was to be done, SPINE would require at least three pipeline passes to be fully executed (i.e., through recirculation). Thus, the throughput of SPINE would be decreased by a factor of three. In contrast, PINOT was executed on the ASIC with a single pipeline pass, and hence, has a higher throughput that the other solutions.

#### 4) Privacy and Anonymity in Switch-based and Legacy Systems

Contemporary approaches that provide privacy and anonymity in the Internet uses special routing overlay networks to hide the physical location of each node from other participants (e.g., Tor). Such approaches have performance limitations as proxy servers (overlays) are maintained by volunteers and have no performance guarantees. Moreover, they often require performing advanced encryption routines to obfuscate from where the packet is originated (e.g., onion routing technique used by Tor involves encapsulating messages in several layers of encryption). On the other hand, approaches that are based on programmable switches often rely on headers modification and simplified encryption and hashing to conceal information (e.g., SPINE [198]).

### D. ACCESS CONTROL

#### 1) Background

The selective restriction to access digital resources is known as access control in cybersecurity. Typically, access control begins with "authentication" in order to verify the identity of a party. Afterwards, "authorization" is enforced through policies to specify access rights to resources. To authenticate parties, methods such as passwords, biometric analysis, cryptographic keys, and others are used. With respect to authorization, methods such as ACL are used to describe what operations are allowed on given objects.

With the advent of programmable switches, it is now possible to delegate authentication and authorization to the data plane. As a result, access can be promptly granted or denied at line rate, before reaching the target server. A clear advantage of this approach is that servers are no longer busy processing access verification routines, which increases their services throughput.

#### 2) Firewalls

Datta et al. [201] presented P4Guard, a stateful P4-based configurable firewall that acts based on predefined policies set by the controller and pushed as entries to data plane tables. Similarly, Cao et al. [202] proposed CoFilter, another stateful firewall that encodes the access control rules in the data plane. Li et al. [203] presented an architecture in SDN-based clouds where P4-based firewalls are provided to the tenants. Almaini et al. [204] proposed delegating the authentication of end hosts to the data plane. The method is based on port knocking, in which hosts deliver a sequence of packets addressed to an ordered list of closed ports. If the ports match the ones configured by the network administrators, then end host is authenticated, and subsequent packets are allowed.

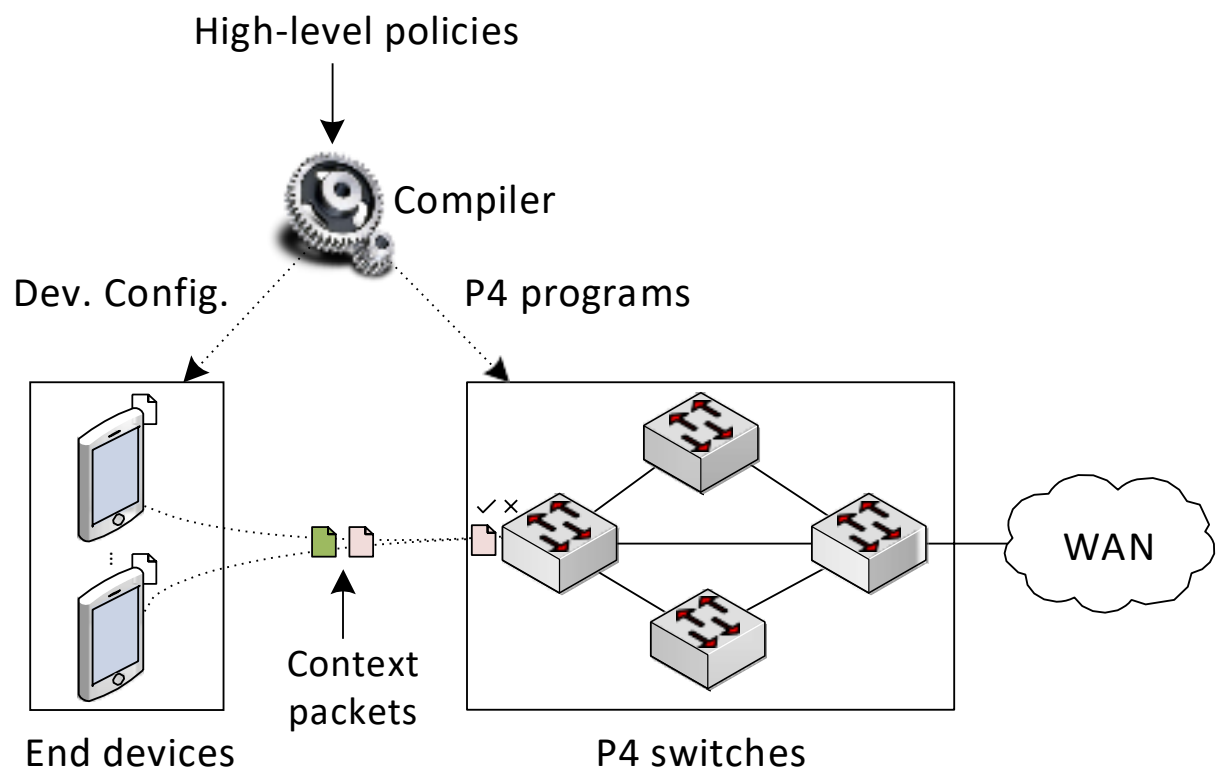

FIGURE 22: Overview of Poise [206]. A compiler translates high-level policies into P4 programs and device configurations. Context packets are continuously sent from the clients to the network, where the switches enforce the policies.

Likewise, Zaballa et al. [205] implemented port knocking in the data plane.

#### 3) Other Access Control

Kang et al. [206] presented a scheme that implements context-aware security policies (see Fig. 22). The policies are applicable to enterprise and campus networks with diverse devices, i.e., Bring Your Own Device (BYOD) (e.g., laptops, mobile devices, tablets, etc.). In context-aware policies, devices are granted access dynamically based on the device's runtime properties. Finally, Bai et al. [207] presented P40f, a tool that performs OS fingerprinting on programmable switches, and consequently, applies security policies (e.g., allow, drop, redirect) at line rate. Almaini et al. [208] implemented an authentication technique based on One Time Passwords (OTP). The technique follows the Leslie Lamport algorithm [318] in which a chain of successive hash functions are verified for authentication.

#### 4) Access Control Comparison, Discussions, and Limitations

Table 31 compares the aforementioned access control schemes. P4Guard provides access control based on security policies translated from high-level security policies to table entries. Note that P4Guard only operates up to the transport layer (e.g., source/destination IP addresses, source/destination ports, protocol, etc.), similar to a traditional firewall. While programmable switches provide increased flexibility in the parser (e.g., parse beyond the transport layer) and the packet processing logic, P4Guard did not leverage such capabilities. It would be interesting to investigate additional capabilities such as those enabled by next-generation firewalls (NGFW).

The solution in [204] controls access by performing authentication in the data plane. The solution has several limitations since it uses on port knocking, a technique that has several security implications. For instance, programmable switches do not use cryptographic hashes, making the solution vulnerable to IP address spoofing attacks. Additionally, unencrypted port knocking is vulnerable to packet sniffing.

**TABLE 31:** Access control schemes comparison.

| | Ref / Name | Core idea | Scope | Limitations | Platform | |
|---|---|---|---|---|---|---|
| | | | | | HW | SW |
| **Firewalls** | P4Guard [201] | Translates from high-level security policies to table entries | Header-based firewall (layer-4) | Lacks NGFW capabilities | | |
| | CoFilter [202] | Stateful packet filter on the switch | Flow ID and state-based firewall | Relies on external servers | | |
| | N/A [203] | P4 firewall as a service for cloud tenants | Header-based firewall (layer-4) | Lacks NGFW capabilities | | |
| | N/A [204] | Uses port knocking technique for authentication | Unencrypted sequence-based authentication | Unencrypted sequence vulnerable to packet sniffing | | |
| | P4Knocking [205] | Uses port knocking technique for authentication | Unencrypted sequence-based authentication | Unencrypted sequence vulnerable to packet sniffing | N/A | N/A |
| **Other Access Control** | Poise [206] | Context-aware policies enforcement | CAS dynamic policies based on runtime contexts | External encryptions are slow; lack of authentication | | |
| | P40f [207] | OS fingerprinting and policy enforcement | Uses p0f to filter connections | Lack of advanced built-in actions (e.g., rate-limiting) | | |
| | N/A [208] | Uses One Time Password technique for authentication | Leslie Lamport algorithm | Continuous interaction with the control plane | | |

Furthermore, port knocking relies on security through obscurity.

In [206], the scheme dynamically enforces access control to users based on contexts (e.g., if the user's device uses Secure Shell (SSH) 2.0 or higher, then the switch forwards the packets of this flow. Otherwise, the switch drops the packets). The scheme requires user devices to run an application which communicates with the switch using a custom protocol (context packets). The context packets are generated on a per-flow basis. The switch tracks flows using a match action table and registers at the data plane. Actions over a packet are dropping, allowing, and forwarding to other appliances for deep packet inspection. Data packets are not modified. Evaluations show that the proposed approach can operate (install new flows in the and update rules) with a minimum latency, even under heavy DoS attacks. On the other hand, such attacks can decimate similar SDN-based systems. One of the main drawbacks of the proposed system is the lack of authentication, integrity, and confidentiality of the context packets. Thus, the system can be subject to attacks such as snooping (i.e., eavesdropping) on communication between user devices and the switch, impersonation, and others.

Finally, [207] proposes fingerprinting OS in the data plane. The main motivation behind this work is that software-based passive fingerprinting tools (e.g., p0f [319]) are not practical nor sufficient with large amounts of traffic on high-speed links. Furthermore, out-of-band monitoring systems cannot promptly take actions (e.g., drop, forward, rate-limit) on traffic at line rate. The main drawback of the solution is that it lacks sophisticated policies that involve rate-limiting traffic.

#### 5) Comparison between Switch-based and Server-based Access Control

Controlling access to resources often starts with authentication. While server-based approaches are more flexible in the methods of authentication they can provide, they typically require client connections to reach the server before the communication starts. In switch-based approaches, the authentication can be done in-network at the edge, eliminating unnecessary latency incurred from traversing the network and from software processing.

Various features are considered when comparing P4-based firewalls to traditional firewalls. First, P4 firewalls are capable of performing headers inspection above the transport layer (also known as deep packet inspection (DPI)), whereas traditional firewalls only reach the transport layer and typically operate on the 5-tuple fields. It is important to note that DPI in P4 switches is limited: if only few bytes are parsed above the transport layer, line rate will be achieved; however, if the packet is deeply parsed, the throughput will start degrading accordingly. Second, in P4, policies and rules can be customized to be activated based on arbitrary information stored in the switch state (e.g., measurements through streaming); such capabilities are not present in traditional firewalls. Third, in P4, access control algorithms' exclusivity and innovation are solely attributed to operators, unlike fixed-function firewalls which are provided by device vendors. Note that non-programmable Next-Generation Firewalls (NGFW) are capable of performing advanced DPI at the cost of having much lower throughput than the line rate.

Access to resources can be controlled after fingerprinting end-hosts OS. Software-based passive fingerprinting tools cannot keep up with the high load (gigabits/s links). The literature has shown that said tools lead to 38% degradation in throughput [320]. Additionally, such tools are out-of-band, meaning that it is not possible to apply policies on traffic (e.g., after fingerprinting an OS). On the other hand, switch hardware is able to perform OS fingerprinting and apply security policies at line rate. Context-aware policies applied on nodes (clients/servers) have local visibility. A newer approach is to use a centralized SDN controller (e.g., [321]), but such scheme is vulnerable to control plane saturation attacks and is subject for delay increases. Switch-based schemes on the other hand are able to provide access control at line rate.

### *E. DEFENSES*

#### 1) Background

DDoS attacks remain among the top security concerns despite the continuous efforts towards the development of their detection and mitigation schemes. This concern is exacerbated not only by the frequency of said attacks, but also by their high volumes and rates. Recent attacks (e.g. [322, 323]) reached the order of terabits per seconds, a rate that existing defense mechanisms cannot keep with.

There are two main concerns with existing defense methods handled by end-hosts or deployed as middlebox functions on x86-based servers. First, they dramatically degrade the throughput and increase latency and jitter, impacting the performance of the network. Second, they present severe consequences on the network operation when they are installed at the last mile (i.e., far from the edge). The escalation of volumetric DDoS attacks and the lack of robust and efficient defense mechanisms motivated the idea of architecting defenses into the network. Up until recently, in-network security methods were restricted to simple access control lists encoded into the switching and routing devices. The main reason is that the data plane was fixed in function, impeding the capabilities of developing customized and dynamic algorithms that can assist in detecting attacks. With the advent of programmable data planes, it is possible to develop systems that detect and mitigate various types of attacks without imposing significant overhead on the network.

#### 2) Attack-specific

Hill et al. [209] presented a system that tracks flows in the data plane using bloom filters. The authors evaluated SYN flooding as a use case for their system. Li et al. [210] presented NETHCF, a Hop-Count Filtering (HCF) defense mechanism that mitigates spoofed IP traffic. HCF schemes filter spoofed traffic with an IP-to-hop-count mapping table. Another attack-specific scheme proposed by Febro et al. [211] mitigates against distributed SIP DDoS in the data plane. Furthermore, Scholz et al. [212, 213] presented a scheme that defends against SYN flood attacks. Ndonda et al. [214] implemented an intrusion detection system in P4 that whitelists and filters Modbus protocol packets in industrial control systems.

#### 3) Generic Attacks

Some schemes are generic and aim at addressing multiple attacks concurrently. For instance, Xing et al. [215] proposed FastFlex, an abstraction that architects defenses into the network paths based on changing attacks. Kang et al. [216] presented an automated approach for discovering sensitivity attacks targeting the data plane programs. Sensitivity attacks in this context are intelligently crafted traffic patterns that exploit the behavior of the P4 program. Lapolli et al. [217] implemented a mechanism to perform real-time DDoS attack detection based on entropy changes. Such changes will be used to compute anomaly detection thresholds. Mi et al. [218] proposed ML-Pushback, a P4-based implementation of the Pushback method [219].

Zhang et al. [220] proposed Poseidon, a system that mitigates against volumetric DDoS attacks through programmable switches. It provides a language where operators can express a range of security policies. Friday et al. [221] proposed a unified in-network DDoS detection and mitigation strategy that considers both volumetric and slow/stealthy DDoS attacks. Xing et al. [222] proposed NetWarden, a broad-spectrum defense against network covert channels in a performance-preserving manner. The method in [223] models a stateful security monitoring function as an Extended Finite State Machine (EFSM) and expresses the EFSM using P4 abstractions. Ripple [224] provides decentralized link-flooding defense against dynamic adversaries.

Ilha et al. [225] presented EUCLID, an extension to [217] where the data plane runs a fine-grained traffic analysis mechanism for DDoS attack detection and mitigation. EUCLID is based on information-theoretic and statistical analysis (entropy) to detect the attacks. Khooi et al. [226] presented a Distributed In-network Defense Architecture (DIDA), a solution that deals with with the sophisticated amplified reflection DDoS. Ding et al. [227] proposed IN-DDoS, an in-network DDoS victim identification system that fingerprints the devices that for which the number of packets exceeds a certain threshold. Musumeci et al. [228] proposed a system where ML algorithms executed on the control plane update the data plane after observing the traffic. Finally, Liu et al. [229] proposed Jaqen, an inline DDoS detection and mitigation scheme that addresses a broad range of attacks in an ISP deployment.

#### 4) Defense Schemes Comparison, Discussions, and Limitations

Table 32 compares the aforementioned defense schemes. Broadly, defense schemes can be grouped into two main categories: attack-specific and generic. Attack-specific category consists of the work that address a specific attack (e.g., NETHCF for IP spoofing, [211] for SIP DDoS, etc.), while the generic category aims at addressing various types of attacks (e.g., FastFlex for various availability attacks, Ripple for link flooding attacks, etc.).

The significant advantage of architecting defenses in the data plane is the performance improvement of the application. For instance, NETHCF is motivated by the fact that traditional HCF-based schemes are implemented on end-hosts, which delays the filtering of spoofed packets and increases the bandwidth overhead. Moreover, since traditional schemes are implemented in server-based middleboxes, low latency and minimal jitter are hard to achieve. Similarly, FastFlex advocates on the need to offload the defenses to the data plane. Specifically, it tackles the following key challenges that are faced when programming defenses in the data plane: 1) resource multiplexing; 2) optimal placement; 3) distributed control; and 4) dynamic scaling.

**TABLE 32:** Defenses schemes comparison.

| | Name & Scheme | Mitigated attack | External Computations | Network-wide | Limitations | Platform HW | Platform SW |
|---|---|---|---|---|---|---|---|
| **Attack-specific** | [209] | SYN flooding | | ✓ | Simple implementation that cannot scale on HW | | |
| | NETHCF [210] | IP spoofing | | ✓ | Hop-counts incorrectness with the presence of NAT | | |
| | [211] | SIP DDoS | | ✓ | No support for encrypted packets (e.g., SIP/TLS) | | |
| | [212, 213] | SYN floods | | ✓ | Lack of cryptographic hash functions | | |
| | [214] | Attacks on Modbus | | ✓ | External server is required to run the Bro engine | | |
| **Generic** | FastFlex [215] | Availability attacks | ✓ | | Cross-domain federation complexity and security | | |
| | [216] | Sensitivity attacks | ✓ | ✓ | Limited evaluation on complex data plane programs | | |
| | [217] | DDoS anomalies | ✓ | ✓ | Lack of self-tuning and can be easily defeated | | |
| | ML-Pushback [218] | DDoS anomalies | | ✓ | Heavily depends on external computation | ✓ | ✓ |
| | Poseidon [220] | Volumetric DDoS | | ✓ | Human intervention for writing the defense policies | | |
| | [221] | Volumetric and stealthy DDoS | | ✓ | Only synthetic evaluations no extensive experimentation | | |
| | NetWarden [222] | Network covert channels | | ✓ | Slowpath/fastpath communication latency | | |
| | [223] | ECN protocol abuse | ✓ | ✓ | Small subset of attack space | | |
| | Ripple [224] | Link-flooding | ✓ | | Lack of comparison with other P4 approaches | | |
| | EUCLID [225] | DDoS anomalies | ✓ | ✓ | Lack of self-tuning for the sensitivity coefficient | | |
| | DIDA [226] | DDoS anomalies | ✓ | | Lack of evaluation on real traces and on an ASIC | | |
| | INDDoS [227] | DDoS anomalies | | ✓ | Highly parametrized with predefined thresholds | | |
| | Jaqen [229] | DDoS anomalies | ✓ | | Requires few seconds to react to attacks | | |

When deploying defenses in the data plane, operators must be aware of the capabilities of the constrained targets. Many operations that require extensive computations cannot be easily implemented on the data plane. The existing work either approximate the computations in the data plane (considering the computation complexity and the measurements accuracy trade-off), or delegate the computations to external processors (e.g., CPU on the switch, external server, SDN controller, etc.). For instance, NETHCF decouples the HCF defense into a *cache* running in the data plane and a *mirror* in the control plane. The cache serves the legitimate packets at line rate, while the mirror processes the missed packets, maintains the IP-to-hop-count mapping table, and adjust the state of the system based on network dynamics. In Poseidon, the defense primitives are partitioned to be executed on switches and on servers, based on their properties. On the other hand, in [217], the authors estimated the entropies of source and destination IP addresses of incoming packets for consecutive partitions (observation windows) in the data plane, without consulting external devices.

Perhaps the most significant state-of-the-art works in the defense schemes are Poseidon and Jaqen. Poseidon provides a modular abstraction that allows operators to express their defense policies. Poseidon requires external modules running on servers, making its deployment challenging, especially in ISP settings. Furthermore, such design incurs additional costs and undesirable latency. Jaqen addressed those limitations and was designed to be executed fully in the switch, without external support from servers. Additionally, Jaqen used universal sketches as data structures; this selection enables detecting a wide range of attacks instead of crafting custom algorithms for specific ones.

Network-wide defenses are those that are not restricted to a single switch, and require multiple switches to co-operate in the attacks detection and mitigation phases. Such co-operation significantly improves the accuracy and the promptness of the detection. More details on network-wide data plane systems is explained in Section XIII-D.

Finally, Table 32 lists some limitations of the existing schemes, which can be explored in future work to advance the state-of-the-art.

#### 5) Comparison between P4-based and Traditional Defense Schemes

Network attacks such as large-scale DDoS and link flooding may have substantial impact on the network operation. For such attacks, server-based defenses deployed at the last mile are problematic and inherently insufficient, especially when attacks target the network core. Moreover, it is not feasible to detect and mitigate large volume of attack traffic (e.g., SYN flood) on end-hosts without impacting the throughput of the network. Other defense schemes are proprietary, and hence are costly and limited to the detection algorithms provided by the vendors. Table 33 highlights the costs and the performance differences between switch-based schemes (Poseidon and Jaqen) and other existing solutions. When defenses are architected into the network (i.e., detection and mitigation are programmed into the forwarding devices), it is easy to detect, throttle, or drop suspicious traffic at any vantage point, at line rate, with significant cost reductions.

**TABLE 33:** Comparison of DDoS defense schemes. Source: [229].

| DDoS Solutions | Mitigation | Performance | Cost/Power |
|---|---|---|---|
| Bohatei [324] | Server-based | 10Gbps (80ms) | $5,600/600W |
| Arbor APS [325] | Cloud-based | 20Gbps (80ms) | $47,746/400W |
| ADS-8000 [326] | Hardware | 40Gbps (<10ms) | $102,550/450W |
| Poseidon [220] | Switch & Servers | 3.3Tbps (12us-80ms) | >$10,500/350W |
| Jaqen [229] | Switch | 3.3/6.5Tbps (12us) | $10,500/350W |

### F. SUMMARY AND LESSONS LEARNED

In the context of cybersecurity, a wide range of works leveraged programmable switches to achieve the following goals: 1) detect heavy hitters and apply countermeasures; 2) execute cryptographic primitives in the data plane to enable further applications; 3) protect the identity and the behavior of end-hosts, as well as obfuscate the network topology; 4) enforce access control policies in the network while considering network dynamics; and 5) architect defenses in the data plane to accelerate the detection and mitigation processes.

Identifying heavy hitters at line rate has several advantages. Recent works considered various data structures and streaming algorithms to detect heavy hitters. Future systems could explore more complex data structures that reduce the amount of state storage required on the switches. Furthermore, novel systems must minimize the false positives and the false negatives compared to both P4-based and legacy heavy hitter detection systems. Finally, new schemes should explore strategies for incremental deployment while maximizing flow visibility across the network.

There is an absolute necessity to implement cryptographic functions (e.g., hash, encrypt, decrypt) in the data plane. Such functions can be used by various applications that require low hashing collisions (e.g., load balancing) and strong data protection. Most existing efforts delegate the complex computations to the control plane. However, recent systems have demonstrated that AES, a well-known symmetric key encryption algorithm, can be implemented in the data plane.

Another interesting line of work provided privacy and anonymity to the network. Recent efforts obfuscated the network topology in order to mitigate topology-centric attacks (e.g., LFA). Such systems must preserve the practicality of path tracing tools, while being robust against obfuscation inversion. Additionally, link failures in the physical topology should remain visible after obfuscation. Furthermore, when randomizing identifiers to achieve session unlinkability, the identifiers must fit into the small fixed header space so that compatibility with legacy networks is preserved. Other efforts considered rewriting source information and headers concealing to protect the identity of Internet users.

Finally, access control methods and in-network defenses were proposed. Future access control schemes should explore further in-network methods to authenticate the users, beyond port knocking. Additionally, since switches are capable of inspecting upper-layer headers, it is worth exploring offloading some next generation firewall functionalities to the data plane (such as in [327]). For instance, in [170], the authors proposed a system that allows searching for keywords in the payload of the packet. Similar techniques could be leveraged to achieve URL filtering at line rate. Additionally, schemes should mitigate against stealthy, slow DDoS attacks.

## XII. NETWORK TESTING

Although programmable switches provide flexibility in defining the packet processing logic, they introduce potential risks of having erroneous and buggy programs. Such bugs may cause fatal damages, especially when they are unexpectedly triggered in production networks. In such scenarios, the network starts experiencing a degradation in performance as well as disruption in its operation. Bugs can occur in various phases in the P4 program development workflow (e.g., in the P4 program itself, in the controller updating data plane table entries, in the target compiler, etc.). Bugs are usually manifested after processing a sequence of packets with certain combinations not envisioned by the designer of the code. This section gives an overview of the troubleshooting and verification schemes for P4 programmable networks.

### A. TROUBLESHOOTING

#### 1) Background

Intensive research interests were drawn on troubleshooting the network. Previous efforts are mainly based on passive packet behavior tracking through the usage of monitoring technologies (e.g., NetSight [328], EverFlow [329]). Other techniques (e.g., Automatic test Packet Generation (ATPG) [330]) send probing packets to proactively detect network bugs. Such techniques have two main problems. First, the number of probe packets increases exponentially as the size of the network increases. Second, the coverage is limited by the number of probes-generating servers. Despite the flexibility that programmable switches offer, writing data plane programs increases the chance of introducing bugs into the network. Programs are inevitably prone to faults which could significantly compromise the performance of the network and incur high penalty costs.

**TABLE 34:** Troubleshooting schemes comparison.

| Name & scheme | Core idea | Fault detection | | Memory requirements | Platform | |
|---|---|---|---|---|---|---|
| | | Passive | Proactive | | HW | SW |
| P4DB [230] | On-the-fly runtime debugging using watch, break, and next primitives | | | High | | |
| P4Tester [231] | Probing-based troubleshooting using BDD | | | Low | | |
| [232] | Targets' behavior examination when undesired actions are triggered | N/A | | N/A | | |
| [233] | Execution paths profiling using Ball-Larus encoding | | | Low | | |
| KeySight [234] | Probing-based troubleshooting using PEC | | | Low | | |

#### 2) Programmable Networks Troubleshooting

Zhang et al. [230] proposed P4DB, an on-the-fly runtime debugging platform. The system debugs P4 programs in three levels of visibility by provisioning operator-friendly primitives: *watch*, *break*, and *next*. Zhou et al. [231] proposed P4Tester, a troubleshooting system for data plane runtime faults. It generates intermediate representation of P4 programs and table rules based on BDD data structure. Dumitru et al. [232] examined how three different targets, BMv2, P4-NetFPGA, and Barefoot's Tofino, behave when undesired behaviours are triggered. Kodeswaran et al. [233] proposed a data plane primitive for detecting and localizing bugs as they occur in real time. Finally, Zhou et al. [234] proposed KeySight, a platform that troubleshoots programmable switches with high scalability and high coverage. It uses Packet Equivalence Class (PEC) abstraction when generating probes.

Some schemes such as Whippersnapper [331], BB-Gen [332], P8 [333], and [334] provide benchmarking for P4 programs and aim at understanding their performance.

#### 3) Troubleshooting Schemes Comparison, Discussions, and Limitations

Table 34 compares the aforementioned troubleshooting schemes. Essentially, the schemes either passively track how packets are processed inside switches (e.g., [230, 233]) or diagnoses faults by injecting probes (e.g., [231, 234]). The main limitation of passive detection is that schemes can only detect rule faults that have been triggered by existing packets, and cannot check the correctness of all table rules. On the other hand, probing-based schemes may incur large control and probes overheads.

Examples of probing-based schemes include P4Tester and KeySight. P4Tester generates intermediate representation of P4 programs and table rules based on BDD data structure. Afterwards, it performs an automated analysis to generate probes. Probes are sent using source routing to achieve high rule coverage while maintaining low overheads. The system was prototyped on a hardware switch (Tofino), and results show that it can check all rules efficiently and that the probes count is smaller than that of server-based probe injection systems (i.e., ATPG and Pronto).

Other schemes that use passive fault detection (e.g., P4DB) assume that packets consistently trigger the runtime bugs. P4DB debugs P4 programs in three levels of visibility by provisioning operator-friendly primitives: *watch*, *break*, and *next*. P4DB does not require modifying the implementation of the data plane. It was implemented and evaluated on a software switch (BMv2), and the results show that it is capable of troubleshooting runtime bugs with a small throughput penalty and little latency increase.

Another important criterion that differentiate the troubleshooting schemes is the memory footprint they require. Some schemes (e.g., P4DB) require more memory than others (e.g., KeySight).

Finally, the work in [232] is different than the others. The authors examined how three different targets, BMv2, P4-NetFPGA, and Barefoot's Tofino, behave when undesired behaviours are triggered. The authors first developed buggy programs in order to observe the actual behavior of targets. Then, they examined the most complex P4 program publicly available, *switch.p4*, and found that it can be exploited when attackers know the specifics of the implementation. In summary, the paper suggests that BMv2 leaks information from previous packets. This behavior is not observed with the other two targets. Furthermore, the authors were able to perform privilege escalation on switch.p4 due to a header destined to ensure communication between the CPU and the P4 data plane.

#### 4) Comparison Legacy vs. P4-based Debugging

In legacy networks, network devices are equipped with fixed-function services that operate on standard protocols. Troubleshooting these networks often involve testing protocols and typical data plane functions (e.g., layer-3 routing) through rigid probing. On the other hand, with programmable networks, since operators have the flexibility of defining custom data plane functions and protocols, testing is more complex and is program-dependent. Probing-based approaches should craft patterns depending on the deployed P4 program. Other approaches proposed primitives that increase the levels of visibility when debugging P4 programs. Research work extracted from the literature show that it is essential to develop flexible mechanisms that operate dynamically on diverse P4 programs and targets.

### B. VERIFICATION

#### 1) Background

Program verification consists of tools and methods that ensure correctness of programs with respect to specifications and properties. Verification of P4 programs is an active area as bugs can cause faults that have drastic impacts on the performance and the security of networking systems. Static P4 verification handles programs before deployment to the network, and hence, cannot detect faults that occur at runtime. On the other hand, runtime verification uses passive measurements and proactive network testing. This section describes the major verification work pertaining to P4 programs.

### 2) Program Verification

Lopes et al. [235] proposed P4NOD, a tool that compiles P4 specifications to Datalog rules. The main motivation behind this work is that existing static checking tools (e.g., Header Space Analysis (HSA) [335], VeriFlow [336]) are not capable of handling changes to forwarding behaviors without reprogramming tool internals. The authors introduced the "well formedness" bugs, a class of bugs arising due to the capabilities of modifying and adding headers.

Another interesting work is ASSERT-P4 [236, 237], a network verification technique that checks at compile-time the correctness and the security properties of P4 programs. ASSERT-P4 offers a language with which programmers express their intended properties with assertions. After annotating the program, a symbolic execution takes place with all the assertions being checked while the paths are tested.

Further, Liu et al. [238] proposed p4v, a practical verification tool for P4. It allows the programmer to annotate the program with Hoare logic clauses in order to perform static verification. To improve scalability, the system suggests adding assumptions about the control plane and domain-specific optimizations. The control plane interface is manually written by the programmer and is not verified, which makes it error-prone and cumbersome. The authors evaluated p4v on both an open source and proprietary P4 programs (e.g., switch.p4) that have different sizes and complexities.

Nötzli et al. [239] proposed p4pktgen, a tool that automatically generates test cases for P4 programs using symbolic execution and concrete paths. The tool accepts as input a JSON representation of the P4 program (output of the p4c compiler for BMv2), and generates test cases. These test cases consist of packets, tables configurations, and expected paths. Similarly, Lukács et al. [240] described a framework for verifying functional and non-functional requirement of protocols in P4. The system translates a P4 program in a versatile symbolic formula to analyze various performance costs. The proposed approach estimates the performance cost of a P4 program prior to its execution.

Stoenescu et al. [241] proposed Vera, a symbolic execution-based verification tool for P4 programs. The authors argue in this paper that a data plane program should be verified before deployment to ensure safe operations. Vera accepts as input a P4 program, and translates it to a network verification language, SEFL. It then relies on SymNet [337], a network static analysis tool based on symbolic execution to analyze the behavior of the resulting program. Essentially, Vera generates all possible packets layouts after inspecting the program's parser and assumes that the header fields can accept any value. Afterwards, it tracks the paths when processing these packets in the program following all branches to completion. For scalability improvements, Vera utilizes a novel match-forest data structure to optimize updates and verification time. Parsing/deparsing errors, invalid memory accesses, loops, among others, can be detected by Vera.

**TABLE 35:** Verification schemes comparison.

| Scheme | Name | Engine, language | Evaluated programs | Inconsistency detection |
|---|---|---|---|---|
| [235] | P4NOD | NOD | 2 | ✓ |
| [236] | ASSERT-P4 | KLEE | 5 | ✓ |
| [238] | p4v | Z3 | 23 | ✓ |
| [239] | p4pktgen | SMT | 4 | ✓ |
| [240] | N/A | Pure | 0 | ✓ |
| [241] | Vera | SEFL | 11 | ✓ |
| [242] | P4RL | DDQN | 1 | |
| [243] | bf4 | Z3 | 21 | ✓ |

A different approach uses reinforcement learning is P4RL [242], a fuzzy testing system that automatically verifies P4 switches at runtime. The authors described a query language *p4q* in which operators express their intended switch behavior. A prototype that executes verification on layer-3 switch was implemented, and results show that PR4L detects various bugs and outperforms the baseline approach.

Finally, Dumitrescu et al. [243] proposed bf4, an end-to-end P4 program verification tool. It aims at guarantying that deployed P4 programs are bug-free. First, bf4 finds potential bugs at compile-time. Second, it automatically generates predicates that must be followed by the controller whenever a rule is to be inserted. Third, it proposes code changes if additional bugs remain reachable. bf4 executes a monitor at runtime that inspects the rules inserted by the controller and raises an exception whenever a predicate is not satisfied. The authors executed bf4 on various data plane programs and interesting bugs that were not detected in state-of-the-art approaches were discovered.

### 3) Verification Schemes Discussions

Table 35 compares the aforementioned verification schemes. Essentially, some schemes translate P4 programs to verification languages and engines. For instance, in [235], P4 programs are translated to Datalog to verify the reachability and well-formedness. Similarly, in [238], P4 programs are converted into Guarded Command Language (GCL) models, and then a theorem prover *Z3* is used to verify that several safety, architectural and program-specific properties hold. Other schemes (e.g., p4pktgen, Vera) use symbolic execution to generate test cases for P4 programs.

The verification schemes were evaluated on different P4 programs from the literature. A program that was evaluated by most schemes is *switch.p4* which implements various networking features needed for typical cloud data centers, including Layer 2/3 functionalities, ACL, QoS, etc. It is recommended for future schemes to evaluate *switch.p4* as well as other programs from the literature. Finally, P4RL detects path-related consistency between data-control planes.

#### 4) P4-based and Traditional Network Verification

Traditional verification techniques that address the security properties in computer networks are mainly related to host reachability, isolation, blackholes, and loop-freedom. Techniques that check for the aforementioned properties include Anteater [338], which models the data plane as boolean functions to be used in a Boolean Satisfiability Problem (SAT) solver, NetPlumber [339] which uses header space algebra [335], and others (e.g., VeriFlow [336], DeltaNet [340], Flover [341], and VMN [342]).

Since P4 programs incorporate customized protocols and processing logic to be used in the data plane, traditional tools are not capable of handling changes to forwarding behaviors without reprogramming their internals. Therefore, verification techniques in programmable networks rely on analyzing the P4 programs themselves since they define the behavior of the data plane.

### C. SUMMARY AND LESSONS LEARNED

Network testing can generally be divided into debugging/troubleshooting network problems and verifying the behavior of forwarding devices. While traditional tools and techniques were adequate for non-programmable networks, they are insufficient for programmable ones due to their inability to handle changes to forwarding behaviors without reprogramming and restructuring their internals. A variety of works were proposed to analyze and model P4 programs in order to troubleshoot and verify the correctness of networks' operations.

Network measurements can be collected through P4 switches and used to troubleshoot and verify the correctness of networks (control loop). Future work could explore methods that make a network more autonomous and capable of healing itself (e.g., self-driving networks, knowledge-defined networking, zero-touch networks) by leveraging the collected inputs from programmable switches.

## XIII. CHALLENGES AND FUTURE TRENDS

In this section, a number of research and operational challenges that correspond to the proposed taxonomy are outlined. The challenges are extracted after comprehensively reviewing and diving into each work in the described literature. Further, the section discusses and pinpoints several initiatives for future work which could be worthy of being pursued in this imperative field of programmable switches. The challenges and the future trends are illustrated in Fig. 23

### A. MEMORY CAPACITY (SRAM AND TCAM)

Stateful processing is a key enabler for programmable data planes as it allows applications to store and retrieve data across different packets. This advantage enabled a wide range of novel applications (e.g., in-network caching, fine grained measurements, stateful load balancing, etc.) that were not possible in non-programmable networks. The amount of data stored in the switch is limited by the size of the on-chip memory which ranges from tens to hundreds of megabytes at most. Consequently, the majority of stateful-based applications suffer have trade-offs between performance and memory usage. For instance, the efficiency of caching which is determined by the *hit rate* is directly affected by the memory size. Furthermore, the vast majority of measurement applications require storing statistics in the data plane (e.g., byte/packet counters). The number of flows to be measured and the richness of measurement information is bound by the size of the memory in the switch.

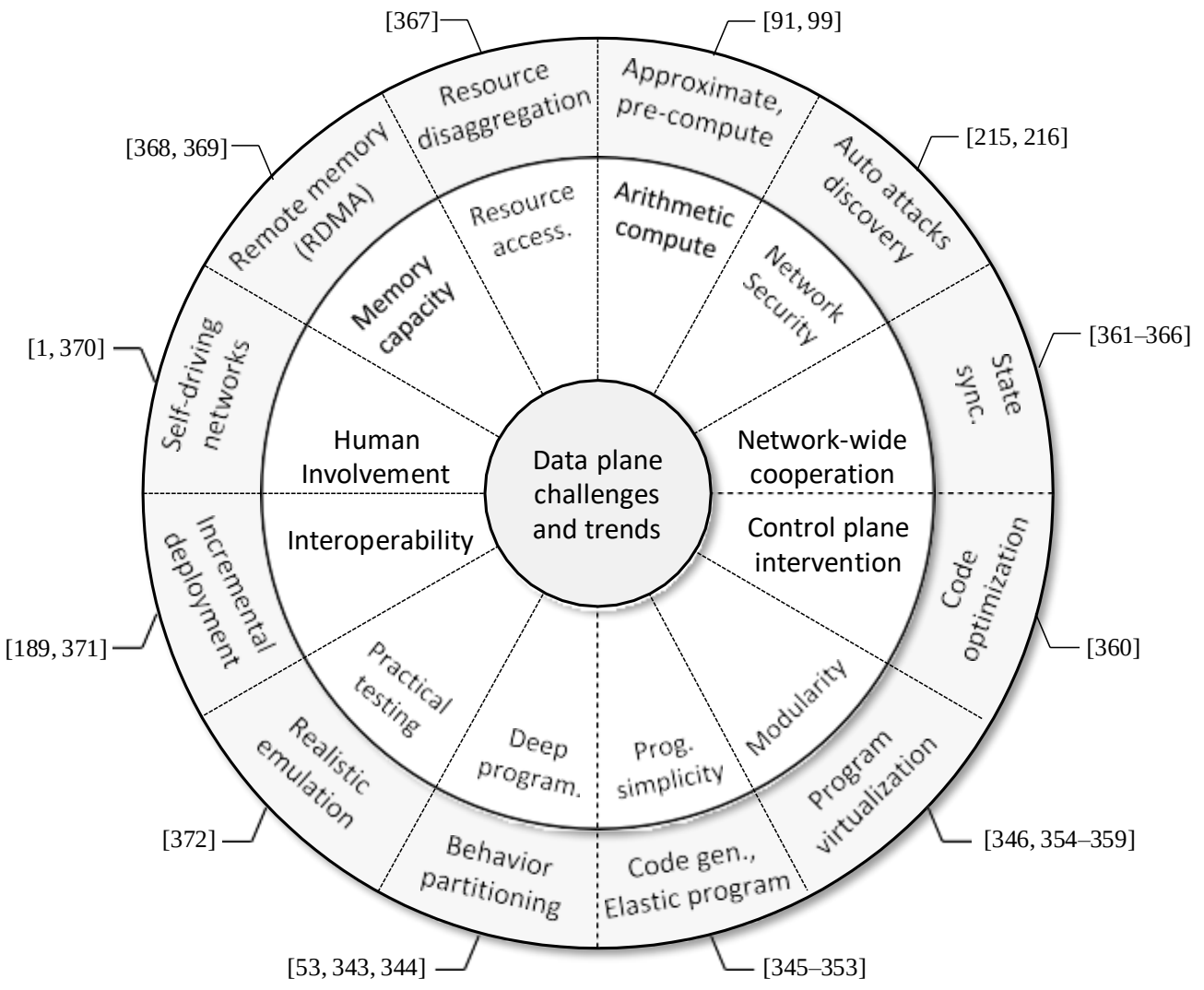

**FIGURE 23:** Challenges and future trends. The references represent examples of existing works that tackle the corresponding future trends.

**Current and future initiatives.** A notable work by Kim et al. [368, 369] suggests accessing remote Dynamic Random Access Memory (DRAM) installed on data center servers purely from data plane to expand the available memory on the switch. The bandwidth of the chip is traded for the bandwidth needed to access the external DRAM. The approach is cheap and flexible since it reuses existing resources in commodity hardware without adding additional infrastructure costs. The system is realized by allowing the data plane to access remote memory through an access channel (RDMA over Converged Ethernet (RoCE)) as shown in Fig. 24. The implementation show that the proposal achieves throughput close to the line rate, and only incur 1-2 extra microseconds latency (Fig. 25). There are some limitations in this approach that can be explored in the future.

- The current implementation only supports address-based memory access, and hence, complicated data layouts and ternary matching in remote memory should be explored.
- Frequent updates in the remote memory requires several packets for fetching and adding. This is common in measurement applications where counters are continuously incremented. A possible solution to the bandwidth overhead is aggregating updates into single operation. This comes with the cost of having delays in the updates.
- Packet loss between the switch and the remote memory should be handled, otherwise, the performance of the ap-

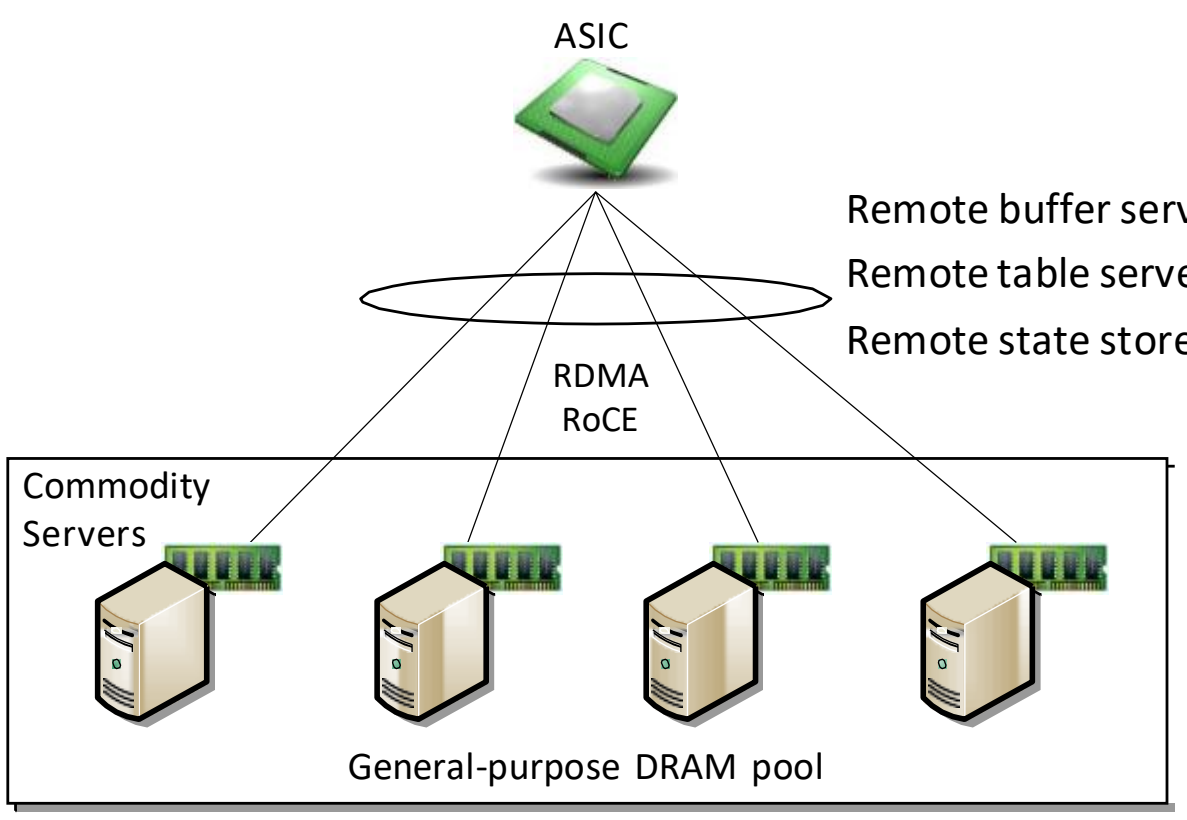

**FIGURE 24:** Expanding switch memory by leveraging remote DRAM on commodity servers [368].

plication and the freshness of the remote values might be affected.
- The interaction between general data plane applications and the remote memory is challenging. A potential improvement is designing well-defined APIs to facilitate the interaction.

### B. RESOURCES ACCESSIBILITY

Beside the size limitation of the on-chip memory, there are other restrictions that data plane developers should take into account [52, 373]. First, since the table memory is local to each stage in the pipeline, other stages cannot reclaim non-utilized memory in other stages. As a result, memory and match/action processing are fuzed, making the placement of tables challenging. Second, the sequential execution of operations in the pipeline lead to poor utilization of resources especially when the matches and the actions are imbalanced (i.e., the presence of default actions that do not need a match).

**Current and Future Initiatives.** An interesting work by Chole et at. [367] explored the idea of disaggregating the memory and compute resources of a programmable switch. The main notion of this work is to centralize the memory as a pool that is accessed by a crossbar. By doing so, each pipeline stage no longer has local memory. Additionally, this work solves the sequential execution limitation by creating a cluster of processors used to execute operations in any order. The main limitation of this approach is the lack of adoption by hardware vendors. Most of the switch vendors (e.g., Cavium's XPliant and Barefoot's Tofino) do not implement the disaggregation model and follow the regular Reconfigurable Match-action Tables (RMT) model. The implementation and analysis of the disaggregation model on hardware targets should be explored in the future.

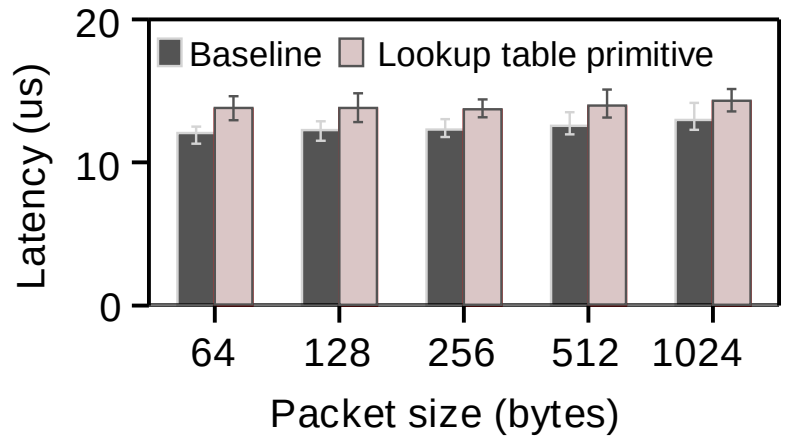

**FIGURE 25:** Accessing remote DRAM latency overhead. Only 1-2us additional latency. Achieved throughput close to the line rate (≈ 37.5 Gbps). Reproduced from [368].

### C. ARITHMETIC COMPUTATIONS

There are several challenges that must be handled when dealing with arithmetic computations in the data plane. First, programmable switches support a small set of simple arithmetic computations that operate on non-floating point values. Second, only few operations are supported per packet to guarantee the execution at line rate. Typically, a packet should only spend tens of nanoseconds in the processing pipeline. Third, computations in the data plane consume significant hardware resources, hampering the possibility of other programs to execute concurrently. A wide range of applications suffer from the lack of complex computations in the data plane. For instance, some operations required by AQMs (e.g., square root function in the CoDel algorithm) are complex to be implemented with P4. Additionally, the majority of machine learning frameworks and models operate on floating point values while the supported arithmetic operations on the switch operate on integer values. In-network model updates aggregation requires calculating the average over a set of floating-point vectors.

**Current and Future Initiatives.** Existing methods to overcome the computation limitations include approximation and pre-computations. In the approximation method, the application designer relies on the small set of supported operations to approximate the desired value, at the cost of sacrificing precision. For example, approximating the square root function can be achieved by counting the number of leading zeros through longest prefix match [99]. It would be beneficial for P4 developers to have access to a community-maintained library which encompasses P4 codes that approximate various complex functions. In the pre-computations method, values are computed by the control plane (e.g., switch CPU) and stored in match-action tables or registers. Future work can explore methods that automatically identify the complex computations that can be pre-evaluated in the control plane. After identification, the data plane code and its corresponding control plane APIs can be automatically generated.

### D. NETWORK-WIDE COOPERATION

The SDN architecture suggests using a centralized controller for network-wide switches management. Through centralization, the state of each programmable switch can be shared with other switches. Consequently, applications will have the ability to make better decisions as network-wide data is available locally on the switch. The problem with such architecture is the requirement of having a continuous exchange of packets with a software-based system. As an alternative, switches can exchange messages to synchronize their states in a decentralized manner.

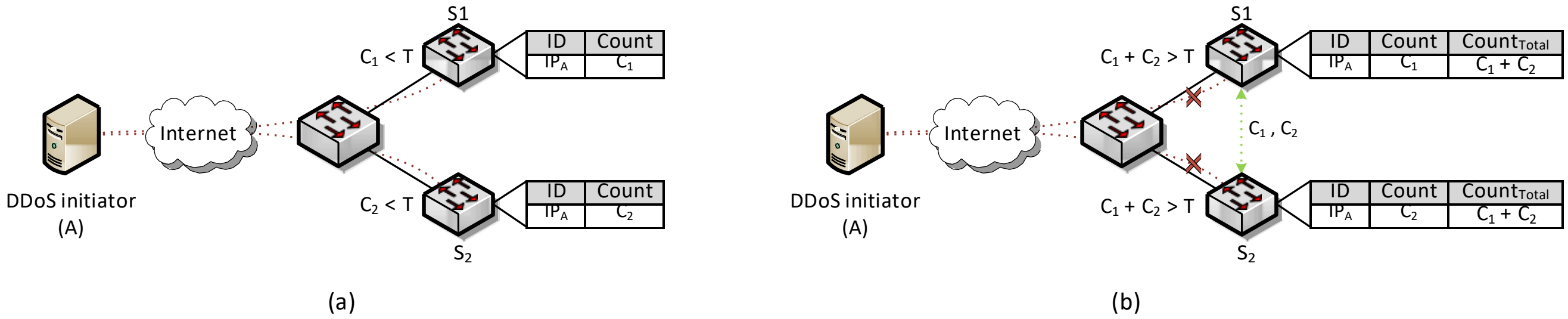

**FIGURE 26:** (a) Local detection of DDoS attacks. (b) network-wide detection of DDoS attack.

Consider Fig. 26 which shows an in-network DDoS defense solution. Each switch maintains a list of senders and their corresponding numbers of bytes. A switch compares the number of bytes transmitted from a given flow to a threshold. When the threshold is crossed, the flow is blocked and the device is identified as a malicious DDoS sender. Assume that the network implements a load balancing mechanism that distributes traffic across the switches. In the scenario where switches do not consider the byte counts of other switches (Fig. 26 (a)), the traffic of a DDoS device might remain under the threshold. On the other hand, when switches synchronize their states by sharing the byte counts (Fig. 26 (b)), the total number of bytes is compared against the threshold. Consequently, the total load of a DDoS device is considered. This example demonstrates an application that heavily depends on network-wide cooperation and hence motivates the need for state synchronization.

**Current and Future Initiatives.** Arashloo et al. [361] proposed SNAP, a centralized stateful programming model that aims at solving the synchronization problem. SNAP introduced the idea of writing programs for "one big switch" instead of many. Essentially, developers write stateful applications without caring about the distribution, placement, and optimization of access to resources. SNAP is limited to one replica of each state in the network. Sviridov et al. [362, 363] proposed LODGE and LOADER to extend SNAP and enable multiple replicas. Luo et al. [364] proposed Swing State, a framework for runtime state migration and management. This approach leverages existing traffic to piggyback state updates between cooperating switches. Swing State overcomes the challenges of the SDN-based architecture by synchronizing the states entirely in the data plane, at line rate, and without intervention from the control plane. There are several limitations with this approach. First, there are no message delivery guarantees (i.e., packets dropped/reordered are not retransmitted), leading to inconsistency in the states among the switches. Second, it does not merge the states if two switches share common states. Third, the overhead can significantly increase if a single state is mirrored several times. Finally, there is no authentication of data or senders. Xing et al. [365] proposed P4Sync, a system that migrates states between switches in the data plane while guaranteeing the authenticity of the senders and the exchanged data. P4Sync addresses the limitations of existing approaches. It guarantees the completeness of the migration, ensuring that the snapshot transfer is completed. Moreover, it solves the overhead of the repeatedly retransmitted updates. An interesting aspect of P4Sync is its ability to control the migration traffic rate depending on the changing network conditions. Zeno et al. [366] presented a design of SwiShmem, a management layer that facilitates the deployment of network functions (NFs) on multiple switches by managing the distributed shared states.

The future work in this area should consider handling *frequent state migrations*. Some systems require migration packets to be generated each RTT, causing increased traffic overhead and additional expensive authentication operations. For instance, P4Sync uses public key cryptography in the control plane to sign and verify the end of the migration sequence chain (2.15ms for signing and 0.07ms to verify using RSA-2048 signature). Frequent migrations would cause this signature to be involved repeatedly. Another major concern that should be handled in future work is *denial of service*. Even with migration updates authentication, changes in the packets cause the receiver to reject updates, leading to state inconsistency among switches.

### *E. CONTROL PLANE INTERVENTION*

Delegating tasks to the control plane incurs latency and affects the application's performance. For instance, in congestion control, rerouting-based schemes often use tables to store alternative routes. Since the data plane cannot directly modify table entries, intervention from the control plane is required. The interaction with the control plane in this application hampers the promptness of rerouting. Another example are methods that use collisions-free hashing. For example, cuckoo hash [374], which rearranges items to solve collisions, uses a complex search algorithm that cannot run on the switch ASIC, and is often executed on the switch CPU. Ideally, the control plane intervention should be minimized when possible. For example, to synchronize the state among switches, in-network cooperation should be considered.

**Current and Future Initiatives.** The design of the interaction between the control plane and the data plane is fully decided by the developer. Experienced developers might have enough background to immediately minimize such interaction. Future work should devise algorithms and tools that automatically determine the excessive interaction between the control/data planes, and suggest alternative workflows (ideally, as generated codes) to minimize such interaction. Operations that could be delegated to the data plane include failure detection and notification and connectivity retrieval [360].

### F. SECURITY

When designing a system for the data plane, the developer must envision the kind of traffic a malicious user can initiate to corrupt the operation of the system. This class of attacks is referred to as *sensitivity attacks* as coined in [216]. Essentially, an attacker can intelligently craft traffic patterns to trigger unexpected behaviors of a system in the data plane. For instance, a load balancer that balances traffic through packet headers hashing without cryptographic support (e.g., modulo operator on the number of available paths) can be tricked by an attacker that craft skewed traffic patterns. This results in traffic being forwarded to a single path, leading to congestion, link saturation, and denial of service. Another example is attacks against in-network caching. Caching in data plane performs well when requests are mostly *reads* rather than *writes*. If an attacker continuously generates high-skewed write requests, the load on the storage servers would be imbalanced. If the system is designed to handle write queries on hot items in the switch, a random failure in the switch causes data to be lost. Further, an attacker can also exploit the memory limitation of switch and request diverse values, causing the pre-cached values to be evicted.

**Current and Future Initiatives.** To mitigate against sensitivity attacks, a developer attempts to discover various unpredicted traffic patterns, and accordingly, develops defense strategies. Such solution is highly unreliable, time consuming, and error-prone. Recent efforts [216] aimed at automatically discovering sensitivity attacks in the data plane. Essentially, the proposed system aims at deriving traffic patterns that would drive the program away from common case behavior as much as possible. Other efforts focused on architecting defenses in the data plane that perform distributed mode changes upon attack discovery [215]. Future work in this direction should consider achieving high assurance by formally verifying the codes. Additionally, the stability of the data plane should be carefully handled with fast mode changes; future work could consider integrating self-stabilizing systems for such purpose. Finally, future work should provide security interfaces for collaborating switches that belong to different domains. It is also worth exposing sensitivity attack patterns for different application types so that data plane developers can avoid the vulnerabilities that trigger those attacks in their codes.

### G. INTEROPERABILITY

Programmable switches pave the way for a wide range of innovative in-network applications. The literature has shown that significant performance improvements are brought when applications offload their processing logic to the network. Despite such facts, it is very unlikely that mobile operators will replace their current infrastructure with programmable switches in one shot. This unlikelihood comes from the fact that major operational and budgeting costs will incur.

**Current and Future Initiatives.** Network operators might deploy programmable switches in an incremental fashion. That is, P4 switches will be added to the network alongside the existing legacy devices. While this solution seems simplistic at first, studies have showed that partial deployment leads to reduced effectiveness [189]. For instance, the accuracy of heavy hitter detection schemes is strongly affected by the flow visibility. The work in [189] devised a greedy algorithm that attempts to strategically position P4 switches in the network, with the goal of monitoring as many distinct network flows as possible. The *F1 score* is used to quantify correctness of switches placement. Other works that focused on incremental deployment include daPIPE [375], TraceILP/TopoILP [371]. Future work in this area should consider generalizing and enhancing this approach to work with any P4 application, and not only heavy hitter detection. For instance, a future work could suggest the positioning of P4 switches in applications such as in-network caching, accelerated consensus, and in-network defenses, while taking into account the current topology consisting of legacy devices.

Amin et al. [376] surveyed the research and development in the field of hybrid SDN networks. Hybrid SDN comprises a mix of SDN and legacy network devices. It is worth noting that the same key concepts and advantages of hybrid SDN networks can be applied to incremental P4 networks.

Recent efforts are also considering *network taps* as a mean to replicate production network's traffic to programmable switches for analysis [88]. Network TAPs and do not alter timing information and packet orders, which may occur with other schemes such as port mirroring operating at layer 2 and layer 3 [377]. ConQuest [88] taps on the ingress and egress links of a legacy router and uses a P4 switch to perform advanced fine-grained queue monitoring techniques. Note that legacy routers only support polling the total queue length statistics at a coarse time interval, and hence, cannot monitor microbursts. By tapping on legacy devices and processing on P4 switches, operators can benefit from the capabilities of P4 switches without the need to fully replace their current infrastructure. This method can be used in a variety of in-network applications (e.g., RTT estimation (see Fig. 27), network-wide telemetry, DDoS detection/mitigation, to name a few). Finally, it is worth mentioning that TAPs are not expensive and a single P4 switch can service many non-programmable devices.

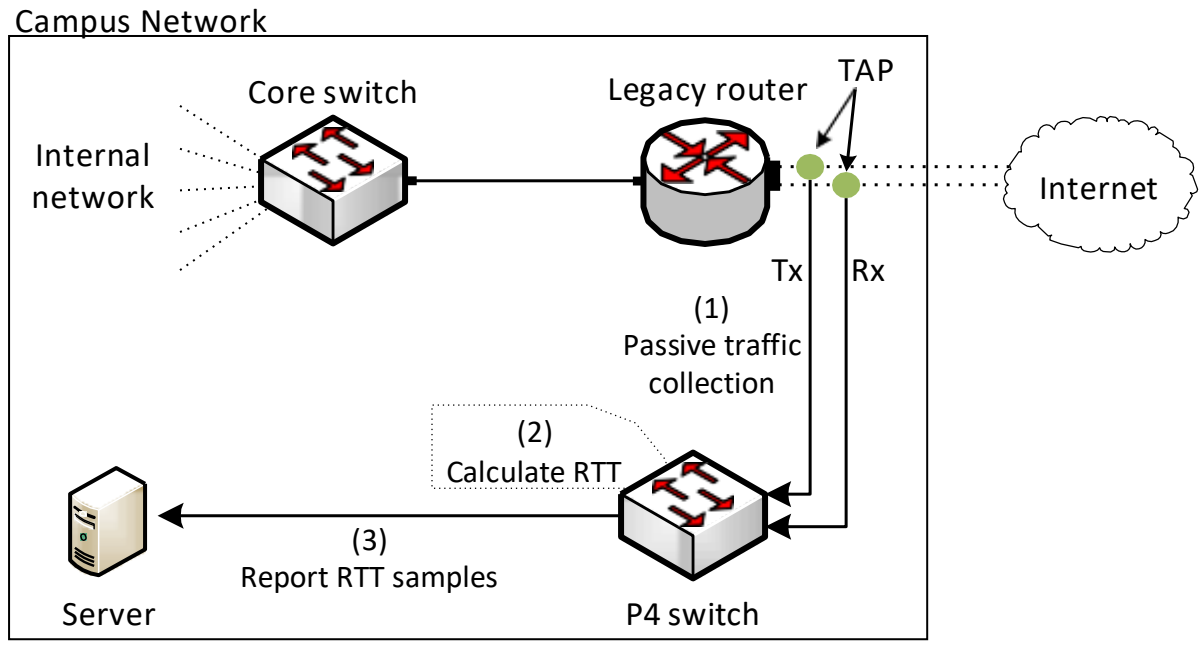

**FIGURE 27:** Example of using taps in a campus network to compute the round-trip time in the data plane. (1) The traffic is passively collected by the P4 switch; (2) the switch calculates the round-trip time by using its high-precision timer (see [95] for details on how to associate the SEQ/ACKs to compute the RTT); (3) the switch report the RTT samples to an external server.

### H. PROGRAMMING SIMPLICITY

Writing in-network applications using the P4 language is not a straightforward task. Recent studies have shown that many existing P4 programs have several bugs that might lead to complete network disruption [232]. Furthermore, since programmable switches have many restrictions on memory and the availability of resources, developers must take into account the low-level hardware limitations when writing the programs. This process is known to be based on trial and error; developers are almost never sure whether their program can "fit" into the ASIC, and hence, they repeatedly try to compile and adjust their codes accordingly. Such problem is exacerbated when the complexity of the in-network application increases, or when multiple functions (e.g., telemetry, monitoring, access control. etc.) are to be executed concurrently in the same P4 program. Additionally, code modularity is not simple in P4; the programmers typically rewrite existing functions depending on the constraints of the current context. All the aforementioned facts affect the cost, stability, and correctness of the network on the long run.

For several decades, the networking industry operated in a bottom-up approach, where switches are equipped with fixed-function ASICs. Consequently, little to no programming skills were needed by network operators. With the advent of programmable switches, operators are now expected to have experience in programming the ASIC[2].

**Current and Future Initiatives.** Since programming the ASIC is not a straightforward task, future research endeavours should consider simplifying the programming workflow for the operators and generating code (e.g., [345–352]). For instance, graphical tools can be developed to translate workflows (e.g., flowcharts) to P4 programs that can fit into the hardware.

A noteworthy work (P4All [353]) proposed an extension to P4 where operators write *elastic* programs. Elastic programs are compact programs that stretch to make use of the hardware resources. P4All extends P4 to support loops. The operator supply the P4All program along with the target specifications (i.e., constraints) to the P4All compiler. Afterwards, the compiler analyzes the dependencies between actions and unrolls the loops. Then, it generate the constraints for the optimization based on the target specification file. Next, the compiler solves an optimization problem that maximizes a linear utility function and generates an output P4 program for the target. The authors considered Tofino target in their evaluations. While P4All offered numerous advantages, it is still far from being ready to be used in practice. First, it assumes that programmers are able to write representative utility functions and assign their weights. Second, it assumes that the programmer is aware of the workload (which is needed to write the utility function). The authors suggested that future work could investigate a dynamic system that uses measurements to change the utility functions. Finally, P4All does not support multivariate and nonlinear functions. All the aforementioned limitations can be explored in the future.

[2]Note that most vendors (e.g., Barefoot Networks) provide a program (*switch.p4*) that expresses the forwarding plane of a switch, with the typical features of an advanced layer-2 and layer-3 switch. If the goal is to simply deploy a switch with no in-network applications, then the operators are not required to program the chip. They just need to install a network operating system (NOS) such as SONIC [378] or FBOSS [379]).

### I. DEEP PROGRAMMABILITY

Disaggregation is enabling network owners and operators to take control of the software running the network. It is possible to program virtual and PISA-based switches, hardware accelerators, smartNICs, and end-hosts' networking stacks. Further, acceleration techniques such as the Express Data Path (XDP) and Berkeley Packet Filter (BPF) are being used to accelerate the packet forwarding in the kernel. Additionally, acceleration techniques are used to address the performance issues of Virtual Network Functions (VNFs) running on servers [380, 381].

The malleability of programming various network components is shifting the trend towards *deep programmability*, as coined by McKeown [53, 382]. In deep programmability, the behavior is described at top and partitioned and executed across elements. The operators will focus on "software" rather than "protocols"; for example, functions like routing and congestion control will be described in programs. Software engineering principles will be routinely used to check the correctness of the network behavior (from unit testing to formal/on-the-fly verification). Fine-grained telemetry and measurements will be used to monitor and troubleshoot network performance. Stream computations will be accelerated by the network (e.g., caching, load balancing, etc.). Further, networks will run autonomously under verifiable, closed-loop control. Finally, McKeown envisioned that networks will be programmed by owners, operators, researchers, etc., while being operated by a lot fewer people than today.

There are many open challenges to realize the vision of deep programmability. Consider Fig. 28. The control plane is managing the pipeline of programmable switches, NICs, and virtual switches, which are programmed by P4 through a runtime API (e.g., P4Runtime). The challenge is how to write a clean code that can be moved around within the hardware pipeline, and can run at line rate.

**Current and Future Initiatives.** Fig. 29 shows an example of congestion control application with deep programmability. In loss-based congestion control (e.g., NewReno, CUBIC), packet drops and duplicate ACKs are used to indicate congestion. Such signal is ideally observed by the kernel of the end-host. In delay-based congestion control (e.g., TCP Vegas, TIMELY), RTT is used as the primary signal for congestion, and thus, high-precision timers must be used to get accurate estimations. This is ideally done in the NIC. Other network-assisted congestion control (e.g., HPCC) rely on the queue occupancy in the switch. Note that such mechanism modifies the packet headers, and therefore, both the NIC and the kernel should be aware of it (hence the green arrows in the figure). To be able to automate the process of partitioning functions into the network, systematic methods and algorithms should be carefully devised. There is an immense expertise in userspace and kernel space programming. However, general purpose code cannot be easily ported to the hardware since it might not fit. Hence, there is a need for methods that constrain the programming so that it will work on feedforward loop-free pipeline. McKeown discussed a strawman solution [53] to the problem where the whole pipeline (sequence of devices) is expressed in a language (e.g., P4); the pipeline specification includes the serial dependencies between the devices. The *externs* of P4 will be used to invoke general purpose C/C++ code running on the CPU. Further, P4 will be used to define the forwarding behavior of the code that will be accelerated by the hardware. Future work should explore more sophisticated methods for solving the partitioning problem, while considering the constraints of the hardware and the current networking landscape. Note that tremendous efforts from academia and the industry are being spent on the Pronto project [343, 344], which can be considered as an example of the deep programmable architecture.

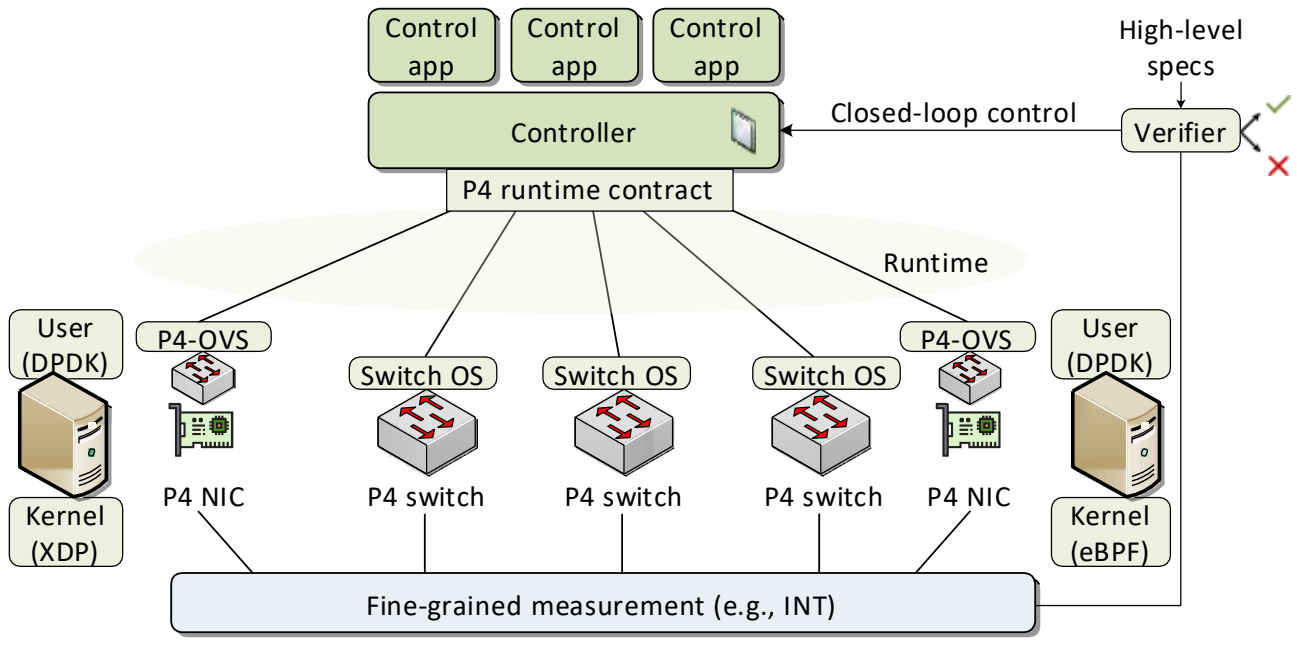

FIGURE 28: Network as a programmable platform. Large cloud or ISP example [53].

## J. MODULARITY AND VIRTUALIZATION

Programmable data plane were originally designed to execute a single program at a given time. However, there is no doubt that in today's networks, operators often require multiple network functions to run simultaneously on a single physical switch. A challenge that operators face when changing data plane programs is the connectivity loss and the service downtime/interruption [383].

Cloud providers are now aiming to offer on-switch network functions as services to a diverse set of cloud customers. Such needs introduce various challenges including resource isolation (memory and resources should be dedicated to a specific function), performance isolation (the performance of a network function must not impact other functions), and security isolation (network function must not read other function's data).

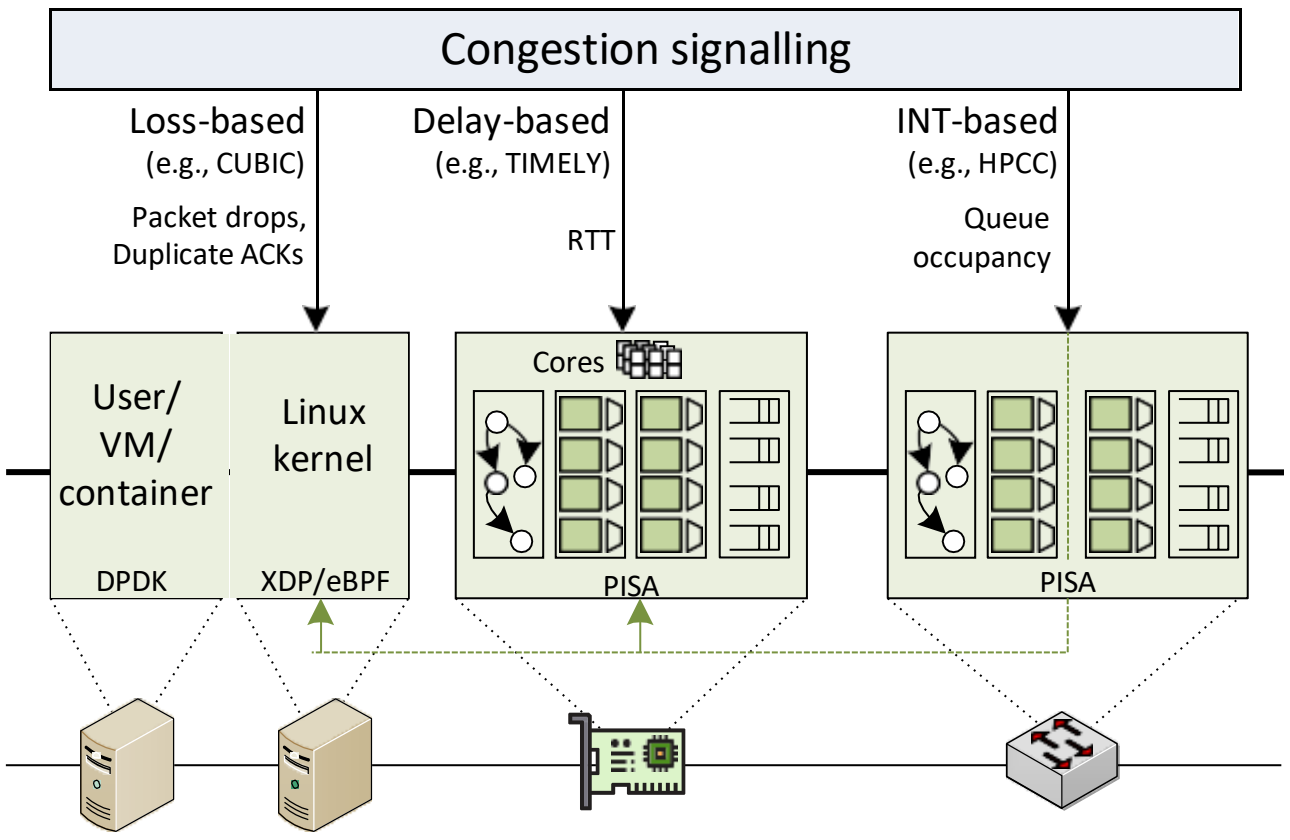

FIGURE 29: Deep programmability, congestion control example. [53].

**Current and Future Initiatives.** P4 programs and functions should become more modular so that programmers can easily integrate multiple services into the hardware pipeline. Current research efforts on data plane virtualization are being proposed in the literature. For instance, Hyper4 [384] provides a general purpose P4 program that can be dynamically configured to adopt new behavior. Essentially, P4 programs are translated into table entries and pushed to the general purpose program, enabling *hot-pluggability*. Hyper4 uses packet recirculation to implement the hot-pluggable parser, and therefore, suffers from performance degradation. Many other data plane virtualization systems have been proposed since then (e.g., HyperV [354], P4VBox [355], P4Visor [346], PRIME [356], P4click [357], MTPSA [358], etc.).

Han et al. [359] performed packet latency measurements on HyperVDP and P4Visor (processor isolation is not supported). Their results show that the overall latency is determined by the P4 program that has the highest latency. To remediate this problem, resource disaggregation methods (e.g., dRMT [367]) can be used. Other challenges that could be explored in the future include performance degradation that result from packet recirculation, lack of flexibility for live reconfiguration, frequent recompilations, loss of states during data plane reconfiguration, etc.

## K. PRACTICAL TESTING

Verifying the correctness of novel protocols and applications in real production networks is of utmost importance for engineers and researchers. Due to the ossification of production networks (cannot run untested systems), engineers typically rely on modeling and mimicking the network behavior in a smaller scale to test their proof-of-concepts. One way to model the network is through *simulations* [385]; while simulations offer flexibility in customizing the scenarios, they cannot achieve the performance of real networks since they typically run on CPUs. Another way to model the network is through *emulations* [386–388]. Emulators run the same software of production networks on CPU and offer flexibility in customization; however, they produce inaccurate measurements with high traffic rates and are bound to the CPU of the machine. Finally, emulating testbeds on a smaller scale might produce results different than production networks.

**Current and Future Initiatives.** TurboNet [372] is a noteworthy approach that leverages the power of programmable switches to emulate production networks at scale while achieving line-rate performance. TurboNet emulates both the data and control planes. Multiple switches can be emulated by slicing a single switch, separating its ports, and dividing the queue resources; this enables TurboNet to scale beyond the number of ports. TurboNet can emulate background traffic, link loss, link delay, etc. Future work in this area could consider further methods that consume less resources than TurboNet. Also, future work should avoid interrupting the emulation whenever the network emulation conditions are being changed.

P4Campus [389] is another promising work that demonstrates how researchers can test and evaluate their novel ideas on campus networks. P4Campus aims at encouraging researchers to migrate from simulation/emulation to an implementation of hardware switches. Second, it advises on replaying campus traffic and run experiments against the production data. The authors of P4Campus are working towards supporting multiple targets, program virtualization, and different topologies. Furthermore, they foresee that their testbed will be expanded to other institutions where P4Campus will be adopted. This will pave the way for more collaboration between researchers, especially since the applications of P4Campus (e.g., microbursts detection, heavy hitter, live traffic anonymization, flow RTT measurement, etc.) are already available to the public [390].

### L. HUMAN INVOLVEMENT

The complexity of managing and configuring today's networks is continuously increasing, especially when the networks are large [391]. Applications are demanding enhanced security, high availability, and high performance. Networks today are opaque and require acquiring some "dials" and configuring "knobs". This is typically done without really understanding what is happening in the network; such process and the complexity of network management inevitably increases the risk of errors (e.g., operator errors). Hence the question, "*If we are operating a large network, can we completely remove the human?*".

**Current and Future Initiatives.** Many techniques and architectures have been proposed to answer this question. In the past few years, the research community started exploring the concepts of "Self-driving networks", "Zero-touch networks", and "Knowledge-Defined Networking" [370, 392–394]. The networking industry in the upcoming years may be shifting towards the closed-loop control architecture (Fig. 30) [1].

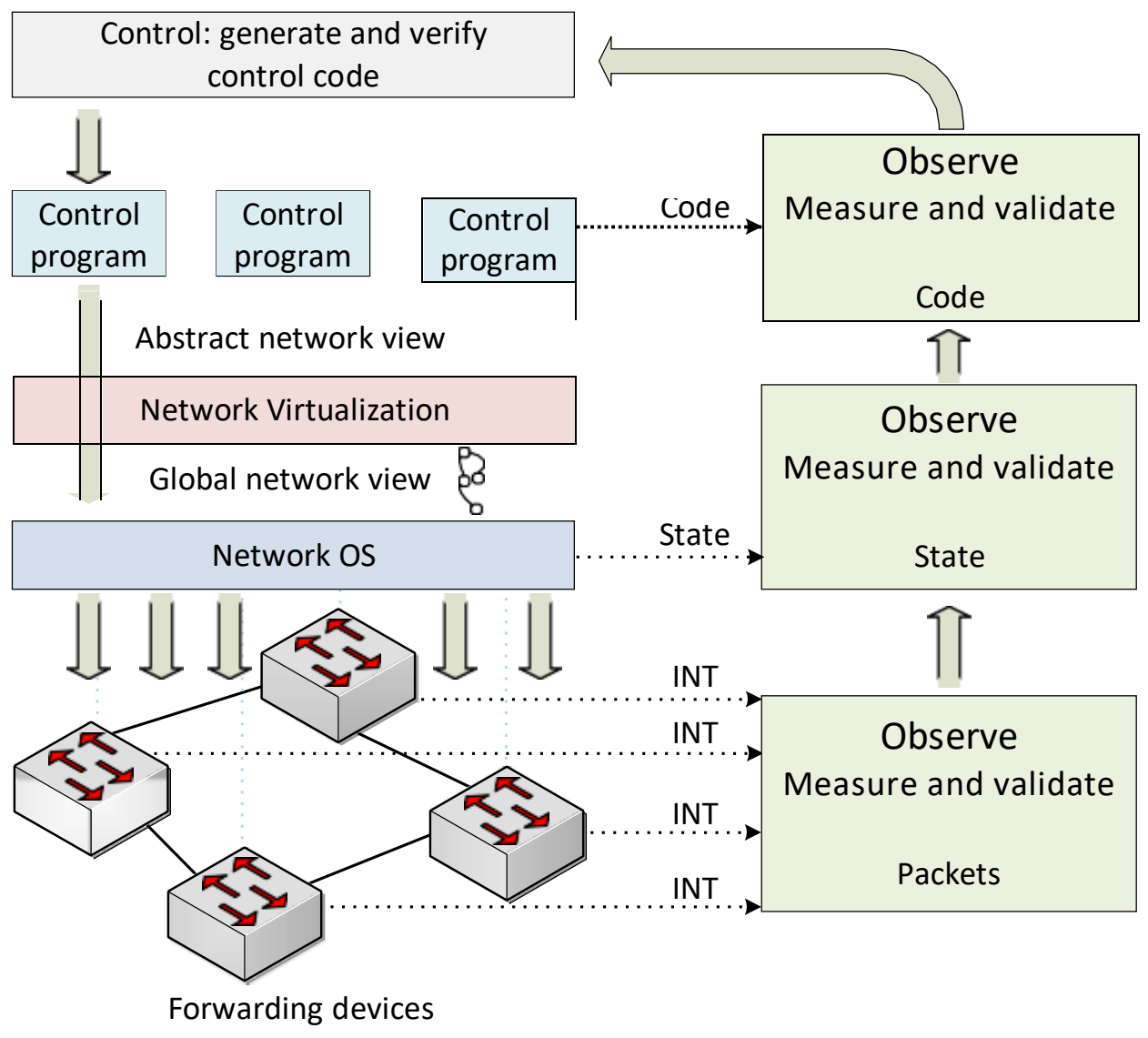

**FIGURE 30:** Closed-loop network. [1, 53, 382]. The packets sent from forwarding devices packets (e.g., through INT), the network state, and the code are being measured and validated. The feedback is used by the control plane to generate new behaviors (new control code, new forwarding code, new states), and to verify that the operation is matching the intentions.

Note that it is not easy to realize the vision of completely automating networks. There are three pieces that need to be addressed to close the loop and make networks more autonomous and intelligent.

- The ability to observe packets, network state and code, in real time (at the nanosecond scale). Observing packets has already started with packet telemetry and measurements (Sections VI, VII-B). It is possible with programmable switches to detect and visualize microbursts; this was not possible in the past. Furthermore, per-packet examination is now possible, giving better visibility into the behavior of the network.
- The ability to generate new control and forwarding behavior on-the-fly to correct errors. Techniques such as header space analysis (HSA) (Section XII) allows building a model of the forwarding behavior of every switch in the network based on the program that describes its behavior, and the state that it currently contains. This allows determining and formally proving if two devices can communicate for instance.
- The ability to verify generated code and deploy it quickly. While the first two pieces already have some progress, this third piece need further advancements. It is advised to explore software engineering techniques to generate, optimize and verify the code.

## XIV. CONCLUSIONS

This article presents an exhaustive survey on programmable data planes. The survey describes the evolution of networking by discussing the traditional control plane and the transition to SDN. Afterwards, the survey motivates the need for programming the data plane and delves into the general architecture of a programmable switch (PISA). A brief description of P4, the de-facto language for programming the data plane was presented. Motivated by the increasing trend in programming the data plane, the survey provides a taxonomy that sheds the light on numerous significant works and compares schemes within each category in the taxonomy and with those in legacy approaches. The survey concludes by discussing challenges and considerations as well as various future trends and initiatives. Evidence indicates that the closed nature of today's networks will diminish in the future, and open-source and the *deep programmability* architecture will dominate.

## ACKNOWLEDGEMENT

This material is based upon work supported by the National Science Foundation under grant numbers 1925484 and 1829698, funded by the Office of Advanced Cyberinfrastructure (OAC).

**TABLE 36:** Abbreviations used in this article.

| Abbreviation | Term |
|---|---|
| ABR | Adaptive Bit Rate |
| ACK | Acknowledgement |
| ACL | Access Control List |
| AFQ | Approximate Fair Queueing |
| AIMD | Additive Increase Multiplicative Decrease |
| ALU | Arithmetic Logical Unit |
| API | Application Programming Interface |
| AQM | Active Queue Management |
| AS | Autonomous System |
| ASIC | Application-specific Integrated Circuit |
| ATPG | Automatic Test Packet Generation |
| ATT | Attribute Protocol |
| BBR | Bottleneck Bandwidth and Round-trip Time |
| BDD | Binary Decision Diagram |
| BFT | Byzantine Fault Tolerance |
| BGP | Border Gateway Protocol |
| BIER | Bit Index Explicit Replication |
| BLE | Bluetooth Low Energy |
| BLESS | Bluetooth Low Energy Service Switch |
| BMv2 | Behavioral Model Version 2 |
| BNN | Binary Neural Network |
| BQPS | Billion Queries Per Second |
| BYOD | Bring Your Own Device |
| CAIDA | Center of Applied Internet Data Analysis |
| CC | Congestion Control |
| CNN | Convolutional Neural Network |
| CoDel | Controlled Delay |
| CPU | Central Processing Unit |
| CRC | Cyclic Redundancy Check |
| CWND | Congestion Window |
| DCQCN | Data Center Quantized Congestion Notification |
| DCTCP | Data Center Transmission Control Protocol |
| DDoS | Distributed Denial-of-Service |
| DIP | Direct Internet Protocol |
| DMA | Direct Memory Access |
| DMZ | Demilitarized Zone |
| DNS | Domain Name Server |
| DPDK | Data Plane Development Kit |
| DRAM | Dynamic Random Access Memory |
| DSP | Digital Signal Processors |
| ECMP | Equal-Cost Multi-Path Routing |
| ECN | Explicit Congestion Notification |
| ESP | Encapsulating Security Payload |
| FAST | Flow-level State Transitions |
| FCT | Flow Completion Time |
| FIB | Forwarding Information Base |
| FPGA | Field-programmable Gate Array |

| Abbreviation | Term |
|---|---|
| FQ | Fair Queueing |
| GPU | Graphics Processing Unit |
| GRE | Generic Routing Encapsulation |
| HCF | Hop-Count Filtering |
| HSA | Header Space Analysis |
| HTCP | Hamilton Transmission Control Protocol |
| HTTP | Hypertext Transfer Protocol |
| IDS | Intrusion Detection System |
| IGMP | Internet Group Management Protocol |
| IKE | Internet Key Exchange |
| ILP | Integer Linear Programming |
| INT | In-band Network Telemetry |
| IoT | Internet of Things |
| IP | Internet Protocol |
| ISP | Internet Service Provider |
| JSON | JavaScript Object Notation |
| KDN | Knowledge-defined Networking |
| KPI | Key Performance Indicator |
| INT | In-band Network Telemetry |
| IoT | Internet of Things |
| IP | Internet Protocol |
| ISP | Internet Service Provider |
| INT | In-band Network Telemetry |
| IoT | Internet of Things |
| INT | In-band Network Telemetry |
| IoT | Internet of Things |
| IP | Internet Protocol |
| ISP | Internet Service Provider |
| JSON | JavaScript Object Notation |
| KDN | Knowledge-defined Networking |
| KPI | Key Performance Indicator |
| LAN | Local Area Network |
| LFA | Link Flooding Attack |
| LPM | Longest Prefix Match |
| LPWAN | Low Power Wide Area Network |
| LTE | Long Term Evolution |
| MAC | Medium Access Control |
| MAU | Match-Action Unit |
| MCM | Multicolor Markers |
| MIMD | Multiplicative Increase Multiplicative Decrease |
| ML | Machine Learning |
| MOS | Mean Opinion Score |
| MPC | Mobile Packet Core |
| MQTT | Message Queueing Telemetry Transport |
| MSS | Maximum Segment Size |
| MPTCP | Multipath Transmission Control Protocol |
| MTU | Maximum Transmission Unit |
| NACK | Negative Acknowledgement |
| NAT | Network Address Translation |
| NDA | Non-disclosure Agreement |
| NDN | Named Data Networking |
| NFV | Network Functions Virtualization |
| NIC | Network Interface Controller |
| NN | Neural Networks |
| NSH | Network Service Header |
| ONOS | Open Network Operating System |
| OSPF | Open Shortest Path First |
| OUM | Ordered Unreliable Multicast |
| OVS | Open Virtual Switch |
| P2P | Peer-to-peer |
| PBT | Postcard-Based Telemetry |
| PCC | Performance-oriented Congestion Control |
| PCC | Per-Connection Consistency |
| PD | Program Dependent |
| PGW | Packet Data Network Gateway |
| PI | Protocol Independent |
| PIE | Proportional Integral Controller Enhanced |
| PISA | Protocol Independent Switch Architecture |
| QoE | Quality of Experience |
| QoS | Quality of Service |
| RAM | Random-Access Memory |
| RDMA | Remote Direct Memory Access |
| RED | Random Early Detection |

| Abbreviation | Term |
|---|---|
| REST | Representational State Transfer |
| RFC | Request for Comments |
| RMT | Reconfigurable Match-action Tables |
| RSS | Really Simple Syndication |
| RTT | Round-trip Time |
| RWND | Receiver Window |
| SAD | Security Association Database |
| SAT | Boolean Satisfiability Problem |
| SDN | Software Defined Networking |
| SHA | Secure Hash Algorithm |
| SIP | Session Initiation Protocol |
| SLA | Service Level Agreement |
| SNMP | Simple Network Management Protocol |
| SPD | Security Policy Database |
| SRAM | Static Random-Access Memory |
| TCAM | Ternary Content-Addressable Memory |
| TM | Traffic Management |
| ToR | The Onion Router |
| TPU | Tensor Processing Unit |
| TTL | Time-to-Live |
| UDP | User Datagram Protocol |
| UE | User Equipment |
| VIP | Virtual Internet Protocol |
| VMN | Verifying Mutable Networks |
| VN | Virtual Network |
| VoLTE | Voice over Long-term Evolution |
| VXLAN | Virtual eXtensible Local Area Network |
| WAN | Wide Area Network |
| XDP | eXpress Data Path |

## REFERENCES

[1] N. McKeown, "How we might get humans out of the way." Open Networking Foundation (ONF) Connect 19, Sep. 2019. [Online]. Available: https://tinyurl.com/y4dnxacz.

[2] RFC Editor, "Number of RFCs published per year." [Online]. Available: https://www.rfc-editor.org/rfcs-per-year/.

[3] B. Trammell and M. Kuehlewind, "Report from the IAB workshop on stack evolution in a middlebox Internet (SEMI)," RFC7663. [Online]. Available: https://tools.ietf.org/html/rfc7663.

[4] G. Papastergiou, G. Fairhurst, D. Ros, A. Brunstrom, K.-J. Grinnemo, P. Hurtig, N. Khademi, M. Tüxen, M. Welzl, D. Damjanovic, and S. Mangiante, "De-ossifying the Internet transport layer: a survey and future perspectives," IEEE Communications Surveys & Tutorials, vol. 19, no. 1, pp. 619–639, 2016.

[5] "VMware, Cisco stretch virtual LANs across the heavens." in The Register, Aug. 2011. [Online]. Available: https://tinyurl.com/y6mxhqzn.

[6] M. Mahalingam, D. Dutt, K. Duda, P. Agarwal, L. Kreeger, T. Sridhar, M. Bursell, and C. Wright, "Virtual eXtensible Local Area Network (VXLAN): a framework for overlaying virtualized layer 2 networks over layer 3 networks," RFC7348. [Online]. Available: http://www.rfc-editor.org/rfc/rfc7348.txt.

[7] M. Casado, M. J. Freedman, J. Pettit, J. Luo, N. McKeown, and S. Shenker, "Ethane: taking control of the enterprise," ACM SIGCOMM computer communication review, vol. 37, no. 4, pp. 1–12, 2007.

[8] D. Kreutz, F. M. Ramos, P. E. Verissimo, C. E. Rothenberg, S. Azodolmolky, and S. Uhlig, "Software-defined networking: a comprehensive survey," Proceedings of the IEEE, vol. 103, no. 1, pp. 14–76, 2014.

[9] P. Bosshart, D. Daly, G. Gibb, M. Izzard, N. McKeown, J. Rexford, C. Schlesinger, D. Talayco, A. Vahdat, and G. Varghese, "P4: programming protocol-independent packet processors," ACM SIGCOMM Computer Communication Review, vol. 44, no. 3, pp. 87–95, 2014.

[10] Barefoot Networks, "Use cases." [Online]. Available: https://www.barefootnetworks.com/use-cases/.

[11] A. Weissberger, "Comcast: ONF Trellis software is in production together with L2/L3 white box switches." [Online]. Available: https://tinyurl.com/y69jc7sv.

[12] N. Akiyama, M. Nishiki, "P4 and Stratum use case for new edge cloud." [Online]. Available: https://tinyurl.com/yxuoo9qv.

[13] Stordis GmbH, "New STORDIS advanced programmable switches (APS) first to unlock the full potential of P4 and next generation software defined networking (NG-SDN)." [Online]. Available: https://tinyurl.com/y3kjnypl.

[14] Open Networking Foundation (ONF), "Stratum – ONF launches major new open source SDN switching platform with support from Google." [Online]. Available: https://tinyurl.com/yy3ykw7g.

[15] P4.org Community, "P4 gains broad adoption, joins Open Networking Foundation (ONF) and Linux Foundation (LF) to accelerate next phase of growth and innovation." [Online]. Available: https://p4.org/p4/p4-joins-onf-and-lf.html.

[16] Facebook engineering, "Disaggregate: networking recap." [Online]. Available: https://tinyurl.com/yxoaj7kw.

[17] Open Compute Project (OCP), "Alibaba DC network evolution with open SONiC and programmable HW." [Online]. Available: https://www.opencompute.org/files/OCP2018.alibaba.pdf.

[18] S. Heule, "Using P4 and P4Runtime for optimal L3 routing." [Online]. Available: https://tinyurl.com/y365gnqy.

[19] N. McKeown, "SDN phase 3: getting the humans out of the way. ONF Connect 19." [Online]. Available: https://tinyurl.com/tp9bxw4.

[20] Edgecore, "Wedge 100BF-32X, 100GbE data center switch," 2020. [Online]. Available: https://tinyurl.com/sy2jkqe.

[21] STORDIS, "The new advanced programmable switches are available." [Online]. Available: https://www.stordis.com/products/.

[22] Cisco, "Cisco Nexus 34180YC and 3464C programmable switches data sheet." [Online]. Available: https://tinyurl.com/y92cbdxe.

[23] Arista, "Arista 7170 series." [Online]. Available: https://www.arista.com/en/products/7170-series.

[24] Juniper Networks, "Juniper advancing disaggregation through P4Runtime integration." [Online]. Available: https://tinyurl.com/yygz547t.

[25] Interface Masters, "Tahoe 2624." [Online]. Available: https://interfacemasters.com/products/switches/10g-40g/tahoe-2624/.

[26] Barefoot Networks, "Tofino ASIC." [Online]. Available: https://www.barefootnetworks.com/products/brief-tofino/.

[27] Xilinx, "Xilinx solutions." [Online]. Available: https://www.xilinx.com/products/silicon-devices.html.

[28] Pensando, "The Pensando distributed services platform." [Online]. Available: https://pensando.io/our-platform/.

[29] Mellanox, "Empowering the next generation of secure cloud SmartNICs." [Online]. Available: https://www.mellanox.com/products/smartnic.

[30] Innovium, "Teralynx switch silicon." [Online]. Available: https://www.innovium.com/teralynx/.

[31] I. Baldin, J. Griffioen, K. Wang, I. Monga, A. Nikolich, "Mid-Scale RI-1 (M1:IP): FABRIC: adaptive programmable research infrastructure for computer science and science applications." [Online]. Available: https://tinyurl.com/y463v9z9.

[32] FABRIC, "About FABRIC." [Online]. Available: https://fabric-testbed.net/about/.

[33] J. Mambretti, J. Chen, F. Yeh, and S. Y. Yu, "International P4 networking testbed," in SC19 Network Research Exhibition, 2019.

[34] 2STiC, "A national programmable infrastructure to experiment with next-generation networks." [Online]. Available: https://www.2stic.nl/national-programmable-infrastructure.html.

[35] H. Stubbe, "P4 compiler & interpreter: a survey," Future Internet (FI) and Innovative Internet Technologies and Mobile Communication (IITM), vol. 47, 2017.

[36] T. Dargahi, A. Caponi, M. Ambrosin, G. Bianchi, and M. Conti, "A survey on the security of stateful SDN data planes," IEEE Communications Surveys & Tutorials, vol. 19, no. 3, pp. 1701–1725, 2017.

[37] W. L. da Costa Cordeiro, J. A. Marques, and L. P. Gaspary, "Data plane programmability beyond OpenFlow: opportunities and challenges for network and service operations and management," Journal of Network and Systems Management, vol. 25, no. 4, pp. 784–818, 2017.

[38] A. Satapathy, "Comprehensive study of P4 programming language and software-defined networks," 2018. [Online]. Available: https://tinyurl.com/y4d4zma9.

[39] R. Bifulco and G. Rétvári, "A survey on the programmable data plane: abstractions, architectures, and open problems," in 2018 IEEE 19th International Conference on High Performance Switching and Routing (HPSR), pp. 1–7, IEEE, 2018.

[40] E. Kaljic, A. Maric, P. Njemcevic, and M. Hadzialic, "A survey on data plane flexibility and programmability in software-defined

networking," IEEE Access, vol. 7, pp. 47804–47840, 2019.
[41] P. G. Kannan and M. C. Chan, "On programmable networking evolution," CSI Transactions on ICT, vol. 8, no. 1, pp. 69–76, 2020.
[42] L. Tan, W. Su, W. Zhang, J. Lv, Z. Zhang, J. Miao, X. Liu, and N. Li, "In-band network telemetry: A survey," Computer Networks, p. 107763, 2020.
[43] X. Zhang, L. Cui, K. Wei, F. P. Tso, Y. Ji, and W. Jia, "A survey on stateful data plane in software defined networks," Computer Networks, p. 107597, 2020.
[44] G. Bianchi, M. Bonola, A. Capone, and C. Cascone, "OpenState: programming platform-independent stateful OpenFlow applications inside the switch," ACM SIGCOMM Computer Communication Review, vol. 44, no. 2, pp. 44–51, 2014.
[45] M. Moshref, A. Bhargava, A. Gupta, M. Yu, and R. Govindan, "Flow-level state transition as a new switch primitive for SDN," in Proceedings of the third workshop on Hot topics in software defined networking, pp. 61–66, 2014.
[46] P4 Language Consortium, "P4Runtime." [Online]. Available: https://github.com/p4lang/PI/.
[47] Y. Rekhter, T. Li, and S. Hares, "A border gateway protocol 4 (bgp-4)," RFC4271. http://www.rfc-editor.org/rfc/rfc4271.txt.
[48] N. McKeown, T. Anderson, H. Balakrishnan, G. Parulkar, L. Peterson, J. Rexford, S. Shenker, and J. Turner, "Openflow: enabling innovation in campus networks," ACM SIGCOMM Computer Communication Review, vol. 38, no. 2, pp. 69–74, 2008.
[49] N. McKeown, "Why does the Internet need a programmable forwarding plane." [Online]. Available: https://tinyurl.com/y6x7qqpm.
[50] C. Kim, "Evolution of networking, Networking Field Day 21, 2:01," 2019. [Online]. Available: https://tinyurl.com/y9fkj7qx.
[51] A. Shapiro, "P4-programming data plane use-cases." in P4 Expert Roundtable Series, April 28-29, 2020. [Online]. Available: https://tinyurl.com/y5n4k83h.
[52] P. Bosshart, G. Gibb, H.-S. Kim, G. Varghese, N. McKeown, M. Izzard, F. Mujica, and M. Horowitz, "Forwarding metamorphosis: fast programmable match-action processing in hardware for SDN," ACM SIGCOMM Computer Communication Review, vol. 43, no. 4, pp. 99–110, 2013.
[53] Nick McKeown, "Creating an End-to-End Programming Model for Packet Forwarding." [Online]. Available: https://www.youtube.com/watch?v=fiBuao6YZl0&t=4216s.
[54] Z. Liu, J. Bi, Y. Zhou, Y. Wang, and Y. Lin, "Netvision: towards network telemetry as a service," in 2018 IEEE 26th International Conference on Network Protocols (ICNP), pp. 247–248, IEEE, 2018.
[55] J. Hyun, N. Van Tu, and J. W.-K. Hong, "Towards knowledge-defined networking using in-band network telemetry," in NOMS 2018-2018 IEEE/IFIP Network Operations and Management Symposium, pp. 1–7, IEEE, 2018.
[56] Y. Kim, D. Suh, and S. Pack, "Selective in-band network telemetry for overhead reduction," in 2018 IEEE 7th International Conference on Cloud Networking (CloudNet), pp. 1–3, IEEE, 2018.
[57] T. Pan, E. Song, Z. Bian, X. Lin, X. Peng, J. Zhang, T. Huang, B. Liu, and Y. Liu, "Int-path: Towards optimal path planning for in-band network-wide telemetry," in IEEE INFOCOM 2019-IEEE Conference on Computer Communications, pp. 487–495, IEEE, 2019.
[58] J. A. Marques, M. C. Luizelli, R. I. T. da Costa Filho, and L. P. Gaspary, "An optimization-based approach for efficient network monitoring using in-band network telemetry," Journal of Internet Services and Applications, vol. 10, no. 1, p. 12, 2019.
[59] B. Niu, J. Kong, S. Tang, Y. Li, and Z. Zhu, "Visualize your IP-over-optical network in realtime: a P4-based flexible multilayer in-band network telemetry (ML-INT) system," IEEE Access, vol. 7, pp. 82413–82423, 2019.
[60] A. Karaagac, E. De Poorter, and J. Hoebeke, "In-band network telemetry in industrial wireless sensor networks," IEEE Transactions on Network and Service Management, vol. 17, no. 1, pp. 517–531, 2019.
[61] R. Ben Basat, S. Ramanathan, Y. Li, G. Antichi, M. Yu, and M. Mitzenmacher, "PINT: probabilistic in-band network telemetry," in Proceedings of the Annual conference of the ACM Special Interest Group on Data Communication on the applications, technologies, architectures, and protocols for computer communication, pp. 662–680, 2020.
[62] Y. Lin, Y. Zhou, Z. Liu, K. Liu, Y. Wang, M. Xu, J. Bi, Y. Liu, and J. , "Netview: Towards on-demand network-wide telemetry in the data center," Computer Networks, vol. 180, p. 107386, 2020.
[63] N. Van Tu, J. Hyun, and J. W.-K. Hong, "Towards ONOS-based SDN monitoring using in-band network telemetry," in 2017 19th Asia-Pacific Network Operations and Management Symposium (APNOMS), pp. 76–81, IEEE, 2017.
[64] Serkant, "Prometheus INT exporter." [Online]. Available: https://github.com/serkantul/prometheus_int_exporter/.
[65] N. Van Tu, J. Hyun, G. Y. Kim, J.-H. Yoo, and J. W.-K. Hong, "IntCollector: a high-performance collector for in-band network telemetry," in 2018 14th International Conference on Network and Service Management (CNSM), pp. 10–18, IEEE, 2018.
[66] Barefoot Networks, "Barefoot Deep Insight - product brief." [Online]. Available: https://tinyurl.com/u2ncvry.
[67] Broadcom, "BroadView Analytics, Trident 3 in-band telemetry." [Online]. Available: https://tinyurl.com/yxr2qydb.
[68] M. Handley, C. Raiciu, A. Agache, A. Voinescu, A. W. Moore, G. Antichi, and M. Wójcik, "Re-architecting datacenter networks and stacks for low latency and high performance," in Proceedings of the Conference of the ACM Special Interest Group on Data Communication, pp. 29–42, 2017.
[69] A. Feldmann, B. Chandrasekaran, S. Fathalli, and E. N. Weyulu, "P4-enabled network-assisted congestion feedback: a case for NACKs," 2019.
[70] Y. Li, R. Miao, H. H. Liu, Y. Zhuang, F. Feng, L. Tang, Z. Cao, M. Zhang, F. Kelly, and M. Y. Alizadeh, Mohammad, "HPCC: high precision congestion control," in Proceedings of the ACM Special Interest Group on Data Communication, pp. 44–58, 2019.
[71] E. F. Kfoury, J. Crichigno, E. Bou-Harb, D. Khoury, and G. Srivastava, "Enabling TCP pacing using programmable data plane switches," in 2019 42nd International Conference on Telecommunications and Signal Processing (TSP), pp. 273–277, IEEE, 2019.
[72] S. Shahzad, E.-S. Jung, J. Chung, and R. Kettimuthu, "Enhanced explicit congestion notification (eecn) in tcp with p4 programming," in 2020 International Conference on Green and Human Information Technology (ICGHIT), pp. 35–40, IEEE, 2020.
[73] B. Turkovic, F. Kuipers, N. van Adrichem, and K. Langendoen, "Fast network congestion detection and avoidance using P4," in Proceedings of the 2018 Workshop on Networking for Emerging Applications and Technologies, pp. 45–51, 2018.
[74] B. Turkovic and F. Kuipers, "P4air: Increasing fairness among competing congestion control algorithms," 2020.
[75] M. Apostolaki, L. Vanbever, and M. Ghobadi, "Fab: Toward flow-aware buffer sharing on programmable switches," in Proceedings of the 2019 Workshop on Buffer Sizing, pp. 1–6, 2019.
[76] J. Geng, J. Yan, and Y. Zhang, "P4qcn: Congestion control using p4-capable device in data center networks," Electronics, vol. 8, no. 3, p. 280, 2019.
[77] Y. Li, R. Miao, C. Kim, and M. Yu, "Flowradar: A better NetFlow for data centers," in 13th USENIX Symposium on Networked Systems Design and Implementation (NSDI), pp. 311–324, 2016.
[78] Z. Liu, A. Manousis, G. Vorsanger, V. Sekar, and V. Braverman, "One sketch to rule them all: rethinking network flow monitoring with UnivMon," in Proceedings of the 2016 ACM SIGCOMM Conference, pp. 101–114, 2016.
[79] S. Narayana, A. Sivaraman, V. Nathan, P. Goyal, V. Arun, M. Alizadeh, V. Jeyakumar, and C. Kim, "Language-directed hardware design for network performance monitoring," in Proceedings of the Conference of the ACM Special Interest Group on Data Communication, pp. 85–98, 2017.
[80] M. Ghasemi, T. Benson, and J. Rexford, "Dapper: data plane performance diagnosis of TCP," in Proceedings of the Symposium on SDN Research, pp. 61–74, 2017.
[81] T. Yang, J. Jiang, P. Liu, Q. Huang, J. Gong, Y. Zhou, R. Miao, X. Li, and S. Uhlig, "Elastic sketch: adaptive and fast network-wide measurements," in Proceedings of the 2018 Conference of the ACM Special Interest Group on Data Communication, pp. 561–575, 2018.
[82] N. Yaseen, J. Sonchack, and V. Liu, "Synchronized network snapshots," in Proceedings of the 2018 Conference of the ACM Special Interest Group on Data Communication, pp. 402–416, 2018.
[83] R. Joshi, T. Qu, M. C. Chan, B. Leong, and B. T. Loo, "Burstradar: practical real-time microburst monitoring for datacenter networks," in Proceedings of the 9th Asia-Pacific Workshop on Systems, pp. 1–8, 2018.
[84] M. Lee and J. Rexford, "Detecting violations of service-level agree-

ments in programmable switches," 2018. [Online]. Available: https://p4campus.cs.princeton.edu/pubs/mackl_thesis_paper.pdf.

[85] J. Sonchack, O. Michel, A. J. Aviv, E. Keller, and J. M. Smith, "Scaling hardware accelerated network monitoring to concurrent and dynamic queries with* flow," in 2018 USENIX Annual Technical Conference (USENIX ATC 18), pp. 823–835, 2018.

[86] J. Sonchack, A. J. Aviv, E. Keller, and J. M. Smith, "Turboflow: Information rich flow record generation on commodity switches," in Proceedings of the Thirteenth EuroSys Conference, pp. 1–16, 2018.

[87] A. Gupta, R. Harrison, M. Canini, N. Feamster, J. Rexford, and W. Willinger, "Sonata: query-driven streaming network telemetry," in Proceedings of the 2018 Conference of the ACM Special Interest Group on Data Communication, pp. 357–371, 2018.

[88] X. Chen, S. L. Feibish, Y. Koral, J. Rexford, O. Rottenstreich, S. A. Monetti, and T.-Y. Wang, "Fine-grained queue measurement in the data plane," in Proceedings of the 15th International Conference on Emerging Networking Experiments And Technologies, pp. 15–29, 2019.

[89] Z. Liu, S. Zhou, O. Rottenstreich, V. Braverman, and J. Rexford, "Memory-efficient performance monitoring on programmable switches with lean algorithms," in Symposium on Algorithmic Principles of Computer Systems (APoCS), 2020.

[90] T. Holterbach, E. C. Molero, M. Apostolaki, A. Dainotti, S. Vissicchio, and L. Vanbever, "Blink: fast connectivity recovery entirely in the data plane," in 16th USENIX Symposium on Networked Systems Design and Implementation (NSDI 19), pp. 161–176, 2019.

[91] D. Ding, M. Savi, and D. Siracusa, "Estimating logarithmic and exponential functions to track network traffic entropy in P4," in IEEE/IFIP Network Operations and Management Symposium (NOMS), 2019.

[92] W. Wang, P. Tammana, A. Chen, and T. E. Ng, "Grasp the root causes in the data plane: diagnosing latency problems with SpiderMon," in Proceedings of the Symposium on SDN Research, pp. 55–61, 2020.

[93] R. Teixeira, R. Harrison, A. Gupta, and J. Rexford, "PacketScope: monitoring the packet lifecycle inside a switch," in Proceedings of the Symposium on SDN Research, pp. 76–82, 2020.

[94] J. Bai, M. Zhang, G. Li, C. Liu, M. Xu, and H. Hu, "FastFE: accelerating ML-based traffic analysis with programmable switches," in Proceedings of the Workshop on Secure Programmable Network Infrastructure, SPIN '20, p. 1–7, Association for Computing Machinery, 2020.

[95] X. Chen, H. Kim, J. M. Aman, W. Chang, M. Lee, and J. Rexford, "Measuring TCP round-trip time in the data plane," in Proceedings of the Workshop on Secure Programmable Network Infrastructure, pp. 35–41, 2020.

[96] Y. Qiu, K.-F. Hsu, J. Xing, and A. Chen, "A feasibility study on time-aware monitoring with commodity switches," in Proceedings of the Workshop on Secure Programmable Network Infrastructure, pp. 22–27, 2020.

[97] Q. Huang, H. Sun, P. P. Lee, W. Bai, F. Zhu, and Y. Bao, "OmniMon: re-architecting Network telemetry with resource efficiency and full accuracy," in Proceedings of the Annual conference of the ACM Special Interest Group on Data Communication on the applications, technologies, architectures, and protocols for computer communication, pp. 404–421, 2020.

[98] X. Chen, S. Landau-Feibish, M. Braverman, and J. Rexford, "BeauCoup: answering many network traffic queries, one memory update at a time," in Proceedings of the Annual conference of the ACM Special Interest Group on Data Communication on the applications, technologies, architectures, and protocols for computer communication, pp. 226–239, 2020.

[99] R. Kundel, J. Blendin, T. Viernickel, B. Koldehofe, and R. Steinmetz, "P4-CoDel: active queue management in programmable data planes," in 2018 IEEE Conference on Network Function Virtualization and Software Defined Networks (NFV-SDN), pp. 1–4, IEEE, 2018.

[100] F. Schwarzkopf, S. Veith, and M. Menth, "Performance analysis of CoDel and PIE for saturated TCP sources," in 2016 28th International Teletraffic Congress (ITC 28), vol. 1, pp. 175–183, IEEE, 2016.

[101] N. K. Sharma, M. Liu, K. Atreya, and A. Krishnamurthy, "Approximating fair queueing on reconfigurable switches," in 15th USENIX Symposium on Networked Systems Design and Implementation (NSDI), pp. 1–16, 2018.

[102] S. Laki, P. Vörös, and F. Fejes, "Towards an AQM evaluation testbed with P4 and DPDK," in Proceedings of the ACM SIGCOMM 2019 Conference Posters and Demos, pp. 148–150, 2019.

[103] C. Papagianni and K. De Schepper, "PI2 for P4: an active queue management scheme for programmable data planes," in Proceedings of the 15th International Conference on emerging Networking EXperiments and Technologies, pp. 84–86, 2019.

[104] I. Kunze, M. Gunz, D. Saam, K. Wehrle, and J. Rüth, "Tofino+ p4: A strong compound for aqm on high-speed networks?," 2021.

[105] L. Toresson, "Making a packet-value based aqm on a programmable switch for resource-sharing and low latency," 2021.

[106] A. Mushtaq, R. Mittal, J. McCauley, M. Alizadeh, S. Ratnasamy, and S. Shenker, "Datacenter congestion control: identifying what is essential and making it practical," ACM SIGCOMM Computer Communication Review, vol. 49, no. 3, pp. 32–38, 2019.

[107] M. Menth, H. Mostafaei, D. Merling, and M. Häberle, "Implementation and evaluation of activity-based congestion management using p4 (p4-abc)," Future Internet, vol. 11, no. 7, p. 159, 2019.

[108] A. G. Alcoz, A. Dietmüller, and L. Vanbever, "Sp-pifo: approximating push-in first-out behaviors using strict-priority queues," in 17th USENIX Symposium on Networked Systems Design and Implementation (NSDI 20), pp. 59–76, 2020.

[109] K. Kumazoe and M. Tsuru, "P4-based implementation and evaluation of adaptive early packet discarding scheme," in International Conference on Intelligent Networking and Collaborative Systems, pp. 460–469, Springer, 2020.

[110] D. Bhat, J. Anderson, P. Ruth, M. Zink, and K. Keahey, "Application-based QoE support with P4 and OpenFlow," in IEEE INFOCOM 2019-IEEE Conference on Computer Communications Workshops (INFOCOM WKSHPS), pp. 817–823, IEEE, 2019.

[111] Y.-W. Chen, L.-H. Yen, W.-C. Wang, C.-A. Chuang, Y.-S. Liu, and C.-C. Tseng, "P4-Enabled bandwidth management," in 2019 20th Asia-Pacific Network Operations and Management Symposium (AP-NOMS), pp. 1–5, IEEE, 2019.

[112] C. Chen, H.-C. Fang, and M. S. Iqbal, "Qostcp: Provide consistent rate guarantees to tcp flows in software defined networks," in ICC 2020-2020 IEEE International Conference on Communications (ICC), pp. 1–6, IEEE, 2020.

[113] K. Tokmakov, M. Sarker, J. Domaschka, and S. Wesner, "A case for data centre traffic management on software programmable ethernet switches," in 2019 IEEE 8th International Conference on Cloud Networking (CloudNet), pp. 1–6, IEEE, 2019.

[114] S. S. Lee and K.-Y. Chan, "A traffic meter based on a multicolor marker for bandwidth guarantee and priority differentiation in sdn virtual networks," IEEE Transactions on Network and Service Management, vol. 16, no. 3, pp. 1046–1058, 2019.

[115] M. Shahbaz, L. Suresh, J. Rexford, N. Feamster, O. Rottenstreich, and M. Hira, "Elmo: Source routed multicast for public clouds," in Proceedings of the ACM Special Interest Group on Data Communication, pp. 458–471, 2019.

[116] M. Kadosh, Y. Piasetzky, B. Gafni, L. Suresh, M. Shahbaz, S. Banerjee, "Realizing source routed multicast using Mellanox's programmable hardware switches, P4 Expert Roundtable Series, Apr. 2020." [Online]. Available: https://tinyurl.com/y8dfcsum.

[117] W. Braun, J. Hartmann, and M. Menth, "Scalable and reliable software-defined multicast with BIER and P4," in 2017 IFIP/IEEE Symposium on Integrated Network and Service Management (IM), pp. 905–906, IEEE, 2017.

[118] N. Katta, M. Hira, C. Kim, A. Sivaraman, and J. Rexford, "Hula: scalable load balancing using programmable data planes," in Proceedings of the Symposium on SDN Research, pp. 1–12, 2016.

[119] C. H. Benet, A. J. Kassler, T. Benson, and G. Pongracz, "MP-HULA: multipath transport aware load balancing using programmable data planes," in Proceedings of the 2018 Morning Workshop on In-Network Computing, pp. 7–13, 2018.

[120] R. Miao, H. Zeng, C. Kim, J. Lee, and M. Yu, "SilkRoad: making stateful layer-4 load balancing fast and cheap using switching ASICs," in Proceedings of the Conference of the ACM Special Interest Group on Data Communication, pp. 15–28, 2017.

[121] Z. Liu, Z. Bai, Z. Liu, X. Li, C. Kim, V. Braverman, X. Jin, and I. Stoica, "Distcache: provable load balancing for large-scale storage systems with distributed caching," in 17th USENIX Conference on File and Storage Technologies (FAST 19), pp. 143–157, 2019.

[122] K.-F. Hsu, P. Tammana, R. Beckett, A. Chen, J. Rexford, and D. Walker, "Adaptive weighted traffic splitting in programmable data planes," in Proceedings of the Symposium on SDN Research, pp. 103–109, 2020.

[123] K.-F. Hsu, R. Beckett, A. Chen, J. Rexford, and D. Walker, "Contra: A programmable system for performance-aware routing," in 17th USENIX Symposium on Networked Systems Design and Implementation (NSDI 20), pp. 701–721, 2020.
[124] V. Olteanu, A. Agache, A. Voinescu, and C. Raiciu, "Stateless datacenter load-balancing with beamer," in 15th USENIX Symposium on Networked Systems Design and Implementation (NSDI), pp. 125–139, 2018.
[125] B. Pit-Claudel, Y. Desmouceaux, P. Pfister, M. Townsley, and T. Clausen, "Stateless load-aware load balancing in p4," in 2018 IEEE 26th International Conference on Network Protocols (ICNP), pp. 418–423, IEEE, 2018.
[126] J.-L. Ye, C. Chen, and Y. H. Chu, "A weighted ecmp load balancing scheme for data centers using p4 switches," in 2018 IEEE 7th International Conference on Cloud Networking (CloudNet), pp. 1–4, IEEE, 2018.
[127] X. Jin, X. Li, H. Zhang, R. Soulé, J. Lee, N. Foster, C. Kim, and I. Stoica, "Netcache: balancing key-value stores with fast in-network caching," in Proceedings of the 26th Symposium on Operating Systems Principles, pp. 121–136, 2017.
[128] M. Liu, L. Luo, J. Nelson, L. Ceze, A. Krishnamurthy, and K. Atreya, "Incbricks: toward in-network computation with an in-network cache," in Proceedings of the Twenty-Second International Conference on Architectural Support for Programming Languages and Operating Systems, pp. 795–809, 2017.
[129] E. Cidon, S. Choi, S. Katti, and N. McKeown, "AppSwitch: application-layer load balancing within a software switch," in Proceedings of the First Asia-Pacific Workshop on Networking, pp. 64–70, 2017.
[130] Q. Wang, Y. Lu, E. Xu, J. Li, Y. Chen, and J. Shu, "Concordia: Distributed shared memory with in-network cache coherence," in 19th USENIX Conference on File and Storage Technologies ( FAST 21), pp. 277–292, 2021.
[131] J. Li, J. Nelson, E. Michael, X. Jin, and D. R. Ports, "Pegasus: Tolerating skewed workloads in distributed storage with in-network coherence directories," in 14th USENIX Symposium on Operating Systems Design and Implementation (OSDI 20), pp. 387–406, 2020.
[132] S. Signorello, R. State, J. François, and O. Festor, "NDN.p4: programming information-centric data-planes," in 2016 IEEE NetSoft Conference and Workshops (NetSoft), pp. 384–389, IEEE, 2016.
[133] G. Grigoryan and Y. Liu, "PFCA: a programmable FIB caching architecture," in Proceedings of the 2018 Symposium on Architectures for Networking and Communications Systems, pp. 97–103, 2018.
[134] C. Zhang, J. Bi, Y. Zhou, K. Zhang, and Z. Ma, "B-cache: a behavior-level caching framework for the programmable data plane," in 2018 IEEE Symposium on Computers and Communications (ISCC), pp. 00084–00090, IEEE, 2018.
[135] J. Vestin, A. Kassler, and J. Åkerberg, "FastReact: in-network control and caching for industrial control networks using programmable data planes," in 2018 IEEE 23rd International Conference on Emerging Technologies and Factory Automation (ETFA), vol. 1, pp. 219–226, IEEE, 2018.
[136] J. Woodruff, M. Ramanujam, and N. Zilberman, "P4DNS: in-network DNS," in 2019 ACM/IEEE Symposium on Architectures for Networking and Communications Systems (ANCS), pp. 1–6, IEEE, 2019.
[137] R. Ricart-Sanchez, P. Malagon, P. Salva-Garcia, E. C. Perez, Q. Wang, and J. M. A. Calero, "Towards an FPGA-accelerated programmable data path for edge-to-core communications in 5G networks," Journal of Network and Computer Applications, vol. 124, pp. 80–93, 2018.
[138] R. Ricart-Sanchez, P. Malagon, J. M. Alcaraz-Calero, and Q. Wang, "Hardware-accelerated firewall for 5G mobile networks," in 2018 IEEE 26th International Conference on Network Protocols (ICNP), pp. 446–447, IEEE, 2018.
[139] R. Shah, V. Kumar, M. Vutukuru, and P. Kulkarni, "TurboEPC: leveraging dataplane programmability to acccelerate the mobile packet core," in Proceedings of the Symposium on SDN Research, pp. 83–95, 2020.
[140] S. K. Singh, C. E. Rothenberg, G. Patra, and G. Pongracz, "Offloading virtual evolved packet gateway user plane functions to a programmable ASIC," in Proceedings of the 1st ACM CoNEXT Workshop on Emerging in-Network Computing Paradigms, pp. 9–14, 2019.
[141] P. Vörös, G. Pongrácz, and S. Laki, "Towards a hybrid next generation nodeb," in Proceedings of the 3rd P4 Workshop in Europe, pp. 56–58, 2020.
[142] P. Palagummi and K. M. Sivalingam, "SMARTHO: a network initiated handover in NG-RAN using P4-based switches," in 2018 14th International Conference on Network and Service Management (CNSM), pp. 338–342, IEEE, 2018.
[143] F. Paolucci, F. Cugini, P. Castoldi, and T. Osinski, "Enhancing 5g sdn/nfv edge with p4 data plane programmability," IEEE Network, 2021.
[144] Y. Lin, C. Tseng, and M. Wang, "Effects of transport network slicing on 5g applications. future internet 2021, 13, 69," 2021.
[145] E. Kfoury, J. Crichigno, and E. Bou-Harb, "Offloading media traffic to programmable data plane switches," in ICC 2020 IEEE International Conference on Communications (ICC), IEEE, 2020.
[146] B.-M. Andrus, S. A. Sasu, T. Szyrkowiec, A. Autenrieth, M. Chamania, J. K. Fischer, and S. Rasp, "Zero-touch provisioning of distributed video analytics in a software-defined metro-haul network with p4 processing," in 2019 Optical Fiber Communications Conference and Exhibition (OFC), pp. 1–3, IEEE, 2019.
[147] T. Jepsen, M. Moshref, A. Carzaniga, N. Foster, and R. Soulé, "Packet subscriptions for programmable ASICs," in Proceedings of the 17th ACM Workshop on Hot Topics in Networks, pp. 176–183, 2018.
[148] C. Wernecke, H. Parzyjegla, G. Mühl, P. Danielis, and D. Timmermann, "Realizing content-based publish/subscribe with P4," in 2018 IEEE Conference on Network Function Virtualization and Software Defined Networks (NFV-SDN), pp. 1–7, IEEE, 2018.
[149] C. Wernecke, H. Parzyjegla, G. Mühl, E. Schweissguth, and D. Timmermann, "Flexible notification forwarding for content-based publish/subscribe using P4," in 2019 IEEE Conference on Network Function Virtualization and Software Defined Networks (NFV-SDN), pp. 1–5, IEEE, 2019.
[150] R. Kundel, C. Gärtner, M. Luthra, S. Bhowmik, and B. Koldehofe, "Flexible content-based publish/subscribe over programmable data planes," in NOMS 2020-2020 IEEE/IFIP Network Operations and Management Symposium, pp. 1–5, IEEE, 2020.
[151] R. Miguel, S. Signorello, and F. M. Ramos, "Named data networking with programmable switches," in 2018 IEEE 26th International Conference on Network Protocols (ICNP), pp. 400–405, IEEE, 2018.
[152] O. Karrakchou, N. Samaan, and A. Karmouch, "Endn: An enhanced ndn architecture with a p4-programmabie data plane," in Proceedings of the 7th ACM Conference on Information-Centric Networking, pp. 1–11, 2020.
[153] J. Li, E. Michael, N. K. Sharma, A. Szekeres, and D. R. Ports, "Just say NO to paxos overhead: replacing consensus with network ordering," in 12th USENIX Symposium on Operating Systems Design and Implementation (OSDI), pp. 467–483, 2016.
[154] H. T. Dang, M. Canini, F. Pedone, and R. Soulé, "Paxos made switch-y," ACM SIGCOMM Computer Communication Review, vol. 46, no. 2, pp. 18–24, 2016.
[155] J. Li, E. Michael, and D. R. Ports, "Eris: coordination-free consistent transactions using in-network concurrency control," in Proceedings of the 26th Symposium on Operating Systems Principles, pp. 104–120, 2017.
[156] B. Han, V. Gopalakrishnan, M. Platania, Z.-L. Zhang, and Y. Zhang, "Network-assisted raft consensus protocol," Feb. 13 2020. US Patent App. 16/101,751.
[157] X. Jin, X. Li, H. Zhang, N. Foster, J. Lee, R. Soulé, C. Kim, and I. Stoica, "Netchain: scale-free sub-rtt coordination," in 15th USENIX Symposium on Networked Systems Design and Implementation (NSDI18), pp. 35–49, 2018.
[158] H. T. Dang, P. Bressana, H. Wang, K. S. Lee, N. Zilberman, H. Weatherspoon, M. Canini, F. Pedone, and R. Soulé, "Partitioned Paxos via the network data plane," arXiv preprint arXiv:1901.08806, 2019.
[159] E. Sakic, N. Deric, E. Goshi, and W. Kellerer, "P4BFT: hardware-accelerated byzantine-resilient network control plane," arXiv preprint arXiv:1905.04064, 2019.
[160] H. T. Dang, P. Bressana, H. Wang, K. S. Lee, N. Zilberman, H. Weatherspoon, M. Canini, F. Pedone, and R. Soulé, "P4xos: Consensus as a network service," IEEE/ACM Transactions on Networking, 2020.
[161] A. Sapio, I. Abdelaziz, A. Aldilaijan, M. Canini, and P. Kalnis, "In-network computation is a dumb idea whose time has come," in Proceedings of the 16th ACM Workshop on Hot Topics in Networks,

pp. 150–156, 2017.
[162] F. Yang, Z. Wang, X. Ma, G. Yuan, and X. An, “SwitchAgg: a further step towards in-network computation,” arXiv preprint arXiv:1904.04024, 2019.
[163] A. Sapio, M. Canini, C.-Y. Ho, J. Nelson, P. Kalnis, C. Kim, A. Krishnamurthy, M. Moshref, D. R. Ports, and P. Richtárik, “Scaling distributed machine learning with in-network aggregation,” arXiv preprint arXiv:1903.06701, 2019.
[164] G. Siracusano and R. Bifulco, “In-network neural networks,” arXiv preprint arXiv:1801.05731, 2018.
[165] D. Sanvito, G. Siracusano, and R. Bifulco, “Can the network be the AI accelerator?,” in Proceedings of the 2018 Morning Workshop on In-Network Computing, pp. 20–25, 2018.
[166] Z. Xiong and N. Zilberman, “Do switches dream of machine learning? toward in-network classification,” in Proceedings of the 18th ACM Workshop on Hot Topics in Networks, pp. 25–33, 2019.
[167] T. Jepsen, M. Moshref, A. Carzaniga, N. Foster, and R. Soulé, “Life in the fast lane: a line-rate linear road,” in Proceedings of the Symposium on SDN Research, pp. 1–7, 2018.
[168] T. Kohler, R. Mayer, F. Dürr, M. Maaß, S. Bhowmik, and K. Rothermel, “P4CEP: towards in-network complex event processing,” in Proceedings of the 2018 Morning Workshop on In-Network Computing, pp. 33–38, 2018.
[169] L. Chen, G. Chen, J. Lingys, and K. Chen, “Programmable switch as a parallel computing device,” arXiv preprint arXiv:1803.01491, 2018.
[170] T. Jepsen, D. Alvarez, N. Foster, C. Kim, J. Lee, M. Moshref, and R. Soulé, “Fast string searching on PISA,” in Proceedings of the 2019 ACM Symposium on SDN Research, pp. 21–28, 2019.
[171] Y. Qiao, X. Kong, M. Zhang, Y. Zhou, M. Xu, and J. Bi, “Towards in-network acceleration of erasure coding,” in Proceedings of the Symposium on SDN Research, pp. 41–47, 2020.
[172] Z. Yu, Y. Zhang, V. Braverman, M. Chowdhury, and X. Jin, “NetLock: fast, centralized lock management using programmable switches,” in Proceedings of the Annual conference of the ACM Special Interest Group on Data Communication on the applications, technologies, architectures, and protocols for computer communication, pp. 126–138, 2020.
[173] M. Tirmazi, R. Ben Basat, J. Gao, and M. Yu, “Cheetah: Accelerating database queries with switch pruning,” in Proceedings of the 2020 ACM SIGMOD International Conference on Management of Data, pp. 2407–2422, 2020.
[174] S. Vaucher, N. Yazdani, P. Felber, D. E. Lucani, and V. Schiavoni, “Zipline: in-network compression at line speed,” in Proceedings of the 16th International Conference on emerging Networking EXperiments and Technologies, pp. 399–405, 2020.
[175] R. Glebke, J. Krude, I. Kunze, J. Rüth, F. Senger, and K. Wehrle, “Towards executing computer vision functionality on programmable network devices,” in Proceedings of the 1st ACM CoNEXT Workshop on Emerging in-Network Computing Paradigms, pp. 15–20, 2019.
[176] S.-Y. Wang, C.-M. Wu, Y.-B. Lin, and C.-C. Huang, “High-speed data-plane packet aggregation and disaggregation by P4 switches,” Journal of Network and Computer Applications, vol. 142, pp. 98–110, 2019.
[177] S.-Y. Wang, J.-Y. Li, and Y.-B. Lin, “Aggregating and disaggregating packets with various sizes of payload in P4 switches at 100 Gbps line rate,” Journal of Network and Computer Applications, p. 102676, 2020.
[178] Y.-B. Lin, S.-Y. Wang, C.-C. Huang, and C.-M. Wu, “The SDN approach for the aggregation/disaggregation of sensor data,” Sensors, vol. 18, no. 7, p. 2025, 2018.
[179] A. L. R. Madureira, F. R. C. Araújo, and L. N. Sampaio, “On supporting IoT data aggregation through programmable data planes,” Computer Networks, p. 107330, 2020.
[180] M. Uddin, S. Mukherjee, H. Chang, and T. Lakshman, “SDN-based service automation for IoT,” in 2017 IEEE 25th International Conference on Network Protocols (ICNP), pp. 1–10, IEEE, 2017.
[181] M. Uddin, S. Mukherjee, H. Chang, and T. Lakshman, “SDN-based multi-protocol edge switching for IoT service automation,” IEEE Journal on Selected Areas in Communications, vol. 36, no. 12, pp. 2775–2786, 2018.
[182] V. Sivaraman, S. Narayana, O. Rottenstreich, S. Muthukrishnan, and J. Rexford, “Heavy-hitter detection entirely in the data plane,” in Proceedings of the Symposium on SDN Research, pp. 164–176, 2017.
[183] J. Kučera, D. A. Popescu, G. Antichi, J. Kořenek, and A. W. Moore, “Seek and push: detecting large traffic aggregates in the dataplane,” arXiv preprint arXiv:1805.05993, 2018.
[184] R. Ben-Basat, X. Chen, G. Einziger, and O. Rottenstreich, “Efficient measurement on programmable switches using probabilistic recirculation,” in 2018 IEEE 26th International Conference on Network Protocols (ICNP), pp. 313–323, IEEE, 2018.
[185] L. Tang, Q. Huang, and P. P. Lee, “A fast and compact invertible sketch for network-wide heavy flow detection,” IEEE/ACM Transactions on Networking, vol. 28, no. 5, pp. 2350–2363, 2020.
[186] M. V. B. da Silva, J. A. Marques, L. P. Gaspary, and L. Z. Granville, “Identifying elephant flows using dynamic thresholds in programmable ixp networks,” Journal of Internet Services and Applications, vol. 11, no. 1, pp. 1–12, 2020.
[187] R. Harrison, Q. Cai, A. Gupta, and J. Rexford, “Network-wide heavy hitter detection with commodity switches,” in Proceedings of the Symposium on SDN Research, pp. 1–7, 2018.
[188] R. Harrison, S. L. Feibish, A. Gupta, R. Teixeira, S. Muthukrishnan, and J. Rexford, “Carpe elephants: Seize the global heavy hitters,” in Proceedings of the Workshop on Secure Programmable Network Infrastructure, pp. 15–21, 2020.
[189] D. Ding, M. Savi, G. Antichi, and D. Siracusa, “An incrementally-deployable P4-enabled architecture for network-wide heavy-hitter detection,” IEEE Transactions on Network and Service Management, vol. 17, no. 1, pp. 75–88, 2020.
[190] L. Tang, Q. Huang, and P. P. Lee, “Spreadsketch: Toward invertible and network-wide detection of superspreaders,” in IEEE INFOCOM 2020-IEEE Conference on Computer Communications, pp. 1608–1617, IEEE, 2020.
[191] D. Scholz, A. Oeldemann, F. Geyer, S. Gallenmüller, H. Stubbe, T. Wild, A. Herkersdorf, and G. Carle, “Cryptographic hashing in P4 data planes,” in 2019 ACM/IEEE Symposium on Architectures for Networking and Communications Systems (ANCS), pp. 1–6, IEEE, 2019.
[192] F. Hauser, M. Häberle, M. Schmidt, and M. Menth, “P4-IPsec: implementation of IPsec gateways in P4 with SDN control for host-to-site scenarios,” arXiv preprint arXiv:1907.03593, 2019.
[193] L. Malina, D. Smekal, S. Ricci, J. Hajny, P. Cíbik, and J. Hrabovsky, “Hardware-accelerated cryptography for software-defined networks with p4,” in International Conference on Information Technology and Communications Security, pp. 271–287, Springer, 2020.
[194] G. Liu, W. Quan, N. Cheng, D. Gao, N. Lu, H. Zhang, and X. Shen, “Softwarized iot network immunity against eavesdropping with programmable data planes,” IEEE Internet of Things Journal, 2021.
[195] X. Chen, “Implementing AES encryption on programmable switches via scrambled lookup tables,” in Proceedings of the Workshop on Secure Programmable Network Infrastructure, SPIN ’20, p. 8–14, Association for Computing Machinery, 2020.
[196] H. Kim and A. Gupta, “ONTAS: flexible and scalable online network traffic anonymization system,” in Proceedings of the 2019 Workshop on Network Meets AI & ML, pp. 15–21, 2019.
[197] H. M. Moghaddam and A. Mosenia, “Anonymizing masses: practical light-weight anonymity at the network level,” arXiv preprint arXiv:1911.09642, 2019.
[198] T. Datta, N. Feamster, J. Rexford, and L. Wang, “SPINE: surveillance protection in the network elements,” in 9th USENIX Workshop on Free and Open Communications on the Internet (FOCI), 2019.
[199] L. Wang, H. Kim, P. Mittal, and J. Rexford, “Programmable in-network obfuscation of dns traffic,” in NDSS: DNS Privacy Workshop, 2021.
[200] R. Meier, P. Tsankov, V. Lenders, L. Vanbever, and M. Vechev, “NetHide: secure and practical network topology obfuscation,” in 27th USENIX Security Symposium (USENIX Security 18), pp. 693–709, 2018.
[201] R. Datta, S. Choi, A. Chowdhary, and Y. Park, “P4Guard: designing P4 based firewall,” in MILCOM 2018-2018 IEEE Military Communications Conference (MILCOM), pp. 1–6, IEEE, 2018.
[202] J. Cao, Y. Liu, Y. Zhou, C. Sun, Y. Wang, and J. Bi, “Cofilter: A high-performance switch-accelerated stateful packet filter for bare-metal servers,” in 2019 28th International Conference on Computer Communication and Networks (ICCCN), pp. 1–9, IEEE, 2019.
[203] J. Li, H. Jiang, W. Jiang, J. Wu, and W. Du, “Sdn-based stateful firewall for cloud,” in 2020 IEEE 6th Intl Conference on Big Data

Security on Cloud (BigDataSecurity), IEEE Intl Conference on High Performance and Smart Computing,(HPSC) and IEEE Intl Conference on Intelligent Data and Security (IDS), pp. 157–161, IEEE, 2020.
[204] A. Almaini, A. Al-Dubai, I. Romdhani, and M. Schramm, "Delegation of authentication to the data plane in software-defined networks," in 2019 IEEE International Conferences on Ubiquitous Computing & Communications (IUCC) and Data Science and Computational Intelligence (DSCI) and Smart Computing, Networking and Services (SmartCNS), pp. 58–65, IEEE, 2019.
[205] E. O. Zaballa, D. Franco, Z. Zhou, and M. S. Berger, "P4knocking: Offloading host-based firewall functionalities to the network," in 2020 23rd Conference on Innovation in Clouds, Internet and Networks and Workshops (ICIN), pp. 7–12, IEEE, 2020.
[206] Q. Kang, L. Xue, A. Morrison, Y. Tang, A. Chen, and X. Luo, "Programmable in-network security for context-aware BYOD policies," arXiv preprint arXiv:1908.01405, 2019.
[207] S. Bai, H. Kim, and J. Rexford, "Passive OS fingerprinting on commodity switches,"
[208] A. Almaini, A. Al-Dubai, I. Romdhani, M. Schramm, and A. Alsarhan, "Lightweight edge authentication for software defined networks," Computing, vol. 103, no. 2, pp. 291–311, 2021.
[209] J. Hill, M. Aloserij, and P. Grosso, "Tracking network flows with p4," in 2018 IEEE/ACM Innovating the Network for Data-Intensive Science (INDIS), pp. 23–32, IEEE, 2018.
[210] G. Li, M. Zhang, C. Liu, X. Kong, A. Chen, G. Gu, and H. Duan, "NetHCF: enabling line-rate and adaptive spoofed IP traffic filtering," in 2019 IEEE 27th International Conference on Network Protocols (ICNP), pp. 1–12, IEEE, 2019.
[211] A. Febro, H. Xiao, and J. Spring, "Distributed SIP DDoS defense with P4," in 2019 IEEE Wireless Communications and Networking Conference (WCNC), pp. 1–8, IEEE, 2019.
[212] D. Scholz, S. Gallenmüller, H. Stubbe, B. Jaber, M. Rouhi, and G. Carle, "Me love (SYN-) cookies: SYN flood mitigation in programmable data planes," arXiv preprint arXiv:2003.03221, 2020.
[213] D. Scholz, S. Gallenmüller, H. Stubbe, and G. Carle, "Syn flood defense in programmable data planes," in Proceedings of the 3rd P4 Workshop in Europe, pp. 13–20, 2020.
[214] G. K. Ndonda and R. Sadre, "A two-level intrusion detection system for industrial control system networks using p4," in 5th International Symposium for ICS & SCADA Cyber Security Research 2018 5, pp. 31–40, 2018.
[215] J. Xing, W. Wu, and A. Chen, "Architecting programmable data plane defenses into the network with FastFlex," in Proceedings of the 18th ACM Workshop on Hot Topics in Networks, pp. 161–169, 2019.
[216] Q. Kang, J. Xing, and A. Chen, "Automated attack discovery in data plane systems," in 12th USENIX Workshop on Cyber Security Experimentation and Test (CSET), 2019.
[217] Â. C. Lapolli, J. A. Marques, and L. P. Gaspary, "Offloading real-time DDoS attack detection to programmable data planes," in 2019 IFIP/IEEE Symposium on Integrated Network and Service Management (IM), pp. 19–27, IEEE, 2019.
[218] Y. Mi and A. Wang, "ML-pushback: machine learning based push-back defense against DDoS," in Proceedings of the 15th International Conference on emerging Networking EXperiments and Technologies, pp. 80–81, 2019.
[219] J. Ioannidis and S. M. Bellovin, "Implementing pushback: router-based defense against DDoS attacks," in NDSS, 2016.
[220] M. Zhang, G. Li, S. Wang, C. Liu, A. Chen, H. Hu, G. Gu, Q. Li, M. Xu, and J. Wu, "Poseidon: mitigating volumetric DDoS attacks with programmable switches," in Proceedings of NDSS, 2020.
[221] K. Friday, E. Kfoury, E. Bou-Harb, and J. Crichigno, "Towards a unified in-network DDoS detection and mitigation strategy," in 2020 6th IEEE Conference on Network Softwarization (NetSoft), pp. 218–226, 2020.
[222] J. Xing, Q. Kang, and A. Chen, "NetWarden: mitigating network covert channels while preserving performance," in 29th USENIX Security Symposium (USENIX Security 20), 2020.
[223] A. Laraba, J. François, I. Chrisment, S. R. Chowdhury, and R. Boutaba, "Defeating protocol abuse with p4: Application to explicit congestion notification," in 2020 IFIP Networking Conference (Networking), pp. 431–439, IEEE, 2020.
[224] "Ripple: A programmable, decentralized link-flooding defense against adaptive adversaries," in 30th USENIX Security Symposium (USENIX Security 21), (Vancouver, B.C.), USENIX Association, 2021.
[225] A. da Silveira Ilha, Â. C. Lapolli, J. A. Marques, and L. P. Gaspary, "Euclid: A fully in-network, p4-based approach for real-time ddos attack detection and mitigation," IEEE Transactions on Network and Service Management, 2020.
[226] X. Z. Khooi, L. Csikor, D. M. Divakaran, and M. S. Kang, "Dida: Distributed in-network defense architecture against amplified reflection ddos attacks," in 2020 6th IEEE Conference on Network Softwarization (NetSoft), pp. 277–281, IEEE, 2020.
[227] D. Ding, M. Savi, F. Pederzolli, M. Campanella, and D. Siracusa, "In-network volumetric ddos victim identification using programmable commodity switches," IEEE Transactions on Network and Service Management, 2021.
[228] F. Musumeci, V. Ionata, F. Paolucci, F. Cugini, and M. Tornatore, "Machine-learning-assisted ddos attack detection with p4 language," in ICC 2020-2020 IEEE International Conference on Communications (ICC), pp. 1–6, IEEE, 2020.
[229] Z. Liu, H. Namkung, G. Nikolaidis, J. Lee, C. Kim, X. Jin, V. Braverman, M. Yu, and V. Sekar, "Jaqen: A high-performance switch-native approach for detecting and mitigating volumetric ddos attacks with programmable switches," in 30th USENIX Security Symposium (USENIX Security 21), 2021.
[230] C. Zhang, J. Bi, Y. Zhou, J. Wu, B. Liu, Z. Li, A. B. Dogar, and Y. Wang, "P4DB: on-the-fly debugging of the programmable data plane," in 2017 IEEE 25th International Conference on Network Protocols (ICNP), pp. 1–10, IEEE, 2017.
[231] Y. Zhou, J. Bi, Y. Lin, Y. Wang, D. Zhang, Z. Xi, J. Cao, and C. Sun, "P4tester: efficient runtime rule fault detection for programmable data planes," in Proceedings of the International Symposium on Quality of Service, pp. 1–10, 2019.
[232] M. V. Dumitru, D. Dumitrescu, and C. Raiciu, "Can we exploit buggy P4 programs?," in Proceedings of the Symposium on SDN Research, pp. 62–68, 2020.
[233] S. Kodeswaran, M. T. Arashloo, P. Tammana, and J. Rexford, "Tracking P4 program execution in the data plane," in Proceedings of the Symposium on SDN Research, pp. 117–122, 2020.
[234] Y. Zhou, J. Bi, T. Yang, K. Gao, C. Zhang, J. Cao, and Y. Wang, "Keysight: Troubleshooting programmable switches via scalable high-coverage behavior tracking," in 2018 IEEE 26th International Conference on Network Protocols (ICNP), pp. 291–301, IEEE, 2018.
[235] N. Lopes, N. Bjørner, N. McKeown, A. Rybalchenko, D. Talayco, and G. Varghese, "Automatically verifying reachability and well-formedness in P4 networks," Technical Report, Tech. Rep, 2016.
[236] L. Freire, M. Neves, L. Leal, K. Levchenko, A. Schaeffer-Filho, and M. Barcellos, "Uncovering bugs in P4 programs with assertion-based verification," in Proceedings of the Symposium on SDN Research, pp. 1–7, 2018.
[237] M. Neves, L. Freire, A. Schaeffer-Filho, and M. Barcellos, "Verification of P4 programs in feasible time using assertions," in Proceedings of the 14th International Conference on emerging Networking EXperiments and Technologies, pp. 73–85, 2018.
[238] J. Liu, W. Hallahan, C. Schlesinger, M. Sharif, J. Lee, R. Soulé, H. Wang, C. Caşcaval, N. McKeown, and N. Foster, "P4v: practical verification for programmable data planes," in Proceedings of the 2018 Conference of the ACM Special Interest Group on Data Communication, pp. 490–503, 2018.
[239] A. Nötzli, J. Khan, A. Fingerhut, C. Barrett, and P. Athanas, "P4pktgen: automated test case generation for P4 programs," in Proceedings of the Symposium on SDN Research, pp. 1–7, 2018.
[240] D. Lukács, M. Tejfel, and G. Pongrácz, "Keeping P4 switches fast and fault-free through automatic verification," Acta Cybernetica, vol. 24, no. 1, pp. 61–81, 2019.
[241] R. Stoenescu, D. Dumitrescu, M. Popovici, L. Negreanu, and C. Raiciu, "Debugging P4 programs with Vera," in Proceedings of the 2018 Conference of the ACM Special Interest Group on Data Communication, pp. 518–532, 2018.
[242] A. Shukla, K. N. Hudemann, A. Hecker, and S. Schmid, "Runtime verification of P4 switches with reinforcement learning," in Proceedings of the 2019 Workshop on Network Meets AI & ML, pp. 1–7, 2019.
[243] D. Dumitrescu, R. Stoenescu, L. Negreanu, and C. Raiciu, "bf4: towards bug-free P4 programs," in Proceedings of the Annual conference of the ACM Special Interest Group on Data Communication

on the applications, technologies, architectures, and protocols for computer communication, pp. 571–585, 2020.
[244] A. Bas and A. Fingerhut, "P4 tutorial, slide 22." [Online]. Available: https://tinyurl.com/tb4m749.
[245] M. Shahbaz, S. Choi, B. Pfaff, C. Kim, N. Feamster, N. McKeown, and J. Rexford, "PISCES: A programmable, protocol-independent software switch," in Proceedings of the 2016 ACM SIGCOMM Conference, pp. 525–538, 2016.
[246] B. Pfaff, J. Pettit, T. Koponen, E. Jackson, A. Zhou, J. Rajahalme, J. Gross, A. Wang, J. Stringer, P. Shelar, et al., "The design and implementation of open vswitch," in 12th USENIX Symposium on Networked Systems Design and Implementation (NSDI), pp. 117–130, 2015.
[247] Barefoot Networks, "Barefoot Academy," 2020. [Online]. Available: https://www.barefootnetworks.com/barefoot-academy/.
[248] C. Kim, A. Sivaraman, N. Katta, A. Bas, A. Dixit, and L. J. Wobker, "In-band network telemetry via programmable dataplanes," in ACM SIGCOMM, 2015.
[249] C. Hopps et al., "Analysis of an equal-cost multi-path algorithm," tech. rep., RFC 2992, November, 2000.
[250] S. Sinha, S. Kandula, and D. Katabi, "Harnessing TCP's burstiness with flowlet switching," in Proc. 3rd ACM Workshop on Hot Topics in Networks (Hotnets-III), Citeseer, 2004.
[251] C. Kim, P. Bhide, E. Doe, H. Holbrook, A. Ghanwani, D. Daly, M. Hira, and B. Davie, "In-band network telemetry (INT)," technical specification, 2016.
[252] P. Manzanares-Lopez, J. P. Muñoz-Gea, and J. Malgosa-Sanahuja, "Passive in-band network telemetry systems: The potential of programmable data plane on network-wide telemetry," IEEE Access, vol. 9, pp. 20391–20409, 2021.
[253] M. A. Vieira, M. S. Castanho, R. D. Pacífico, E. R. Santos, E. P. C. Júnior, and L. F. Vieira, "Fast packet processing with eBPF and XDP: concepts, code, challenges, and applications," ACM Computing Surveys (CSUR), vol. 53, no. 1, pp. 1–36, 2020.
[254] J. Crichigno, E. Bou-Harb, and N. Ghani, "A comprehensive tutorial on science DMZ," IEEE Communications Surveys & Tutorials, vol. 21, no. 2, pp. 2041–2078, 2018.
[255] J. F. Kurose and K. W. Ross, "Computer networking a top down approach featuring the intel," 2016.
[256] S. Ha, I. Rhee, and L. Xu, "CUBIC: a new TCP-friendly high-speed TCP variant," ACM SIGOPS operating systems review, vol. 42, no. 5, pp. 64–74, 2008.
[257] D. Leith and R. Shorten, "H-TCP: TCP congestion control for high bandwidth-delay product paths," draft-leith-tcp-htcp-06 (work in progress), 2008.
[258] N. Cardwell, Y. Cheng, C. S. Gunn, S. H. Yeganeh, and V. Jacobson, "BBR: congestion-based congestion control," Communications of the ACM, vol. 60, no. 2, pp. 58–66, 2017.
[259] E. F. Kfoury, J. Gomez, J. Crichigno, and E. Bou-Harb, "An emulation-based evaluation of TCP BBRv2 alpha for wired broadband," Computer Communications, 2020.
[260] S. Floyd, "TCP and explicit congestion notification," ACM SIGCOMM Computer Communication Review, vol. 24, no. 5, pp. 8–23, 1994.
[261] R. Mittal, V. T. Lam, N. Dukkipati, E. Blem, H. Wassel, M. Ghobadi, A. Vahdat, Y. Wang, D. Wetherall, and D. Zats, "TIMELY: RTT-based congestion control for the data center," ACM SIGCOMM Computer Communication Review, vol. 45, no. 4, pp. 537–550, 2015.
[262] Y. Zhu, H. Eran, D. Firestone, C. Guo, M. Lipshteyn, Y. Liron, J. Padhye, S. Raindel, M. H. Yahia, and M. Zhang, "Congestion control for large-scale RDMA deployments," ACM SIGCOMM Computer Communication Review, vol. 45, no. 4, pp. 523–536, 2015.
[263] M. Alizadeh, A. Greenberg, D. A. Maltz, J. Padhye, P. Patel, B. Prabhakar, S. Sengupta, and M. Sridharan, "Data Center TCP (DCTCP)," in Proceedings of the ACM SIGCOMM 2010 conference, pp. 63–74, 2010.
[264] M. Alizadeh, S. Yang, M. Sharif, S. Katti, N. McKeown, B. Prabhakar, and S. Shenker, "pFabric: minimal near-optimal datacenter transport," ACM SIGCOMM Computer Communication Review, vol. 43, no. 4, pp. 435–446, 2013.
[265] M. Dong, Q. Li, D. Zarchy, P. B. Godfrey, and M. Schapira, "PCC: Re-architecting congestion control for consistent high performance," in 12th USENIX Symposium on Networked Systems Design and Implementation (NSDI), pp. 395–408, 2015.
[266] A. Langley, A. Riddoch, A. Wilk, A. Vicente, C. Krasic, D. Zhang, F. Yang, F. Kouranov, I. Swett, J. Iyengar, et al., "The QUIC transport protocol: design and Internet-scale deployment," in Proceedings of the Conference of the ACM Special Interest Group on Data Communication, pp. 183–196, 2017.
[267] P. Cheng, F. Ren, R. Shu, and C. Lin, "Catch the whole lot in an action: rapid precise packet loss notification in data center," in 11th USENIX Symposium on Networked Systems Design and Implementation (NSDI), pp. 17–28, 2014.
[268] A. Ramachandran, S. Seetharaman, N. Feamster, and V. Vazirani, "Fast monitoring of traffic subpopulations," in Proceedings of the 8th ACM SIGCOMM conference on Internet measurement, pp. 257–270, 2008.
[269] N. Alon, Y. Matias, and M. Szegedy, "The space complexity of approximating the frequency moments," Journal of Computer and system sciences, vol. 58, no. 1, pp. 137–147, 1999.
[270] V. Braverman and R. Ostrovsky, "Zero-one frequency laws," in Proceedings of the forty-second ACM symposium on Theory of computing, pp. 281–290, 2010.
[271] M. Charikar, K. Chen, and M. Farach-Colton, "Finding frequent items in data streams," in International Colloquium on Automata, Languages, and Programming, pp. 693–703, Springer, 2002.
[272] G. Cormode and S. Muthukrishnan, "An improved data stream summary: the count-min sketch and its applications," Journal of Algorithms, vol. 55, no. 1, pp. 58–75, 2005.
[273] S. Floyd and V. Jacobson, "Random early detection gateways for congestion avoidance," IEEE/ACM Transactions on networking, vol. 1, no. 4, pp. 397–413, 1993.
[274] P. Flajolet, D. Gardy, and L. Thimonier, "Birthday paradox, coupon collectors, caching algorithms and self-organizing search," Discrete Applied Mathematics, vol. 39, no. 3, pp. 207–229, 1992.
[275] R. Dolby, "Noise reduction systems," Nov. 5 1974. US Patent 3,846,719.
[276] S. V. Vaseghi, Advanced digital signal processing and noise reduction. John Wiley & Sons, 2008.
[277] J. Gettys, "Bufferbloat: dark buffers in the Internet," IEEE Internet Computing, no. 3, p. 96, 2011.
[278] M. Allman, "Comments on bufferbloat," ACM SIGCOMM Computer Communication Review, vol. 43, no. 1, pp. 30–37, 2013.
[279] Y. Gong, D. Rossi, C. Testa, S. Valenti, and M. D. Täht, "Fighting the bufferbloat: on the coexistence of AQM and low priority congestion control," Computer Networks, vol. 65, pp. 255–267, 2014.
[280] C. Staff, "Bufferbloat: what's wrong with the Internet?," Communications of the ACM, vol. 55, no. 2, pp. 40–47, 2012.
[281] V. G. Cerf, "Bufferbloat and other internet challenges," IEEE Internet Computing, vol. 18, no. 5, pp. 80–80, 2014.
[282] H. Harkous, C. Papagianni, K. De Schepper, M. Jarschel, M. Dimolianis, and R. Preis, "Virtual queues for p4: A poor man's programmable traffic manager," IEEE Transactions on Network and Service Management, 2021.
[283] K. Nichols, S. Blake, F. Baker, and D. Black, "Definition of the differentiated services field (DS field) in the IPv4 and IPv6 headers," RFC8376. [Online]. Available: https://tools.ietf.org/html/rfc8376.
[284] B. Fenner, M. Handley, H. Holbrook, I. Kouvelas, R. Parekh, Z. Zhang, and L. Zheng, "Protocol independent multicast-sparse mode (PIM-SM): protocol specification (revised).," [Online]. Available: https://tools.ietf.org/html/rfc7761.
[285] H. Holbrook, B. Cain, and B. Haberman, "Using Internet group management protocol version 3 (IGMPv3) and multicast listener discovery protocol version 2 (MLDv2) for source-specific multicast," RFC 4604 (Proposed Standard), Internet Engineering Task Force, 2006.
[286] I. Wijnands, E. C. Rosen, A. Dolganow, T. Przygienda, and S. Aldrin, "Multicast using bit index explicit replication (BIER)," in RFC Editor, 2017.
[287] S. Luo, H. Yu, K. Li, and H. Xing, "Efficient file dissemination in data center networks with priority-based adaptive multicast," IEEE Journal on Selected Areas in Communications, vol. 38, no. 6, pp. 1161–1175, 2020.
[288] B. Carpenter and S. Brim, "Middleboxes: taxonomy and issues," 2002. [Online]. Available: https://tools.ietf.org/html/rfc3234.
[289] J. McCauley, A. Panda, A. Krishnamurthy, and S. Shenker, "Thoughts on load distribution and the role of programmable switches," ACM SIGCOMM Computer Communication Review, vol. 49, no. 1,

pp. 18–23, 2019.
[290] T. Norp, "5G Requirements and key performance indicators," Journal of ICT Standardization, vol. 6, no. 1, pp. 15–30, 2018.
[291] G. Xylomenos, C. N. Ververidis, V. A. Siris, N. Fotiou, C. Tsilopoulos, X. Vasilakos, K. V. Katsaros, and G. C. Polyzos, "A survey of information-centric networking research," IEEE communications surveys & tutorials, vol. 16, no. 2, pp. 1024–1049, 2013.
[292] D. L. Tennenhouse and D. J. Wetherall, "Towards an active network architecture," in Proceedings DARPA Active Networks Conference and Exposition, pp. 2–15, IEEE, 2002.
[293] E. F. Kfoury, J. Gomez, J. Crichigno, E. Bou-Harb, and D. Khoury, "Decentralized distribution of PCP mappings over blockchain for end-to-end secure direct communications," IEEE Access, vol. 7, pp. 110159–110173, 2019.
[294] S. A. Weil, S. A. Brandt, E. L. Miller, D. D. Long, and C. Maltzahn, "Ceph: A scalable, high-performance distributed file system," in Proceedings of the 7th symposium on Operating systems design and implementation, pp. 307–320, 2006.
[295] L. Lamport et al., "Paxos made simple," ACM Sigact News, vol. 32, no. 4, pp. 18–25, 2001.
[296] D. Ongaro and J. Ousterhout, "In search of an understandable consensus algorithm," in 2014 USENIX Annual Technical Conference (USENIX ATC 14), pp. 305–319, 2014.
[297] Huynh Tu Dang, "Consensus as a network service." [Online]. Available: https://tinyurl.com/y2t9plsu.
[298] J. Nelson, "SwitchML scaling distributed machine learning with in network aggregation." [Online]. Available: https://tinyurl.com/y53upm7k.
[299] D. Das, S. Avancha, D. Mudigere, K. Vaidynathan, S. Sridharan, D. Kalamkar, B. Kaul, and P. Dubey, "Distributed deep learning using synchronous stochastic gradient descent," arXiv preprint arXiv:1602.06709, 2016.
[300] S. Farrell, "Low-power wide area network (LPWAN) overview," RFC8376. [Online]. Available: https://tools.ietf.org/html/rfc8376.
[301] A. Koike, T. Ohba, and R. Ishibashi, "IoT network architecture using packet aggregation and disaggregation," in 2016 5th IIAI International Congress on Advanced Applied Informatics (IIAI-AAI), pp. 1140–1145, IEEE, 2016.
[302] J. Deng and M. Davis, "An adaptive packet aggregation algorithm for wireless networks," in 2013 International Conference on Wireless Communications and Signal Processing, pp. 1–6, IEEE, 2013.
[303] Y. Yasuda, R. Nakamura, and H. Ohsaki, "A probabilistic interest packet aggregation for content-centric networking," in 2018 IEEE 42nd Annual Computer Software and Applications Conference (COMPSAC), vol. 2, pp. 783–788, IEEE, 2018.
[304] A. S. Akyurek and T. S. Rosing, "Optimal packet aggregation scheduling in wireless networks," IEEE Transactions on Mobile Computing, vol. 17, no. 12, pp. 2835–2852, 2018.
[305] K. Zhou and N. Nikaein, "Packet aggregation for machine type communications in LTE with random access channel," in 2013 IEEE Wireless Communications and Networking Conference (WCNC), pp. 262–267, IEEE, 2013.
[306] A. Majeed and N. B. Abu-Ghazaleh, "Packet aggregation in multi-rate wireless LANs," in 2012 9th Annual IEEE Communications Society Conference on Sensor, Mesh and Ad Hoc Communications and Networks (SECON), pp. 452–460, IEEE, 2012.
[307] D. SIG, "Bluetooth core specification version 4.2," Specification of the Bluetooth System, 2014.
[308] S. Farahani, ZigBee wireless networks and transceivers. Newnes, 2011.
[309] O. Hersent, D. Boswarthick, and O. Elloumi, The Internet of things: key applications and protocols. John Wiley & Sons, 2011.
[310] J. Shi, W. Quan, D. Gao, M. Liu, G. Liu, C. Yu, and W. Su, "Flowlet-based stateful multipath forwarding in heterogeneous Internet of things," IEEE Access, vol. 8, pp. 74875–74886, 2020.
[311] S. Do, L.-V. Le, B.-S. P. Lin, and L.-P. Tung, "SDN/NFV-based network infrastructure for enhancing IoT gateways," in 2019 International Conference on Internet of Things (iThings) and IEEE Green Computing and Communications (GreenCom) and IEEE Cyber, Physical and Social Computing (CPSCom) and IEEE Smart Data (SmartData), pp. 1135–1142, IEEE, 2019.
[312] A. Metwally, D. Agrawal, and A. El Abbadi, "Efficient computation of frequent and top-k elements in data streams," in International Conference on Database Theory, pp. 398–412, Springer, 2005.
[313] S. Heule, M. Nunkesser, and A. Hall, "HyperLogLog in practice: algorithmic engineering of a state of the art cardinality estimation algorithm," in Proceedings of the 16th International Conference on Extending Database Technology, pp. 683–692, 2013.
[314] F. Hauser, M. Schmidt, M. Häberle, and M. Menth, "P4-MACsec: dynamic topology monitoring and data layer protection with MACsec in P4-based SDN," IEEE Access, 2020.
[315] M. G. Reed, P. F. Syverson, and D. M. Goldschlag, "Anonymous connections and onion routing," IEEE Journal on Selected areas in Communications, vol. 16, no. 4, pp. 482–494, 1998.
[316] V. Liu, S. Han, A. Krishnamurthy, and T. Anderson, "Tor instead of IP," in Proceedings of the 10th ACM Workshop on Hot Topics in Networks, pp. 1–6, 2011.
[317] C. Chen, D. E. Asoni, D. Barrera, G. Danezis, and A. Perrig, "HORNET: high-speed onion routing at the network layer," in Proceedings of the 22nd ACM SIGSAC Conference on Computer and Communications Security, pp. 1441–1454, 2015.
[318] L. Lamport, "Password authentication with insecure communication," Communications of the ACM, vol. 24, no. 11, pp. 770–772, 1981.
[319] M. Zalewski and W. Stearns, "p0f," see http://lcamtuf. coredump.cx/p0f3, 2006.
[320] J. Barnes and P. Crowley, "k-p0f: A high-throughput kernel passive OS fingerprinter," in Architectures for Networking and Communications Systems, pp. 113–114, IEEE, 2013.
[321] S. Hong, R. Baykov, L. Xu, S. Nadimpalli, and G. Gu, "Towards SDN-defined programmable BYOD (bring your own device) security," in NDSS, 2016.
[322] S. Hilton, "Dyn analysis summary of Friday October 21 Attack, 2016.." [Online]. Available: https://dyn.com/blog/dyn-analysis-summary-of-friday-october-21-attack/.
[323] S. Kottler, "February 28th DDoS incident report, March, 2018." [Online]. Available: https://githubengineering.com/ddos-incident-report/.
[324] S. K. Fayaz, Y. Tobioka, V. Sekar, and M. Bailey, "Bohatei: Flexible and elastic ddos defense," in 24th USENIX Security Symposium (USENIX Security 15), pp. 817–832, 2015.
[325] Arbor Networks, "Arbor Networks APS Datasheet." [Online]. Available: https://www.netscout.com/sites/default/files/2018-04/DS_APS_EN.pdf.
[326] NSFOCUS, "NSFOCUS Anti-DDoS System Datasheet." [Online]. Available: https://nsfocusglobal.com/wp-content/uploads/2018/05/Anti-DDoS-Solution.pdf.
[327] J. Hypolite, J. Sonchack, S. Hershkop, N. Dautenhahn, A. DeHon, and J. M. Smith, "Deepmatch: practical deep packet inspection in the data plane using network processors," in Proceedings of the 16th International Conference on emerging Networking EXperiments and Technologies, pp. 336–350, 2020.
[328] N. Handigol, B. Heller, V. Jeyakumar, D. Mazières, and N. McKeown, "I know what your packet did last hop: using packet histories to troubleshoot networks," in 11th USENIX Symposium on Networked Systems Design and Implementation (NSDI 14), pp. 71–85, 2014.
[329] Y. Zhu, N. Kang, J. Cao, A. Greenberg, G. Lu, R. Mahajan, D. Maltz, L. Yuan, M. Zhang, B. Y. Zhao, and H. Zheng, "Packet-level telemetry in large datacenter networks," in Proceedings of the 2015 ACM Conference on Special Interest Group on Data Communication, pp. 479–491, 2015.
[330] H. Zeng, P. Kazemian, G. Varghese, and N. McKeown, "Automatic test packet generation," in Proceedings of the 8th international conference on Emerging networking experiments and technologies, pp. 241–252, 2012.
[331] H. T. Dang, H. Wang, T. Jepsen, G. Brebner, C. Kim, J. Rexford, R. Soulé, and H. Weatherspoon, "Whippersnapper: A p4 language benchmark suite," in Proceedings of the Symposium on SDN Research, pp. 95–101, 2017.
[332] F. Rodriguez, P. G. K. Patra, L. Csikor, C. Rothenberg, P. V. S. Laki, and G. Pongrácz, "Bb-gen: A packet crafter for p4 target evaluation," in Proceedings of the ACM SIGCOMM 2018 Conference on Posters and Demos, pp. 111–113, 2018.
[333] H. Harkous, M. Jarschel, M. He, R. Pries, and W. Kellerer, "P8: P4 with predictable packet processing performance," IEEE Transactions on Network and Service Management, 2020.
[334] H. Harkous, M. Jarschel, M. He, R. Priest, and W. Kellerer, "Towards understanding the performance of p4 programmable hardware," in

2019 ACM/IEEE Symposium on Architectures for Networking and Communications Systems (ANCS), pp. 1–6, IEEE, 2019.
[335] P. Kazemian, G. Varghese, and N. McKeown, "Header space analysis: static checking for networks," in Presented as part of the 9th USENIX Symposium on Networked Systems Design and Implementation (NSDI 12), pp. 113–126, 2012.
[336] A. Khurshid, X. Zou, W. Zhou, M. Caesar, and P. B. Godfrey, "Veriflow: verifying network-wide invariants in real time," in Presented as part of the 10th USENIX Symposium on Networked Systems Design and Implementation (NSDI), pp. 15–27, 2013.
[337] R. Stoenescu, M. Popovici, L. Negreanu, and C. Raiciu, "Symnet: scalable symbolic execution for modern networks," in Proceedings of the 2016 ACM SIGCOMM Conference, pp. 314–327, 2016.
[338] H. Mai, A. Khurshid, R. Agarwal, M. Caesar, P. B. Godfrey, and S. T. King, "Debugging the data plane with Anteater," ACM SIGCOMM Computer Communication Review, vol. 41, no. 4, pp. 290–301, 2011.
[339] P. Kazemian, M. Chang, H. Zeng, G. Varghese, N. McKeown, and S. Whyte, "Real time network policy checking using header space analysis," in Presented as part of the 10th USENIX Symposium on Networked Systems Design and Implementation (NSDI), pp. 99–111, 2013.
[340] A. Horn, A. Kheradmand, and M. Prasad, "Delta-net: real-time network verification using atoms," in 14th USENIX Symposium on Networked Systems Design and Implementation (NSDI), pp. 735–749, 2017.
[341] S. Son, S. Shin, V. Yegneswaran, P. Porras, and G. Gu, "Model checking invariant security properties in OpenFlow," in 2013 IEEE international conference on communications (ICC), pp. 1974–1979, IEEE, 2013.
[342] A. Panda, O. Lahav, K. Argyraki, M. Sagiv, and S. Shenker, "Verifying reachability in networks with mutable datapaths," in 14th USENIX Symposium on Networked Systems Design and Implementation (NSDI), pp. 699–718, 2017.
[343] N. Foster, N. McKeown, J. Rexford, G. Parulkar, L. Peterson, and O. Sunay, "Using deep programmability to put network owners in control," ACM SIGCOMM Computer Communication Review, vol. 50, no. 4, pp. 82–88, 2020.
[344] "Pronto Project." [Online]. Available: https://prontoproject.org/.
[345] Y. Zhou and J. Bi, "Clickp4: Towards modular programming of p4," in Proceedings of the SIGCOMM Posters and Demos, pp. 100–102, 2017.
[346] P. Zheng, T. Benson, and C. Hu, "P4visor: Lightweight virtualization and composition primitives for building and testing modular programs," in Proceedings of the 14th International Conference on Emerging Networking EXperiments and Technologies, pp. 98–111, 2018.
[347] X. Chen, D. Zhang, X. Wang, K. Zhu, and H. Zhou, "P4sc: Towards high-performance service function chain implementation on the p4-capable device," in 2019 IFIP/IEEE Symposium on Integrated Network and Service Management (IM), pp. 1–9, IEEE, 2019.
[348] M. Riftadi and F. Kuipers, "P4i/o: Intent-based networking with p4," in 2019 IEEE Conference on Network Softwarization (NetSoft), pp. 438–443, IEEE, 2019.
[349] E. O. Zaballa and Z. Zhou, "Graph-to-p4: A p4 boilerplate code generator for parse graphs," in 2019 ACM/IEEE Symposium on Architectures for Networking and Communications Systems (ANCS), pp. 1–2, IEEE, 2019.
[350] M. Riftadi, J. Oostenbrink, and F. Kuipers, "Gp4p4: Enabling self-programming networks," arXiv preprint arXiv:1910.00967, 2019.
[351] X. Gao, T. Kim, M. D. Wong, D. Raghunathan, A. K. Varma, P. G. Kannan, A. Sivaraman, S. Narayana, and A. Gupta, "Switch code generation using program synthesis," in Proceedings of the Annual conference of the ACM Special Interest Group on Data Communication on the applications, technologies, architectures, and protocols for computer communication, pp. 44–61, 2020.
[352] J. Gao, E. Zhai, H. H. Liu, R. Miao, Y. Zhou, B. Tian, C. Sun, D. Cai, M. Zhang, and M. Yu, "Lyra: A cross-platform language and compiler for data plane programming on heterogeneous asics," in Proceedings of the Annual conference of the ACM Special Interest Group on Data Communication on the applications, technologies, architectures, and protocols for computer communication, pp. 435–450, 2020.
[353] M. Hogan, S. Landau-Feibish, M. Tahmasbi Arashloo, J. Rexford, D. Walker, and R. Harrison, "Elastic switch programming with p4all," in Proceedings of the 19th ACM Workshop on Hot Topics in Networks, pp. 168–174, 2020.
[354] C. Zhang, J. Bi, Y. Zhou, A. B. Dogar, and J. Wu, "Hyperv: A high performance hypervisor for virtualization of the programmable data plane," in 2017 26th International Conference on Computer Communication and Networks (ICCCN), pp. 1–9, IEEE, 2017.
[355] M. Saquetti, G. Bueno, W. Cordeiro, and J. R. Azambuja, "P4vbox: Enabling p4-based switch virtualization," IEEE Communications Letters, vol. 24, no. 1, pp. 146–149, 2019.
[356] R. Parizotto, L. Castanheira, F. Bonetti, A. Santos, and A. Schaeffer-Filho, "Prime: Programming in-network modular extensions," in NOMS 2020-2020 IEEE/IFIP Network Operations and Management Symposium, pp. 1–9, IEEE, 2020.
[357] E. O. Zaballa, D. Franco, M. S. Berger, and M. Higuero, "A perspective on p4-based data and control plane modularity for network automation," in Proceedings of the 3rd P4 Workshop in Europe, pp. 59–61, 2020.
[358] R. Stoyanov and N. Zilberman, "Mtpsa: Multi-tenant programmable switches," in Proceedings of the 3rd P4 Workshop in Europe, pp. 43–48, 2020.
[359] S. Han, S. Jang, H. Choi, H. Lee, and S. Pack, "Virtualization in programmable data plane: A survey and open challenges," IEEE Open Journal of the Communications Society, vol. 1, pp. 527–534, 2020.
[360] E. C. Molero, S. Vissicchio, and L. Vanbever, "Hardware-accelerated network control planes," in Proceedings of the 17th ACM Workshop on Hot Topics in Networks, pp. 120–126, 2018.
[361] M. T. Arashloo, Y. Koral, M. Greenberg, J. Rexford, and D. Walker, "SNAP: stateful network-wide abstractions for packet processing," in Proceedings of the 2016 ACM SIGCOMM Conference, pp. 29–43, 2016.
[362] G. Sviridov, M. Bonola, A. Tulumello, P. Giaccone, A. Bianco, and G. Bianchi, "LODGE: Local decisions on global states in programmable data planes," in 2018 4th IEEE Conference on Network Softwarization and Workshops (NetSoft), pp. 257–261, IEEE, 2018.
[363] G. Sviridov, M. Bonola, A. Tulumello, P. Giaccone, A. Bianco, and G. Bianchi, "Local decisions on replicated states (LOADER) in programmable data planes: programming abstraction and experimental evaluation," arXiv preprint arXiv:2001.07670, 2020.
[364] S. Luo, H. Yu, and L. Vanbever, "Swing state: consistent updates for stateful and programmable data planes," in Proceedings of the Symposium on SDN Research, pp. 115–121, 2017.
[365] J. Xing, A. Chen, and T. E. Ng, "Secure state migration in the data plane," in Proceedings of the Workshop on Secure Programmable Network Infrastructure, pp. 28–34, 2020.
[366] L. Zeno, D. R. Ports, J. Nelson, and M. Silberstein, "Swishmem: Distributed shared state abstractions for programmable switches," in Proceedings of the 19th ACM Workshop on Hot Topics in Networks, pp. 160–167, 2020.
[367] S. Chole, A. Fingerhut, S. Ma, A. Sivaraman, S. Vargaftik, A. Berger, G. Mendelson, M. Alizadeh, S.-T. Chuang, I. Keslassy, et al., "drmt: Disaggregated programmable switching (extended version),"
[368] D. Kim, Y. Zhu, C. Kim, J. Lee, and S. Seshan, "Generic external memory for switch data planes," in Proceedings of the 17th ACM Workshop on Hot Topics in Networks, pp. 1–7, 2018.
[369] D. Kim, Z. Liu, Y. Zhu, C. Kim, J. Lee, V. Sekar, and S. Seshan, "TEA: enabling state-intensive network functions on programmable switches," in Proceedings of the 2020 ACM SIGCOMM Conference, 2020.
[370] T. Mai, S. Garg, H. Yao, J. Nie, G. Kaddoum, and Z. Xiong, "In-network intelligence control: Toward a self-driving networking architecture," IEEE Network, vol. 35, no. 2, pp. 53–59, 2021.
[371] Y. Shi, M. Wen, and C. Zhang, "Incremental deployment of programmable switches for sketch-based network measurement," in 2020 IEEE Symposium on Computers and Communications (ISCC), pp. 1–7, IEEE, 2020.
[372] J. Cao, Y. Zhou, Y. Liu, M. Xu, and Y. Zhou, "Turbonet: Faithfully emulating networks with programmable switches," in 2020 IEEE 28th International Conference on Network Protocols (ICNP), pp. 1–11, IEEE, 2020.
[373] S. Chole, A. Fingerhut, S. Ma, A. Sivaraman, S. Vargaftik, A. Berger, G. Mendelson, M. Alizadeh, S.-T. Chuang, I. Keslassy, et al., "dRMT: disaggregated programmable switching," in Proceedings of the Conference of the ACM Special Interest Group on Data Communication, pp. 1–14, 2017.

[374] R. Pagh and F. F. Rodler, "Cuckoo hashing," J. Algorithms, vol. 51, p. 122–144, May 2004.
[375] M. Baldi, "dapipe a data plane incremental programming environment," in 2019 ACM/IEEE Symposium on Architectures for Networking and Communications Systems (ANCS), pp. 1–6, IEEE, 2019.
[376] R. Amin, M. Reisslein, and N. Shah, "Hybrid SDN networks: a survey of existing approaches," IEEE Communications Surveys & Tutorials, vol. 20, no. 4, pp. 3259–3306, 2018.
[377] J. Zhang and A. Moore, "Traffic trace artifacts due to monitoring via port mirroring," in 2007 Workshop on End-to-End Monitoring Techniques and Services, pp. 1–8, IEEE, 2007.
[378] SONiC, "Software for open networking in the cloud," 2020. [Online]. Available: https://azure.github.io/SONiC/.
[379] S. Choi, B. Burkov, A. Eckert, T. Fang, S. Kazemkhani, R. Sherwood, Y. Zhang, and H. Zeng, "FBOSS: building switch software at scale," in Proceedings of the 2018 Conference of the ACM Special Interest Group on Data Communication, pp. 342–356, 2018.
[380] L. Linguaglossa, S. Lange, S. Pontarelli, G. Rétvári, D. Rossi, T. Zinner, R. Bifulco, M. Jarschel, and G. Bianchi, "Survey of performance acceleration techniques for network function virtualization," Proceedings of the IEEE, vol. 107, no. 4, pp. 746–764, 2019.
[381] P. Shantharama, A. S. Thyagaturu, and M. Reisslein, "Hardware-accelerated platforms and infrastructures for network functions: A survey of enabling technologies and research studies," IEEE Access, vol. 8, pp. 132021–132085, 2020.
[382] Nick McKeown, "Creating an End-to-End Programming Model for Packet Forwarding." [Online]. Available: https://lwn.net/Articles/828056/.
[383] J. Krude, J. Hofmann, M. Eichholz, K. Wehrle, A. Koch, and M. Mezini, "Online reprogrammable multi tenant switches," in Proceedings of the 1st ACM CoNEXT Workshop on Emerging in-Network Computing Paradigms, pp. 1–8, 2019.
[384] D. Hancock and J. Van der Merwe, "Hyper4: Using p4 to virtualize the programmable data plane," in Proceedings of the 12th International on Conference on emerging Networking EXperiments and Technologies, pp. 35–49, 2016.
[385] T. Issariyakul and E. Hossain, "Introduction to network simulator 2 (ns2)," in Introduction to network simulator NS2, pp. 1–18, Springer, 2009.
[386] Stanford, "Reproducing Network Research." [Online]. Available: https://reproducingnetworkresearch.wordpress.com/.
[387] Mininet, "An Instant Virtual Network on your Laptop (or other PC)." [Online]. Available: http://mininet.org/.
[388] N. Handigol, B. Heller, V. Jeyakumar, B. Lantz, and N. McKeown, "Reproducible network experiments using container-based emulation," in Proceedings of the 8th international conference on Emerging networking experiments and technologies, pp. 253–264, 2012.
[389] H. Kim, X. Chen, J. Brassil, and J. Rexford, "Experience-driven research on programmable networks," ACM SIGCOMM Computer Communication Review, vol. 51, no. 1, pp. 10–17, 2021.
[390] Princeton, "P4Campus: framework, applications, and artifacts.)." [Online]. Available: https://p4campus.cs.princeton.edu/.
[391] H. Kim and N. Feamster, "Improving network management with software defined networking," IEEE Communications Magazine, vol. 51, no. 2, pp. 114–119, 2013.
[392] N. Feamster and J. Rexford, "Why (and how) networks should run themselves," arXiv preprint arXiv:1710.11583, 2017.
[393] D. D. Clark, C. Partridge, J. C. Ramming, and J. T. Wroclawski, "A knowledge plane for the internet," in Proceedings of the 2003 conference on Applications, technologies, architectures, and protocols for computer communications, pp. 3–10, 2003.
[394] A. Mestres, A. Rodriguez-Natal, J. Carner, P. Barlet-Ros, E. Alarcón, M. Solé, V. Muntés-Mulero, D. Meyer, S. Barkai, M. J. Hibbett, et al., "Knowledge-defined networking," ACM SIGCOMM Computer Communication Review, vol. 47, no. 3, pp. 2–10, 2017.

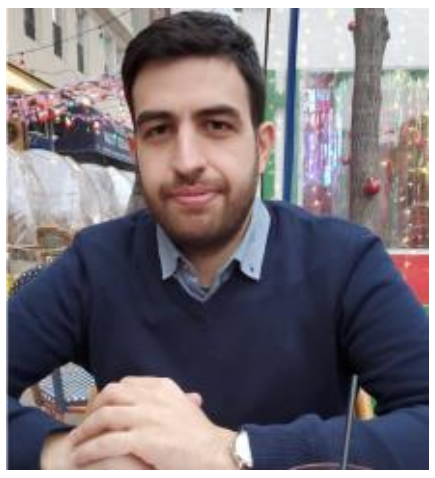

ELIE F. KFOURY is currently a PhD student in the College of Engineering and Computing at the University of South Carolina, USA. He is a member of the CyberInfrastructure Lab (CI Lab), where he developed training materials for virtual labs on high-speed networks (TCP congestion control, WAN, performance measuring, buffer sizing), cybersecurity, and routing protocols. He previously worked as a research and teaching assistant in the computer science and ICT departments at the American University of Science and Technology in Beirut. His research focuses on telecommunications, network security, Blockchain, Internet of Things (IoT), and P4 programmable switches.

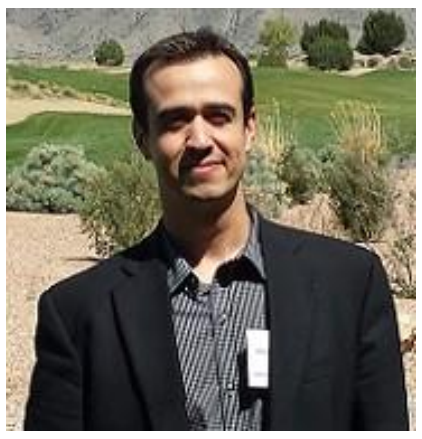

JORGE CRICHIGNO is an Associate Professor in the College of Engineering and Computing at the University of South Carolina (USC) and the director of the Cyberinfrastructure Lab at USC. He has over 15 years of experience in the academic and industry sectors. Dr. Crichigno's research focuses on P4 programmable switches, implementation of high-speed networks, network security, TCP optimization, offloading functionality to programmable switches, and IoT devices. His work has been funded by private industry and U.S. agencies such as the National Science Foundation (NSF), the Department of Energy, and the Office of Naval Research (ONR). He received his Ph.D. in Computer Engineering from the University of New Mexico in Albuquerque, USA, in 2009.

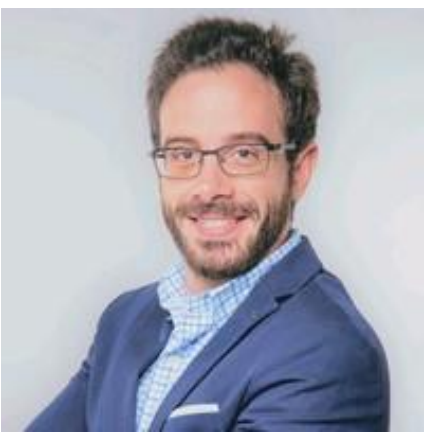

ELIAS BOU-HARB is currently the Director of the Cyber Center For Security and Analytics at UTSA, where he leads, co-directs and co-organizes university-wide innovative cyber security research, development and training initiatives. He is also an Associate Professor at the department of Information Systems and Cyber Security specializing in operational cyber security and data science as applicable to national security challenges. Previously, he was a senior research scientist at Carnegie Mellon University (CMU) where he contributed to federally-funded projects related to critical infrastructure security and worked closely with the Software Engineering Institute (SEI). He is also a permanent research scientist at the National Cyber Forensic and Training Alliance (NCFTA) of Canada; an international organization which focuses on the investigation of cyber-crimes impacting citizens and businesses. Dr. Bou-Harb holds a Ph.D. degree in computer science from Concordia University in Montreal, Canada, which was executed in collaboration with Public Safety Canada, Industry Canada and NCFTA Canada. His research and development activities and interests focus on operational cyber security, attacks' detection and characterization, malware investigation, cyber security for critical infrastructure and big data analytics. Dr. Bou-Harb has authored more than 90 refereed publications in leading security and data science venues, has acquired state and federal cyber security research grants valued at more than $4M, and is the recipient of 5 best research paper awards, including the prestigious ACM's best digital forensics research paper.